\documentclass[useAMS,usenatbib]{mn2e}
\usepackage{graphicx,times}

\title[Direct imaging with highly diluted
apertures II]{Direct imaging with highly diluted apertures. II:
Properties of the point spread function of a
hypertelescope}
\author[F. Patru et al.]{F. Patru$^{1,2}$, N. Tarmoul$^{1}$, D. Mourard$^{1}$
and O. Lardi\`ere$^{3}$ \\
 $^{1}$Laboratoire H. FIZEAU, UMR CNRS 6525 - UNS, OCA -
 Avenue Copernic, 06130 Grasse, France\\
   $^{2}$Laboratoire d'Astrophysique de Grenoble (LAOG), 414 Rue de la Piscine, Domaine Universitaire, 38400 Saint-Martin d'H\`eres,
   France\\
  $^{3}$Adaptive Optics Lab, Engineering Lab Wing B133, University of Victoria, PO Box 3055 STN CSC, Victoria, BC,
Canada V8W 3P6}

\begin{document}
\date{Accepted. Received}

\pagerange{\pageref{firstpage}--\pageref{lastpage}} \pubyear{2008}

\maketitle

\label{firstpage}

\begin{abstract}
In the future, optical stellar interferometers will provide true
images thanks to larger number of telescopes and to advanced
cophasing subsystems. These conditions are required to have
sufficient resolution elements (resel) in the image and to provide
direct images in the hypertelescope mode. It has already been shown
\citep{Lardiere 2007} that hypertelescopes provide snapshot images
with a significant gain in sensitivity without inducing any loss of
the useful field of view for direct imaging applications. This paper aims at studying the properties of the point spread
functions of future large arrays using the hypertelescope mode.
Numerical simulations have been performed and criteria have been
defined to study the image properties. It is shown that the choice
of the configuration of the array is a trade-off between the
resolution, the halo level and the field of view. A regular pattern of the array of telescopes optimizes the image quality (low halo level and maximum encircled energy in the central peak), but decreases the useful field of view.
Moreover, a non redundant array is less sensitive to the space aliasing effect than a redundant array.
\end{abstract}


\begin{figure}
\begin{center}
\includegraphics[width=85mm]{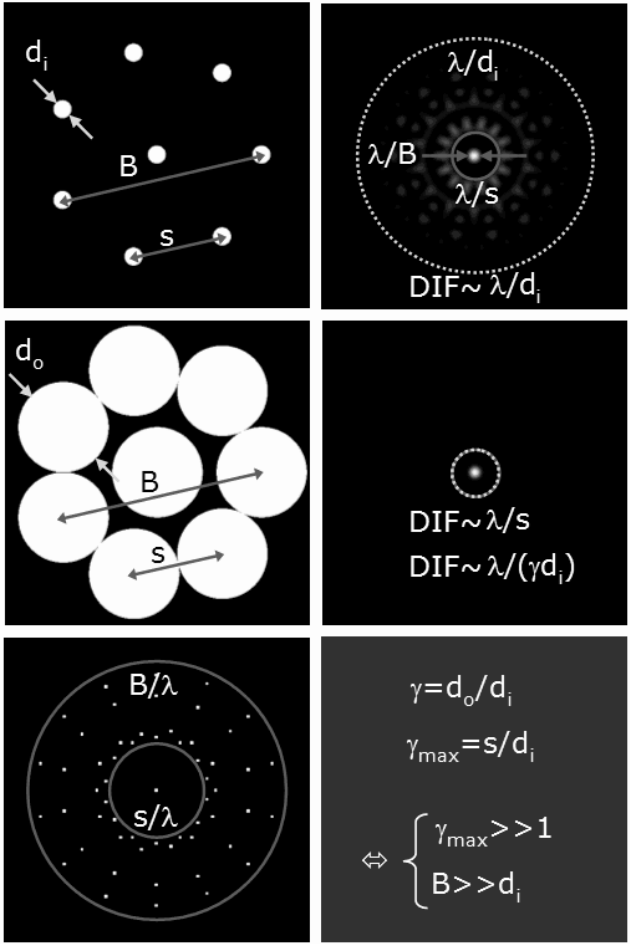}
\end{center}\caption{Field of view and direct imaging in hypertelescope mode.
An example of an input pupil (top-left) and a densified pupil
(middle-left) is shown. We have represented the (u,v) plane coverage
(bottom-left) considering that the finite size of the individual
telescopes has been neglected. It can be seen that the high spatial
frequencies are distributed over the interval $[s/\lambda,
B/\lambda]$, with $s$ the smallest baseline and $B$ the largest
baseline of the array. The maximum densification factor
$\gamma_{max}$ equals to $s/d_i$ (bottom-right). The Fizeau
(top-right) and densified (middle-right) point spread functions (PSF) are
shown. As predicted, we observe a central peak inside a clean zone,
called the Clean Field ($CLF$), where the contribution of the
side-lobes'halo is minor (see \citep{Lardiere 2007} for the
different definitions of fields). It is surrounded by a
non-negligible halo of speckles outside the $CLF$ diameter
($CLF=\lambda/s$). The Fizeau PSF is limited by the Airy envelope of
an input sub-aperture, which corresponds to the coupled field
($CF=\lambda/d_i$). The pupil densification reduces the Direct
Imaging Field (DIF) to match the CLF. The central peak is
intensified by a factor $\gamma^2$. } \label{fig:dif_cf_clf}
\end{figure}

\section{Introduction}

Future large interferometers \citep{Labeyrie 2008} need a large
number of telescopes and an active cophasing system, to provide
images with sufficient sensitivity. If both conditions are met,
snapshot imaging can be used in the hypertelescope mode
\citep{labey96} and a multi-axial beam combiner seems to be the best
solution. If the entrance pupil is highly diluted, the
hypertelescope mode improves the Fizeau mode, with a high
sensitivity gain without any loss of the useful field of
view for direct imaging applications. This useful field,
where a direct image can be correctly recovered, is called the Clean
Field \citep{Lardiere 2007}.

Direct imaging has two main features. In a conventional sense, the
goal is to provide snapshot images that could then be post-processed
by deconvolution techniques. These images give valuable
regularization constraints for the a posteriori astrophysical
analysis process. Furthermore, direct imaging is well suitable to
feed the entrance plane of focal instruments such as coronagraphic
devices or integral field spectrometer. In this paper, we
concentrate our analysis on the raw images without considering the
deconvolution techniques or the coupling with a focal instrument.

Keeping in mind the researches on the ways to optimize the
imaging capabilities of a hypertelescope, this paper aims at
characterizing the point spread functions of typical future large
arrays. For this purpose, we have developed a numerical simulation,
called HYPERTEL, which first simulates direct images (Sect.
\ref{hypertel}) and then analyses the densified PSF properties by
defining different quantitative criteria (Sect. \ref{psf}).
Then, we study the impact of the array configuration
(geometry of the array and number of apertures) and of the
recombination mode (Sect. \ref{psf_ppty}). Finally, we
establish the relations between the astrophysical parameters of the
science object and the main parameters of the hypertelescope (Sect.
\ref{psf biases}).

\section{Simulating direct images}\label{hypertel}

\subsection{The input parameters}

The input parameters are the wavelength, the characteristics of the science object, the array configuration and the recombination mode.\\
We assume a perfectly cophased array, without any degradation in the
image due to atmospheric turbulence or instrumental bias. To
simplify the study, we restrict ourselves to the monochromatic case.

\subsubsection{The array configuration}

The array is made of $N_T$ identical sub-apertures of index $k$, defined by their
positions $(u_p(k),v_p(k))$ in the input pupil plane and by their diameter
$d_i$. We note $s$ (resp. $B$) the smallest (resp. largest) baseline of the array. The maximum angular resolution of the array, following the Rayleigh criterion, is given by the highest baseline :

\begin{equation}\label{eq_resel}
resel \simeq \frac{\lambda}{B}
\end{equation}

\subsubsection{The science object}\label{hypertel_object}

The object is defined by a monochromatic brightness map, $N_{pxl}$ pixels wide.
The angular extent of this map is equal to the object diameter $\theta_{obj}$.
The angular size of a pixel $\theta_{pxl}$, i.e. the smallest angular
element seen on the sky, should respect the Shannon criterion:

\begin{equation}
\theta_{pxl}=\frac{\theta_{obj}}{N_{pxl}}<\frac{resel}{2}
\end{equation}

In practice, we choose $\theta_{pxl}<resel/6$. 

Thus, $N_{pxl}>\frac{6B}{\lambda} \cdot \theta_{obj}$.

The object is considered as composed of elementary incoherent
sources, corresponding to the $N_{pxl}^2$ pixels of the brightness
map. Each elementary source of index $m$ is defined by its
coordinates $(X_{obj}(m),Y_{obj}(m))$ and by its intensity
$I_{obj}(m)$.

For each elementary source, the off-axis position is
defined on the two axis by:

\begin{eqnarray}
\theta_{X\,obj}(m)=\left(X_{obj}(m)-\frac{N_{pxl}}{2}\right)\,\theta_{pxl}\nonumber\\
\theta_{Y\,obj}(m)=\left(Y_{obj}(m)-\frac{N_{pxl}}{2}\right)\,\theta_{pxl}
\end{eqnarray}

The image is obtained as the sum of the sub-images of each incoherent elementary source.
As the input sub-pupils are very
diluted, we neglect the variation of the object in the spatial frequency domains [$\frac{B-d_i}{\lambda}$;$\frac{B+d_i}{\lambda}$] accessible in certain recombination schemes.

\subsubsection{The recombination mode}

Pupil densification increases by a factor
$\gamma$ the relative size of the beams and keeps the relative
positions of the sub-pupil centres (Fig. \ref{fig:dif_cf_clf}). This pseudo-homothetic transformation does not affect the interferometric pattern in the image and correctly recovers the high resolution information. The diffraction envelope is reduced, so as to concentrate all the flux in the useful field.

The alternative concept of IRAN \citep{Vakili 2004a} combines the beams by superimposing the images of the sub-pupils with small tilts in the image plane. A direct image is obtained in the recombined pupil plane.

The envelope shape is a Bessel function in Pupil
Densification (DP mode) and is a flat field in IRAN mode. The
envelope width decreases as the densification factor increases. The
value of the latter is chosen between 1 (Fizeau mode) and
$\gamma_{max}$ (maximum densification).

The maximum densification factor depends on the smallest baseline $s$ and on the aperture diameter $d_i$:
\begin{equation}
\gamma_{max}=\frac{s}{d_i}  
\end{equation}

The pupil densification amplifies the intensity of the  signal by a
factor $\gamma^2$. In the IRAN mode, the densification
factor can not be larger than $\gamma_{max}/2$ due to the
diffraction of the sub-pupils \citep{Lardiere 2007}. Thus the
sensitivity gain is reduced by a factor 4 and the DIF is enlarged by
a factor 2, compared to the DP mode.

We mainly focus in this paper on the pupil densification
(DP) for the recombination mode. We also make comparison with the
image densification (IRAN).

\subsection{The image calculation}\label{im_calc}

\begin{figure*}
\begin{center}
\includegraphics[width=120mm]{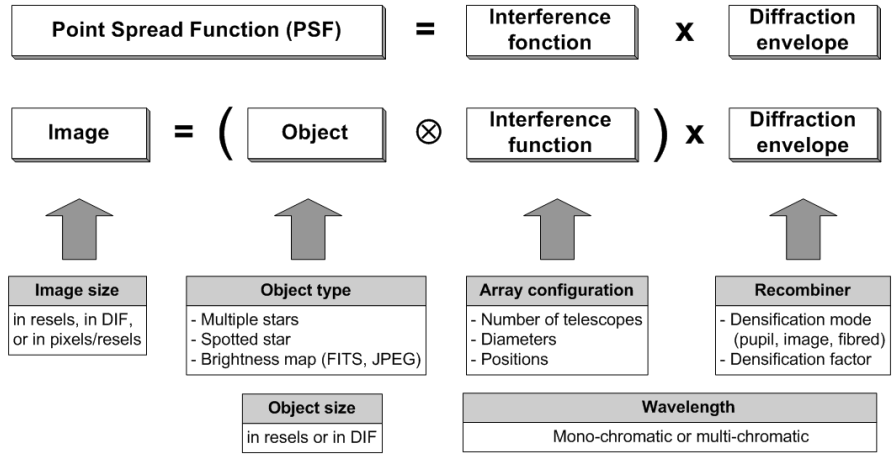}
\end{center}\caption{Main structure of HYPERTEL.
The Point Spread Function is defined as the product of the interference function with the diffraction envelope. The direct image is defined as the convolution of the object brightness distribution by the interference function, the result being multiplied by the diffraction envelope.}
\label{fig:prog_bloc_diag}
\end{figure*}

The principle of the image calculation is schematically described in
Fig.~\ref{fig:prog_bloc_diag}. In the image plane, each pixel of coordinates $(x,y)$ has an intensity of $I(x,y)$.
\\

The Point Spread Function \citep{Lardiere 2007} is defined as the product of the interference function $I_0(x,y)$ (function of the array pattern) with the diffraction envelope $A_0(x,y,\gamma)$ (function of the recombination mode and of the densification factor).

\begin{equation}
I_{PSF}(x,y)\approx A_0(x,y,\gamma) \times I_0(x,y)\ ,
\label{equ:pseudo_conv}
\end{equation}

with:

\begin{equation}
I_0(x,y)=\bigg| \sum_{k=1}^{N_T} \,
e^{\frac{-2 i \pi}{\lambda} (x \cdot u_p(k) + y \cdot v_p(k))} \bigg|^2
\label{equ:interf_ft}
\\
\end{equation}

The direct image of an astrophysical object is defined as the object brightness distribution $O(\theta_{X\,obj},\theta_{Y\,obj})$ convolved by the interference function, the result being multiplied by the diffraction envelope.

\begin{equation}
I(x,y)\approx A_0(x,y,\gamma) \cdot O(\theta_{X\,obj},\theta_{Y\,obj}) \otimes I_0(x,y)\,
\label{equ:pseudo_conv}
\\
\end{equation}

Finally, the image computed by HYPERTEL is the sum of the sub-images of
each incoherent elementary source of the object (as defined in Sect. \ref{hypertel_object}).

\begin{eqnarray}
I(x,y)= A_0(x,y,\gamma) \cdot \bigg|
\sum_{k=1}^{N_T} \,
e^{\frac{-2 i \pi}{\lambda} (x \cdot u_p(k) + y \cdot v_p(k))} \cdot \nonumber\\
\sum_{m=1}^{N_{pxl}^2} \,
I_{obj}(m) \cdot
e^{\frac{2 i \pi}{\lambda} \psi(m,k)}
\bigg|^2
\end{eqnarray}


We denote $\psi(m,k)$ the phase delay seen by the $k^{th}$ sub-aperture due to the
position of the elementary source $m$. Indeed each elementary source
is seen under a different angle by each sub-aperture of the
interferometer. Hence, $\psi(m,k)$ is given by:

\begin{equation}
\psi(m,k)=\theta_{X\,obj}(m).u_p(k)+\theta_{Y\,obj}(m).v_p(k)       
\end{equation}

\section{Definition of the characteristics of the point spread function (PSF)}\label{psf}

\subsection{Input and output pupils' parameters}

As the computed images are a function of the characteristics
of the entrance pupil, we first define two parameters related to the
interferometer configuration. The entrance (resp. densified) pupil
filling rate $\tau_i$ (resp.$\tau_o$) is defined as the ratio
between the total surface area of the input (resp. output)
pupil and the surface area of an input (resp. output)
sub-pupil, $d_i$ (resp $d_o$) being the diameter of the latter:

\begin{equation}
\tau_i=\frac{S_{input~sub-pupils}}{S_{input~pupil}}=N_T\frac{d_i^2}{(B+d_i)^2}
\end{equation}

\begin{equation}
\tau_o=\frac{S_{output~sub-pupils}}{S_{output~pupil}}=N_T
\frac{d_o^2}{(B+d_o)^2}=N_T\left(\frac{\gamma\,d_i}{B+\gamma\,d_i}\right)^2
\label{tau}
\end{equation}

\subsection{Field of view parameters}

The definitions of the different fields of view for a hypertelescope
have been extensively studied by \cite{Lardiere 2007}. We just
recall here the important definitions of these different fields. We
distinguish the CLean Field of view (CLF), the Direct Imaging Field
of view (DIF) and the Coupled field of view (CF). They are
illustrated on figure \ref{fig:dif_cf_clf}.

\begin{equation}
CLF=\frac{\lambda}{s} \,(radians)\,=\frac{B}{s} \,(resels) \\
\end{equation}

\begin{equation}\label{equ:dif}
DIF \simeq \frac{\lambda}{(\gamma-1)\:d_i}     \\
\end{equation}

\begin{equation}
CF=\frac{\lambda}{d_i}          \\
\end{equation}

The Clean Field is physically related to the sampling of the (u,v)
plane and according to Nyquist-Shannon sampling theorem, it is
defined by the smallest baseline $s$. This definition is also
interesting on calculating the number of resels in the Clean Field.

The coupled field (CF) is imposed by the size of a sub-aperture.
The Direct Imaging Field (DIF) depends on the densification mode.
The DIF width is still obviously smaller than the CF width.

\begin{figure}
\begin{center}
\includegraphics[width=55mm]{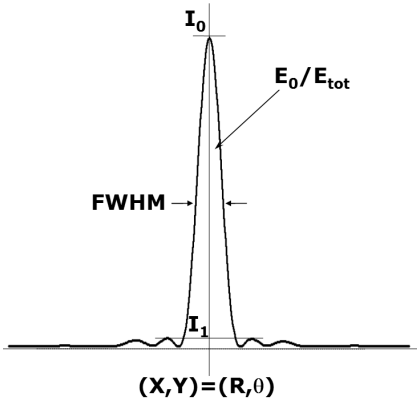}
\end{center}\caption{Imaging parameters of the densified point spread function.
Coordinates ($(X,Y)=(R,\theta)$), on-axis intensity ($I_0$),
encircled energy ($E_0/E_{tot}$), full-width at half-maximum
($FWHM$) of the central peak, maximum halo level ($I_1/I_0$).}
\label{fig:critere_psf}
\end{figure}

\subsection{Astrometric criteria}

In the direct image, the position of each interference peak (central
peak and side-lobes) is given by the coordinates of its photocentre
(Fig. \ref{fig:critere_psf}). The full width at half maximum (FWHM)
of the central peak corresponds to the smallest resolution element
(resel), given by equation \ref{eq_resel}. It characterizes the
sharpness of the image. It depends not only on the
wavelength and on the maximum baseline, but also on the geometry of
the array.

\subsection{Photometric criteria}

The on-axis intensity $I_0$ is equal to the height of the central
peak. The encircled energy is defined as the ratio of the fraction
of energy contained in the central peak $E_0$ to the total energy in
the image $E_{tot}$:

\begin{equation} \label{eq:p encircled}
\frac{E_0}{E_{tot}}=\frac{2\pi}{E_{tot}} \int_0^{\theta_0} I(\rho)
\,\rho \,d\rho
\end{equation}

where $\theta_0$ corresponds to the first minimum from the center of
the field (Fig. \ref{fig:critere_psf}).

\subsection{Halo level criteria}
We also define a criterion to estimate the contribution of
the halo surrounding the central peak. The maximum of the halo level
is defined as the ratio between $I_1$ the intensity of the highest
side-lobe inside the CLF and $I_0$ the intensity of the central peak
(Fig. \ref{fig:critere_psf}).

\begin{equation} \label{eq:halo_max}
Maximum~halo~level\,=\,\frac{I_1}{I_0}
\end{equation}

\section[]{Densified PSF properties}\label{psf_ppty}

\subsection{Influence of the array geometry}\label{influ_array}

\begin{figure*}
\begin{center}
\begin{tabular}{cccccc}

CARLINA & KEOPS & OVLA & ELSA\\

\includegraphics[width=39mm]{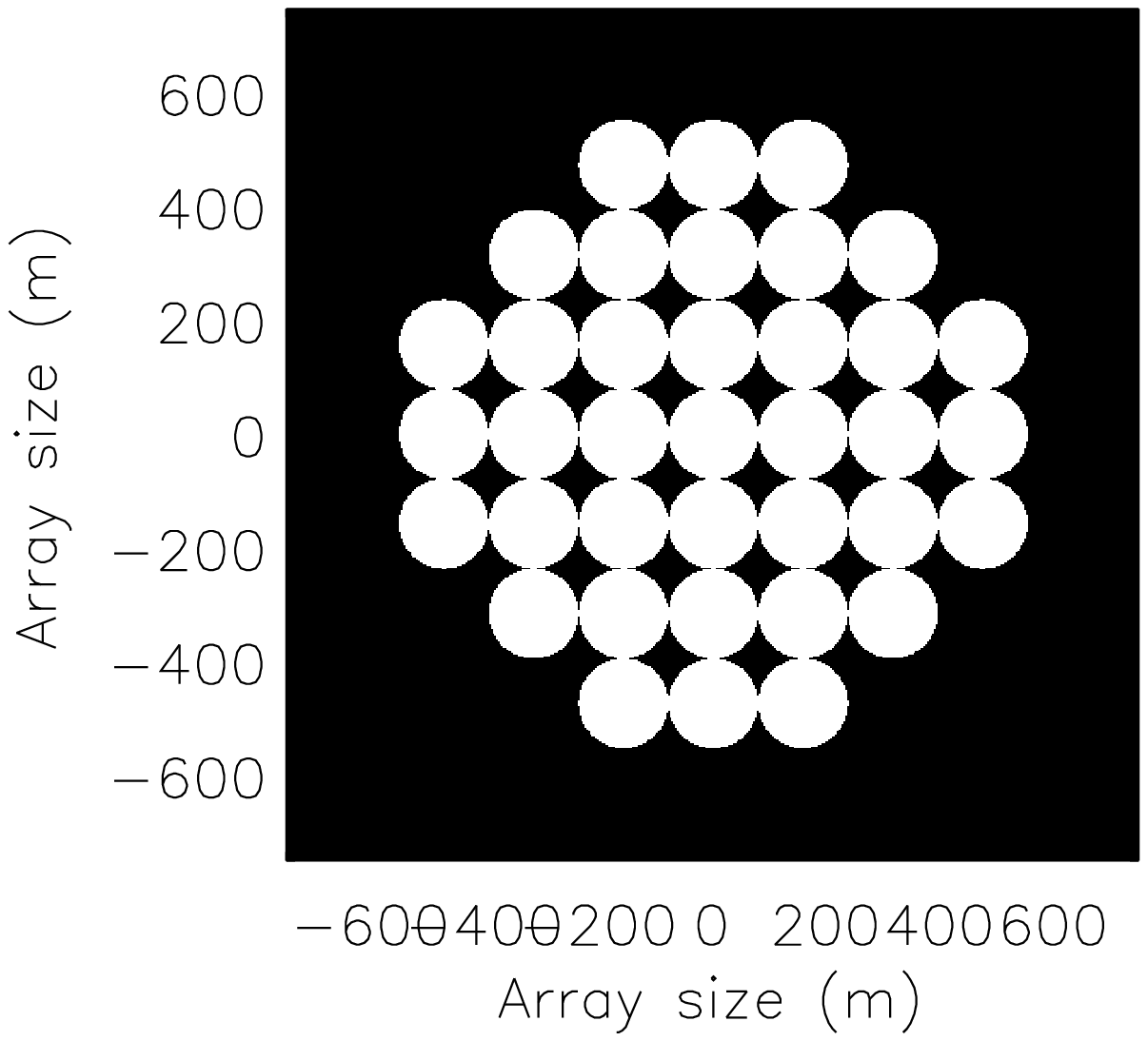} &
\includegraphics[width=39mm]{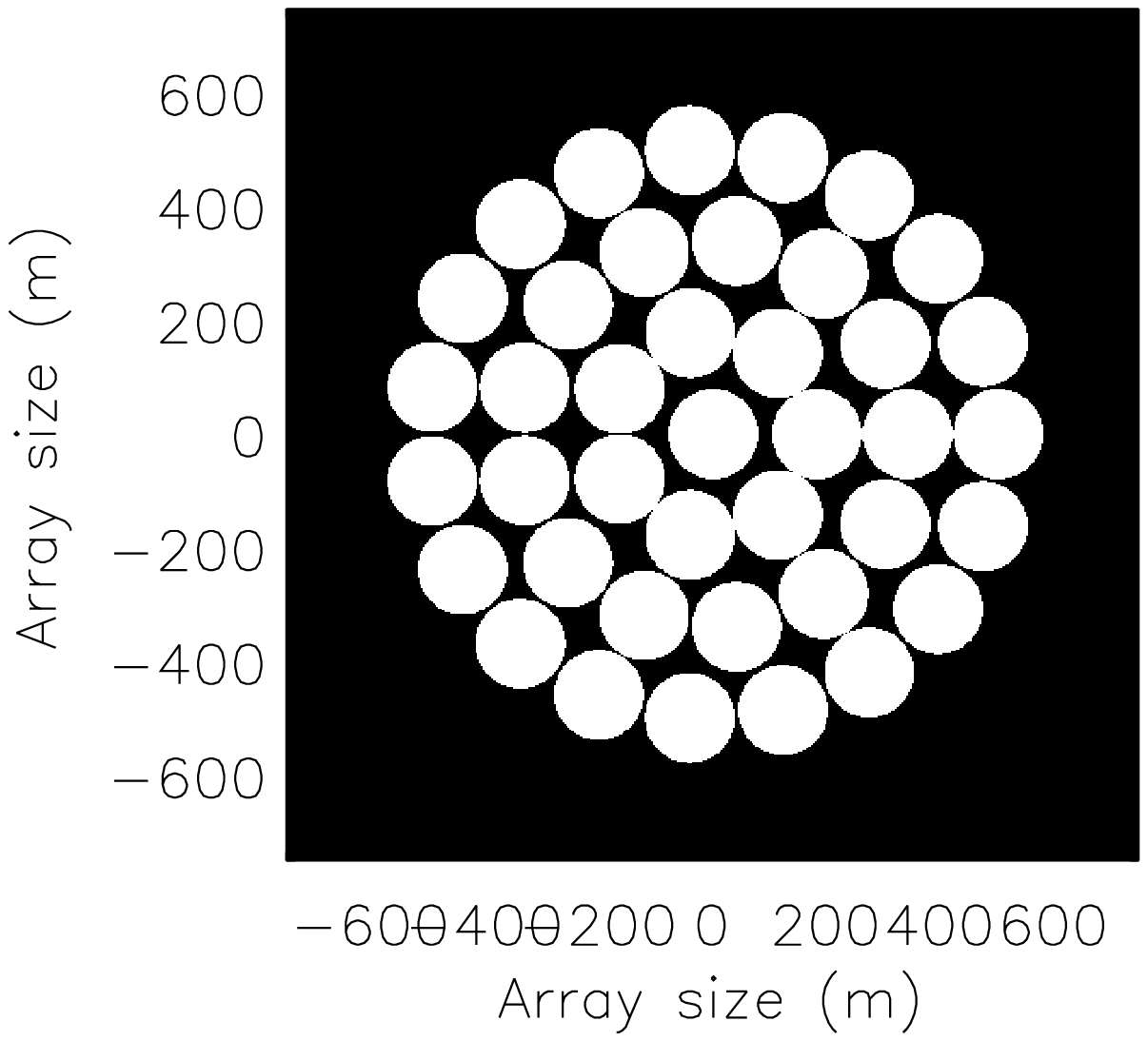} &
\includegraphics[width=39mm]{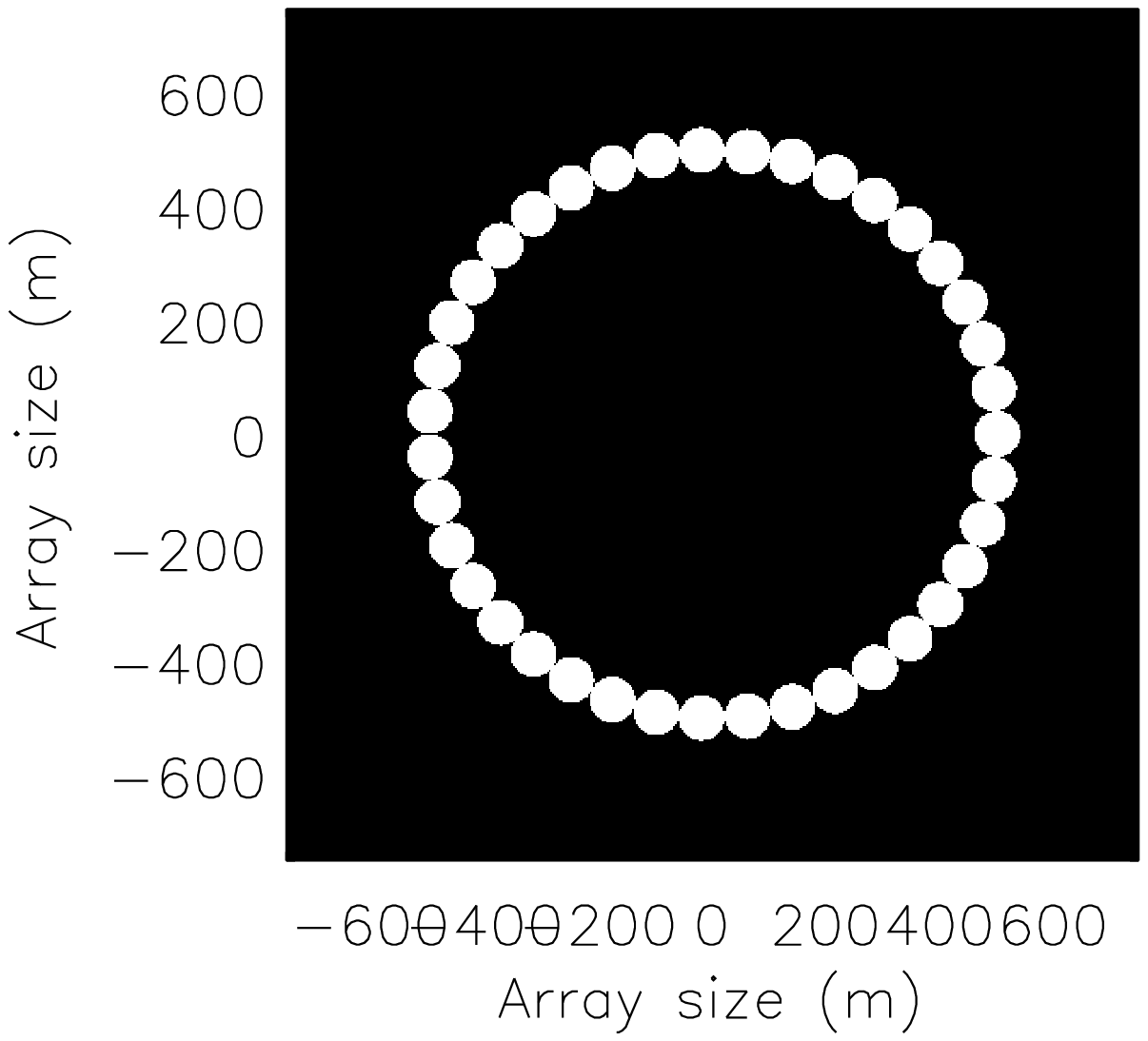} &
\includegraphics[width=39mm]{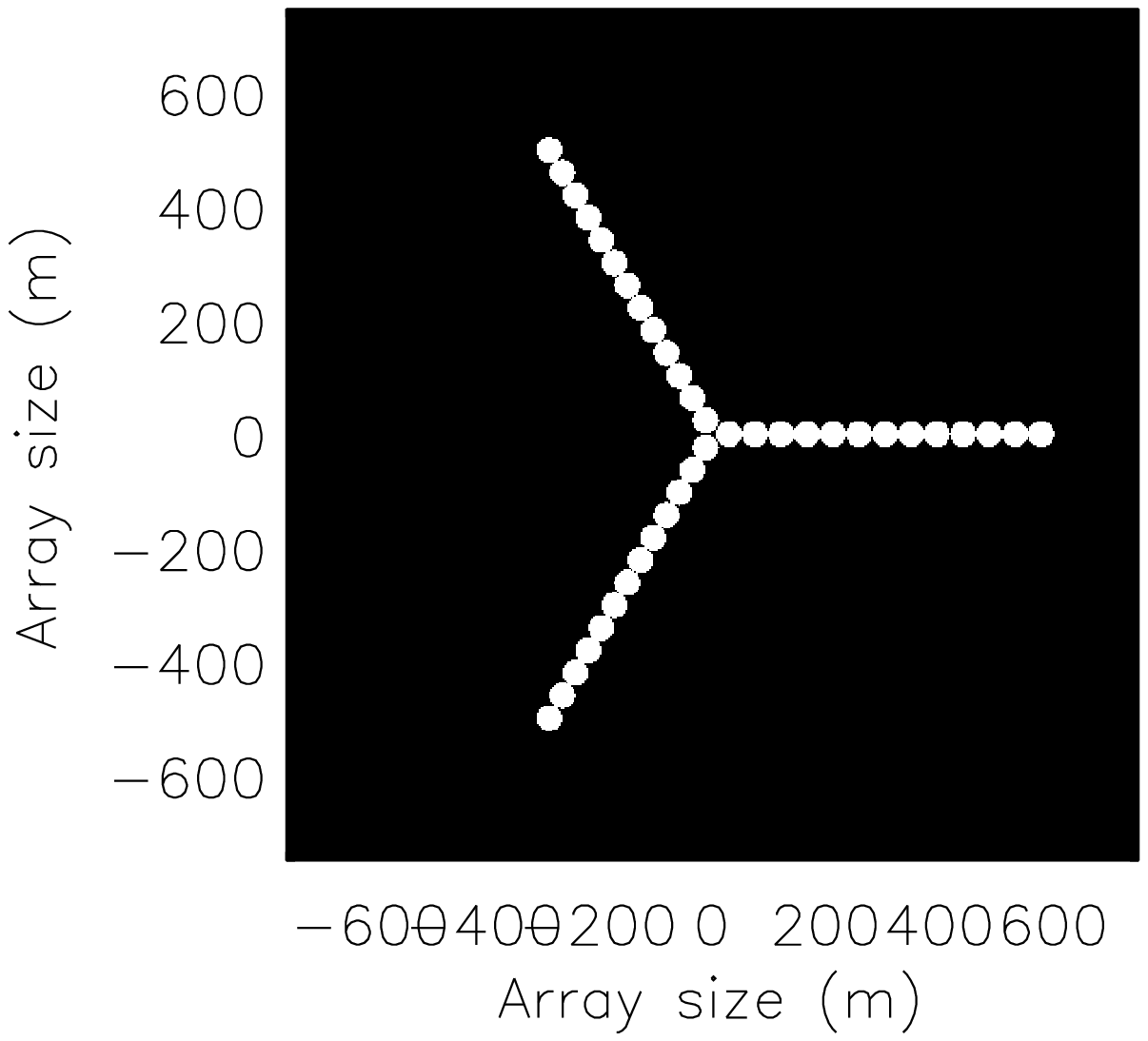} \\


\includegraphics[width=39mm]{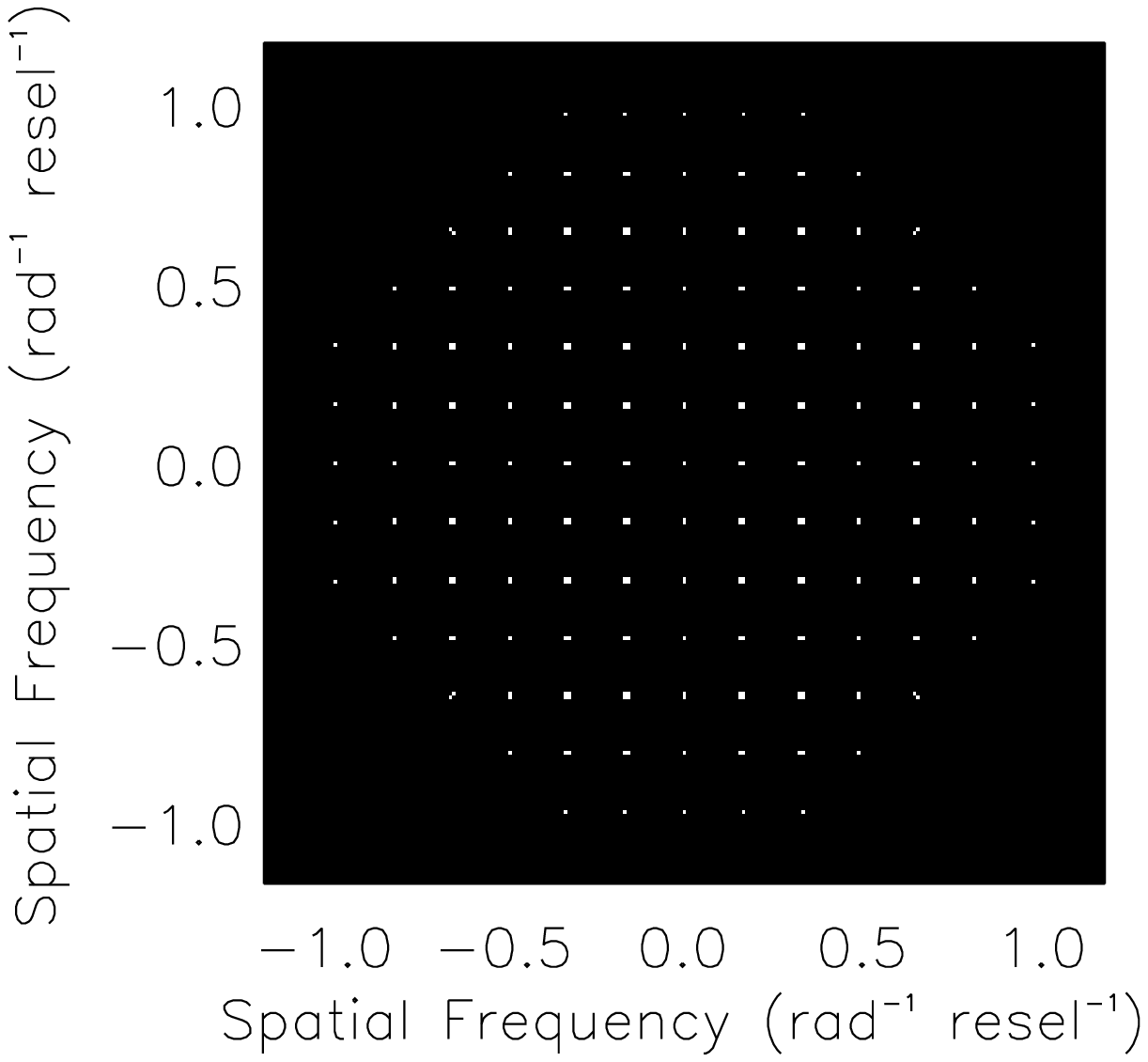} &
\includegraphics[width=39mm]{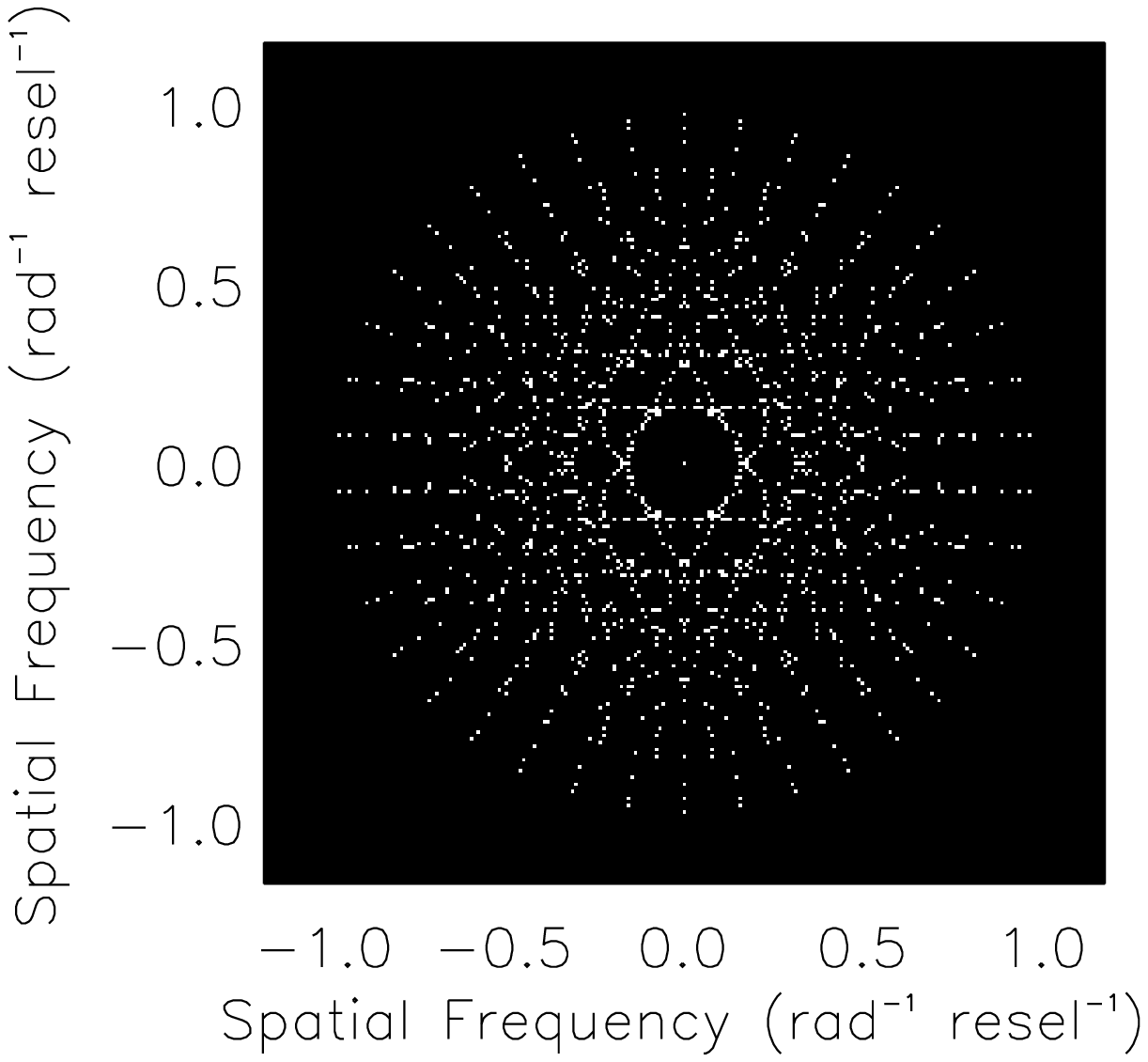} &
\includegraphics[width=39mm]{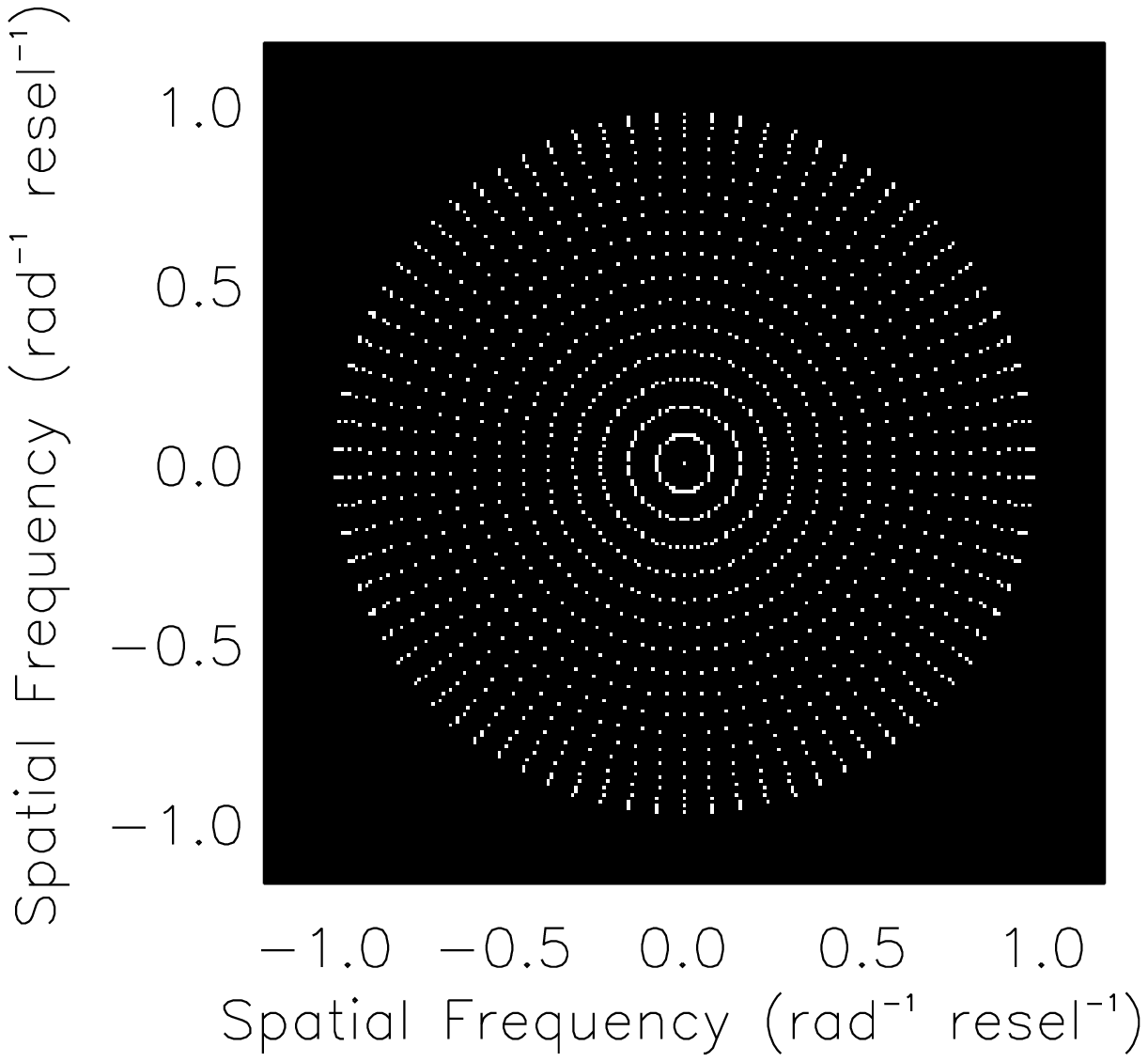} &
\includegraphics[width=39mm]{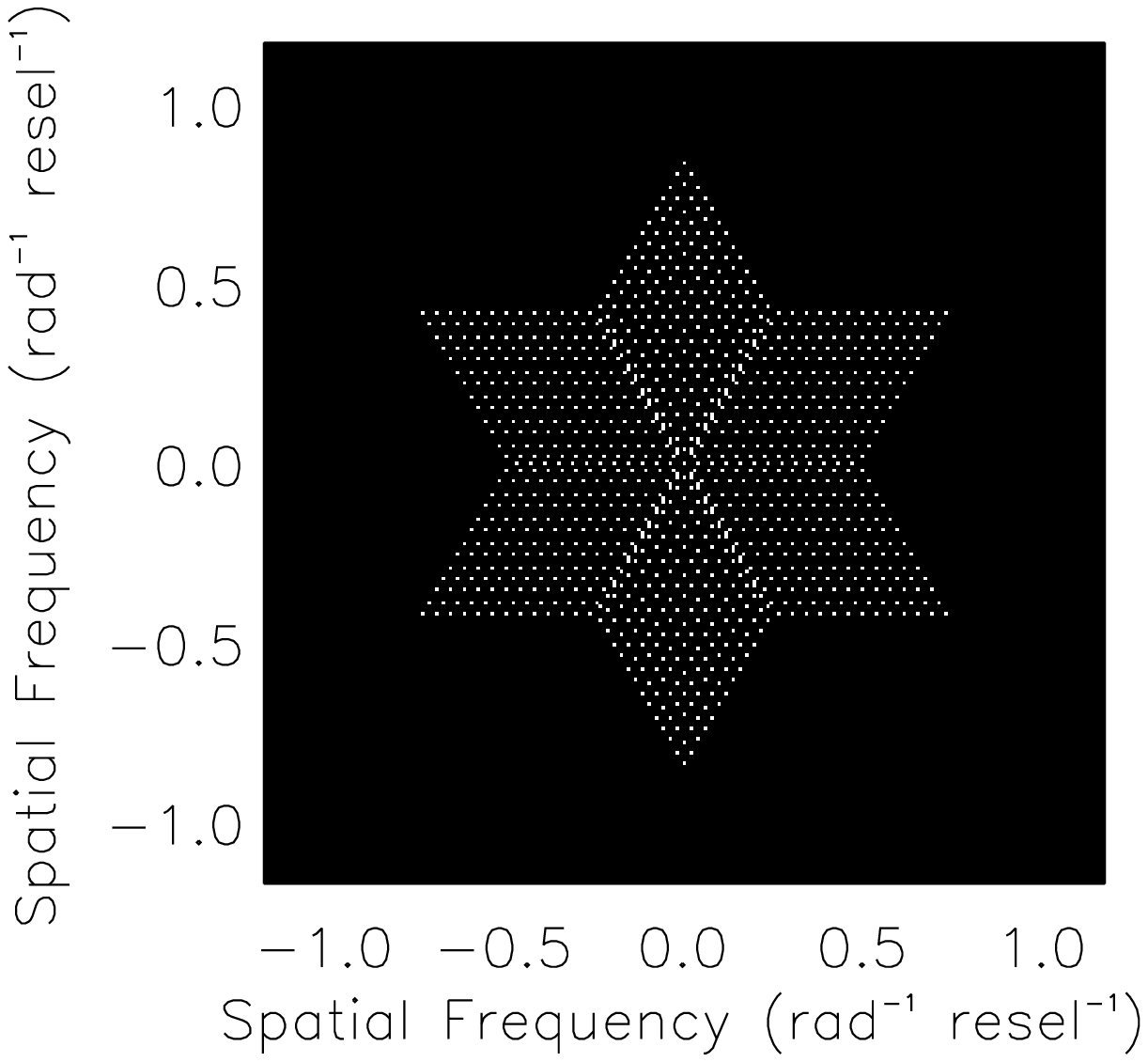} \\

\hline

\includegraphics[width=39mm]{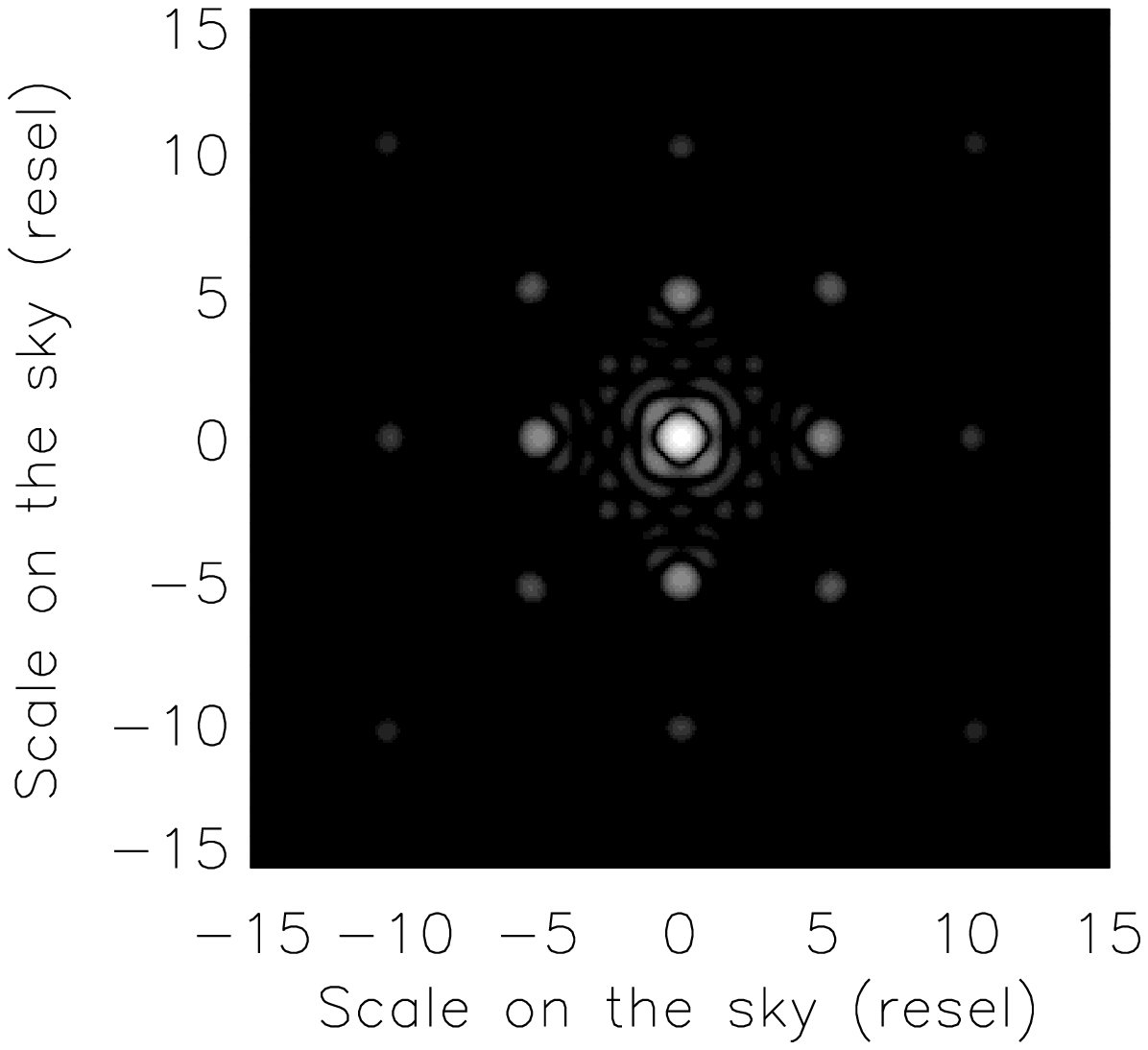} &
\includegraphics[width=39mm]{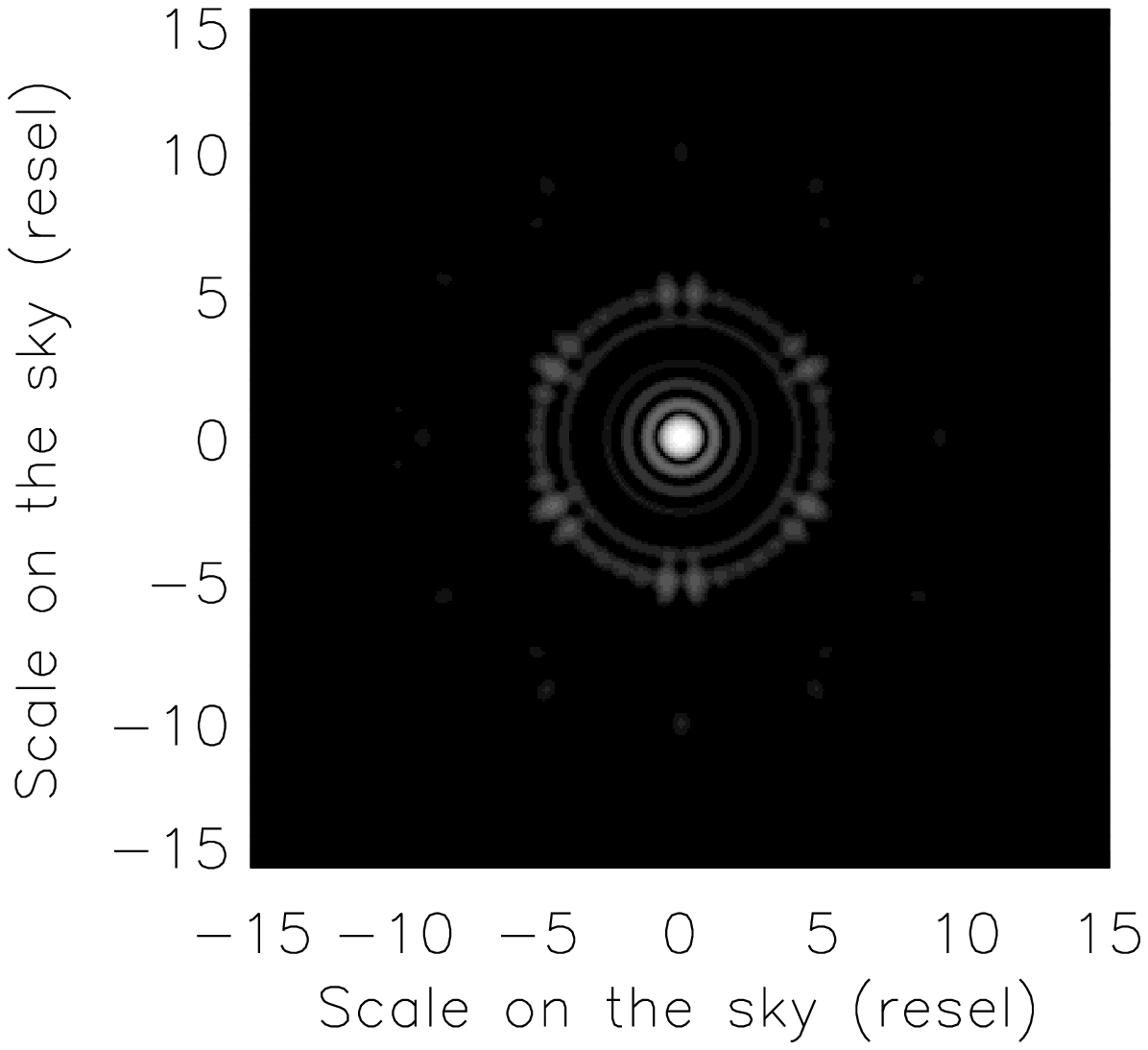} &
\includegraphics[width=39mm]{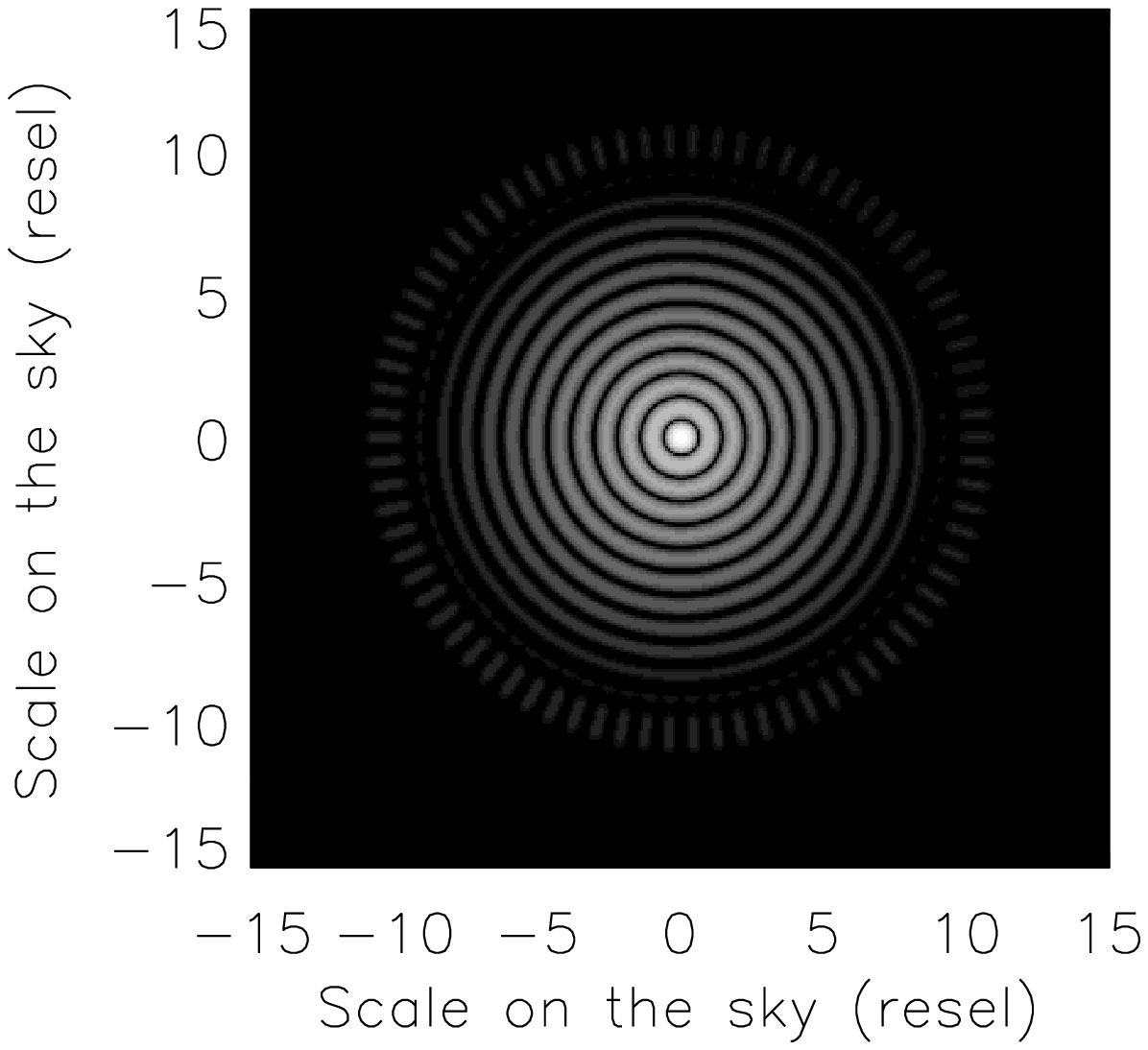} &
\includegraphics[width=39mm]{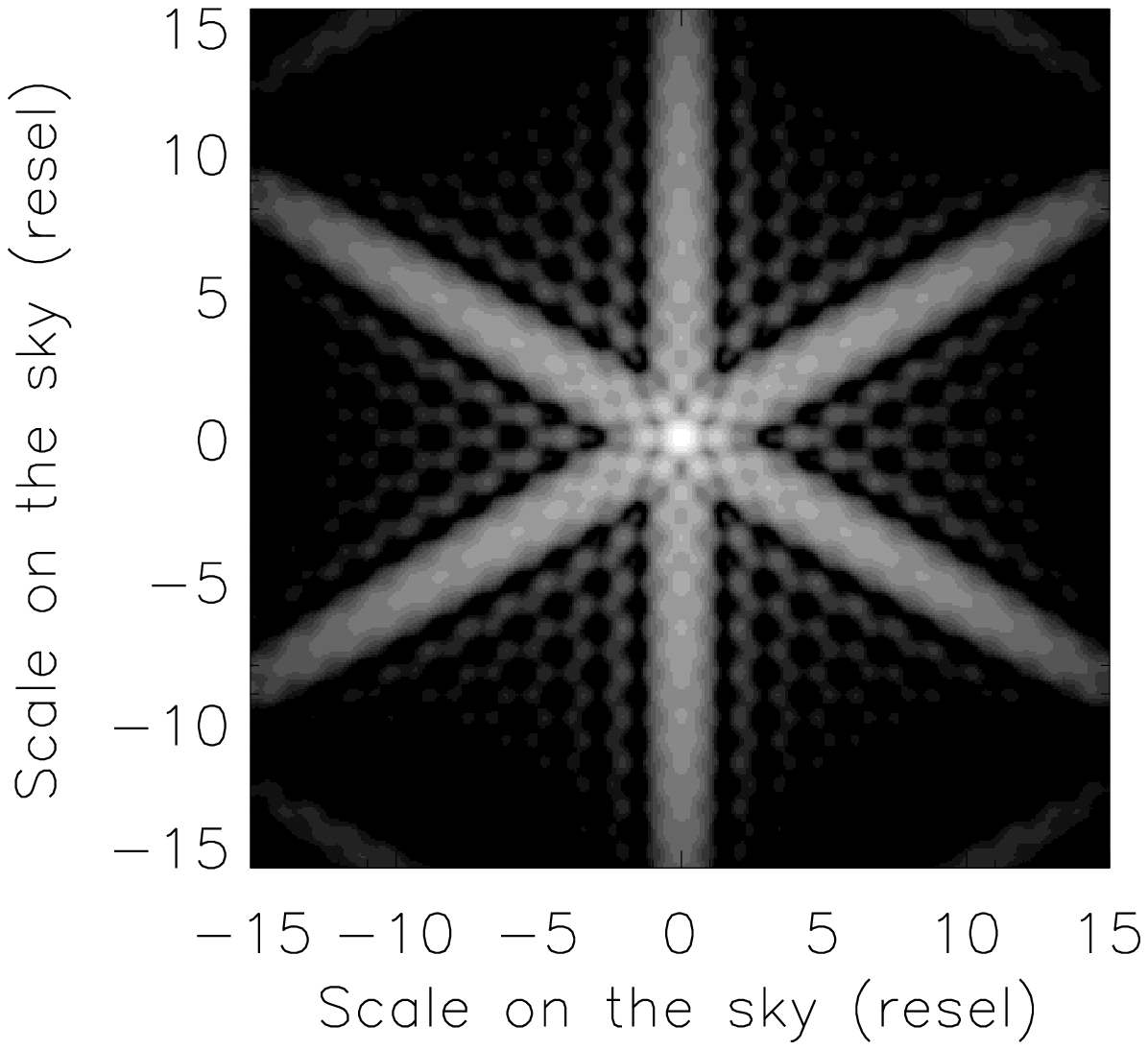} \\

\includegraphics[width=39mm]{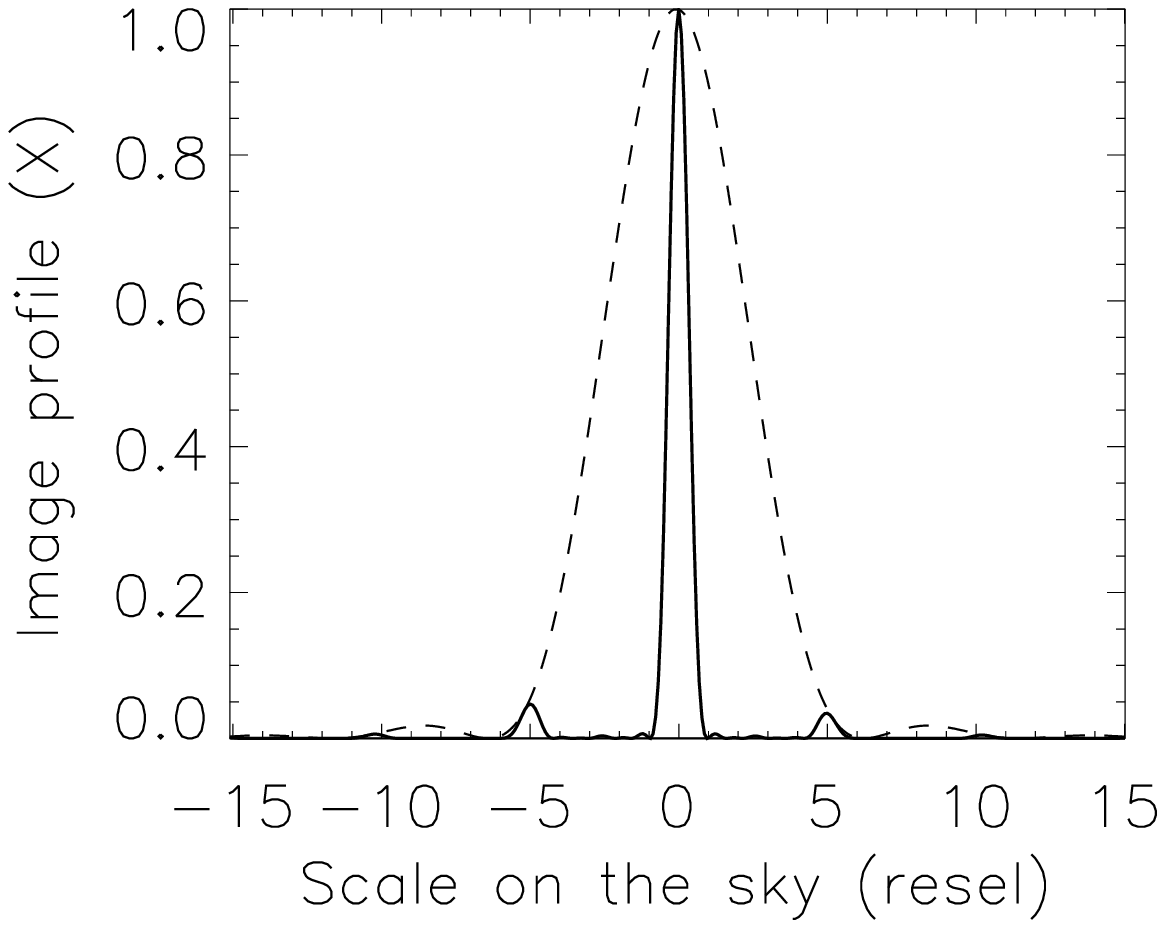} &
\includegraphics[width=39mm]{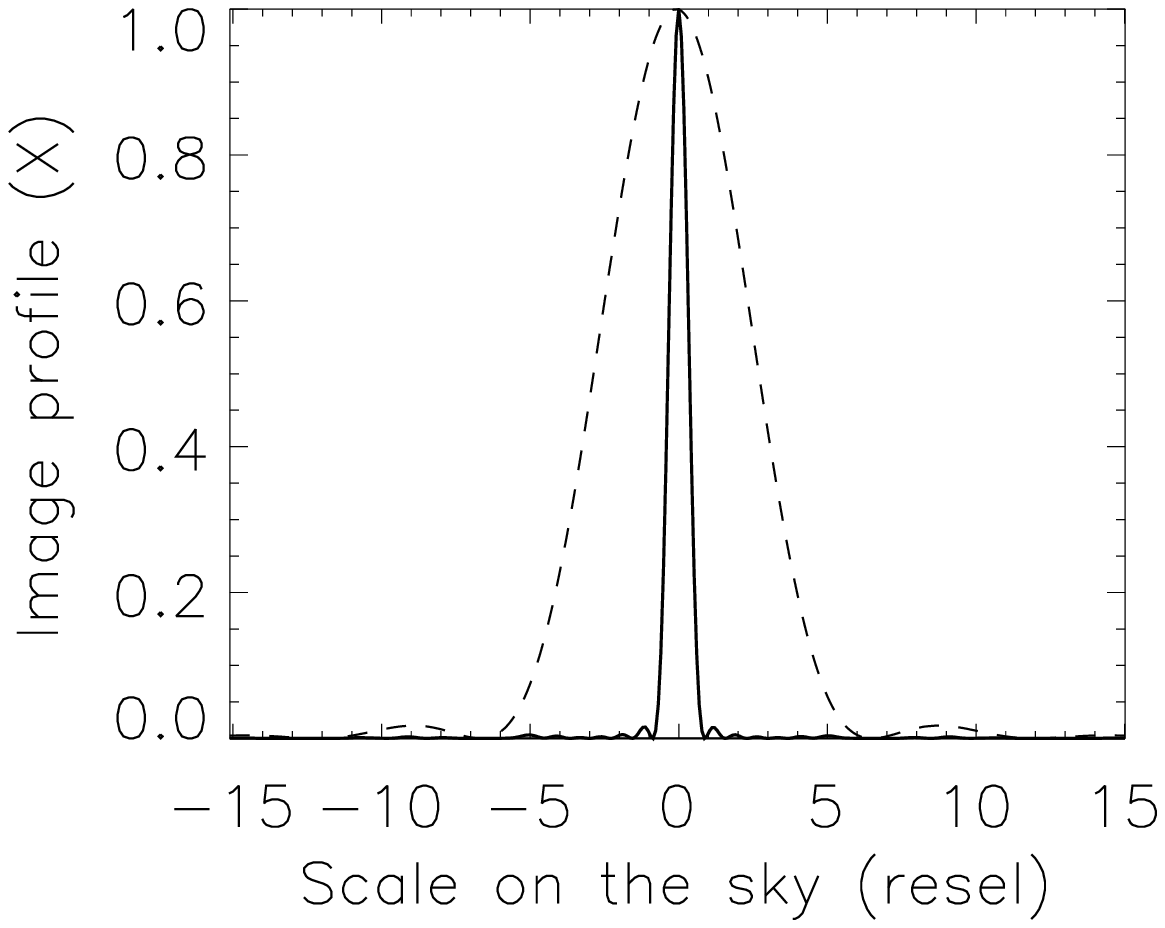} &
\includegraphics[width=39mm]{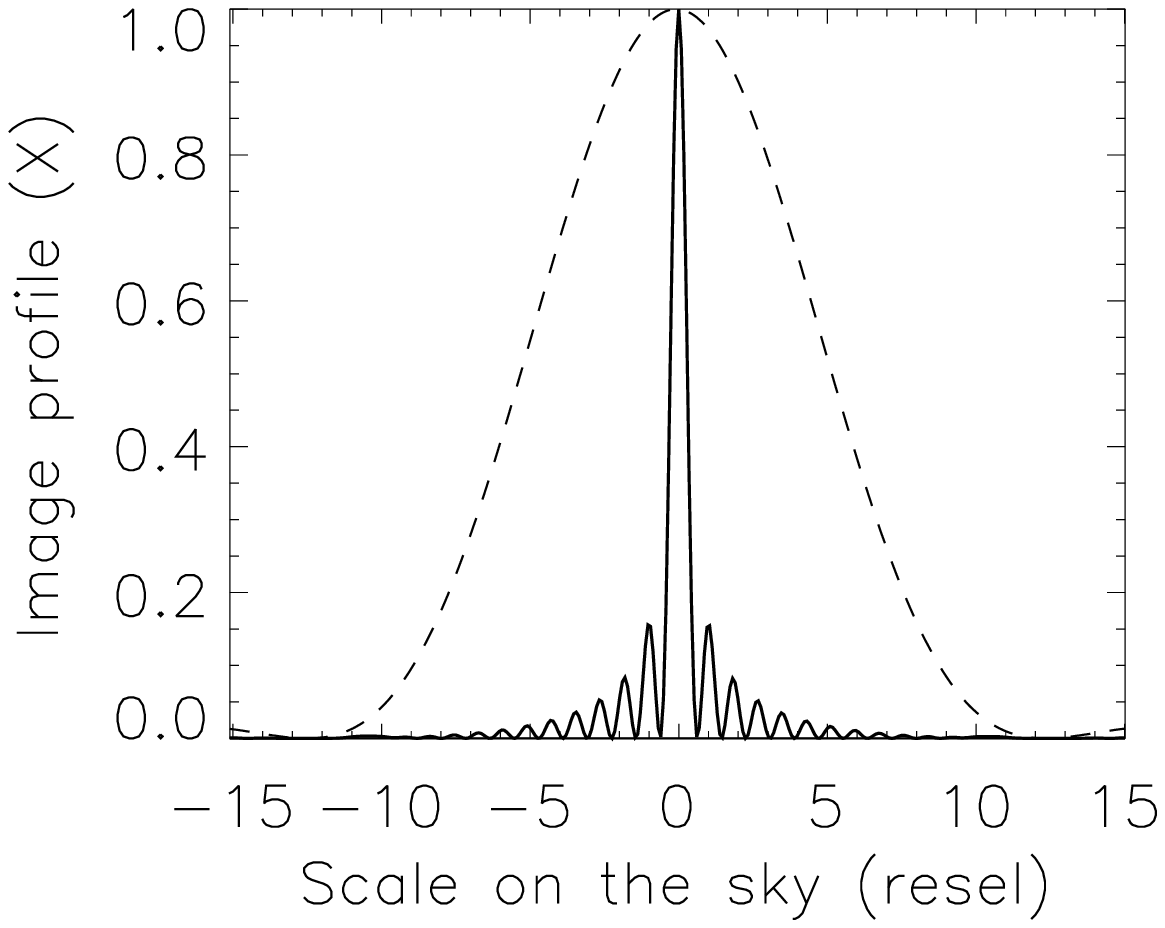} &
\includegraphics[width=39mm]{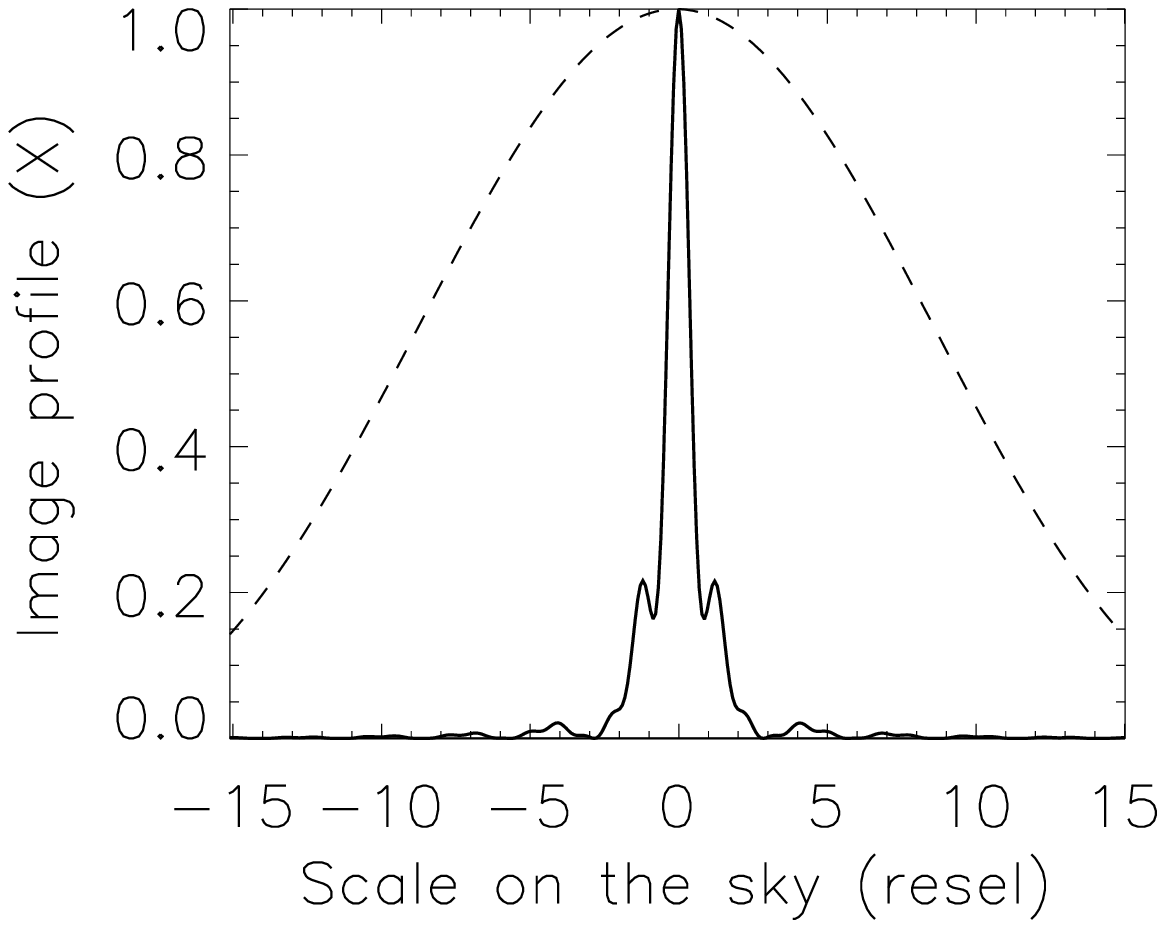} \\

\hline

\includegraphics[width=39mm]{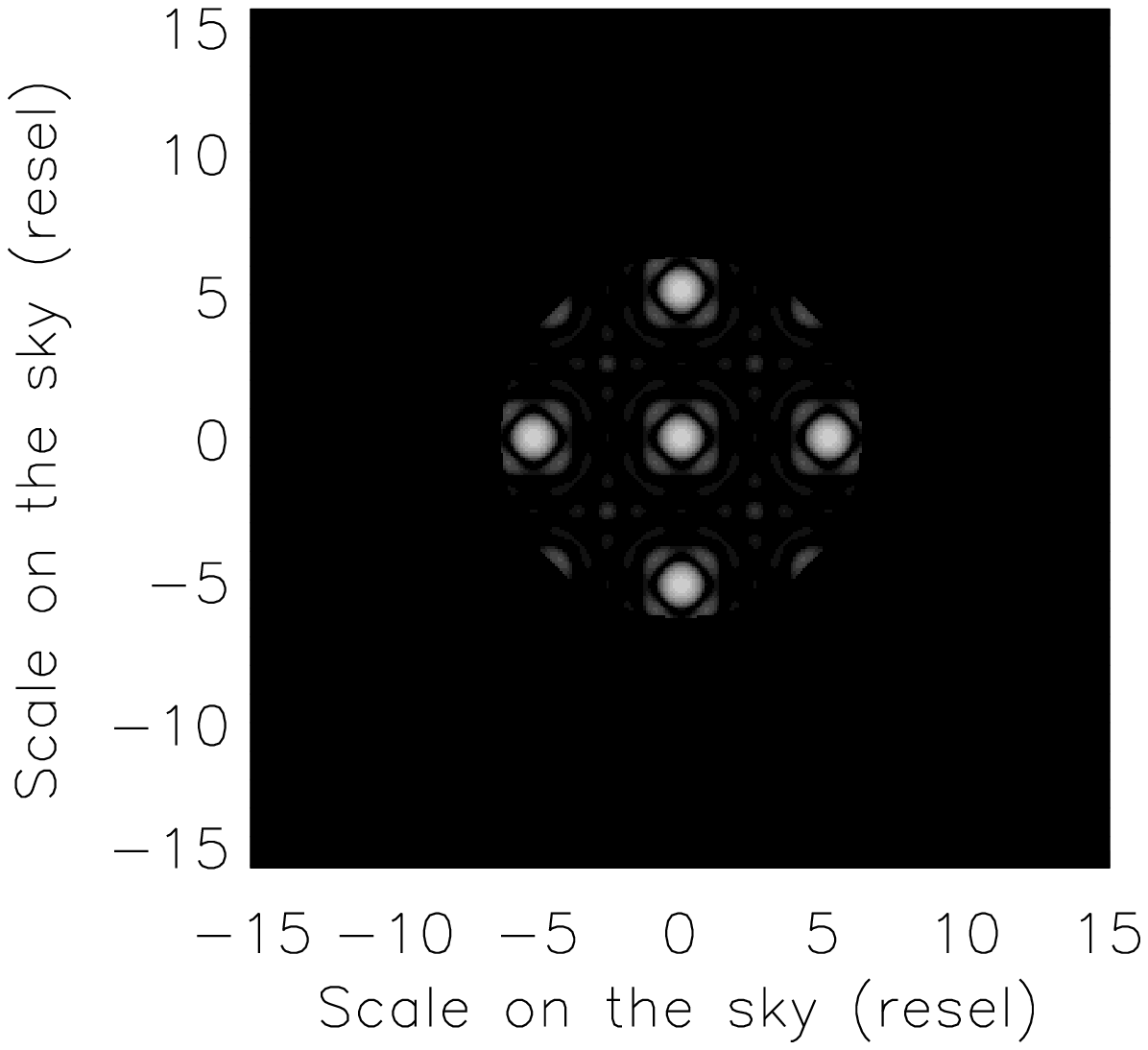} &
\includegraphics[width=39mm]{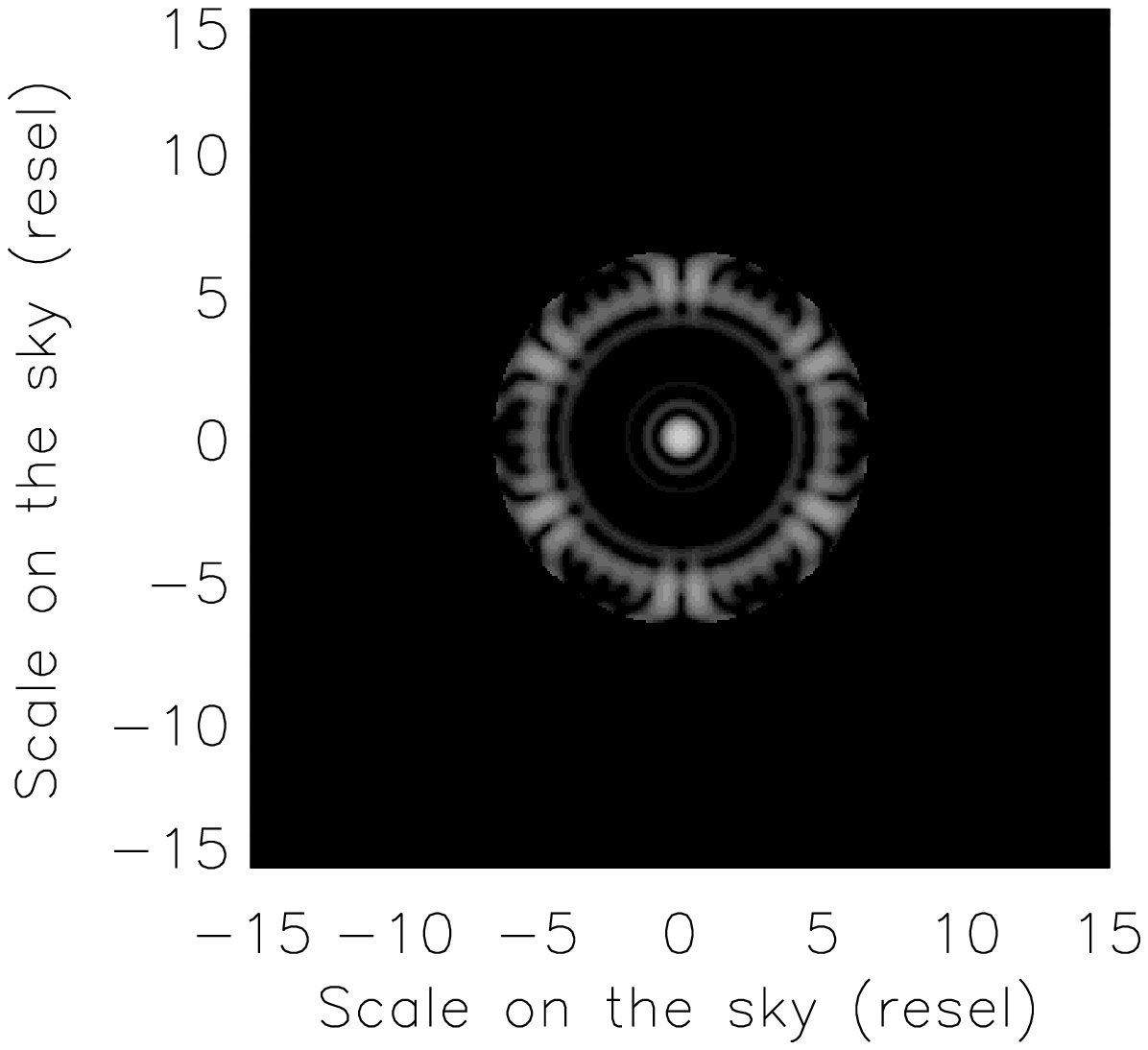} &
\includegraphics[width=39mm]{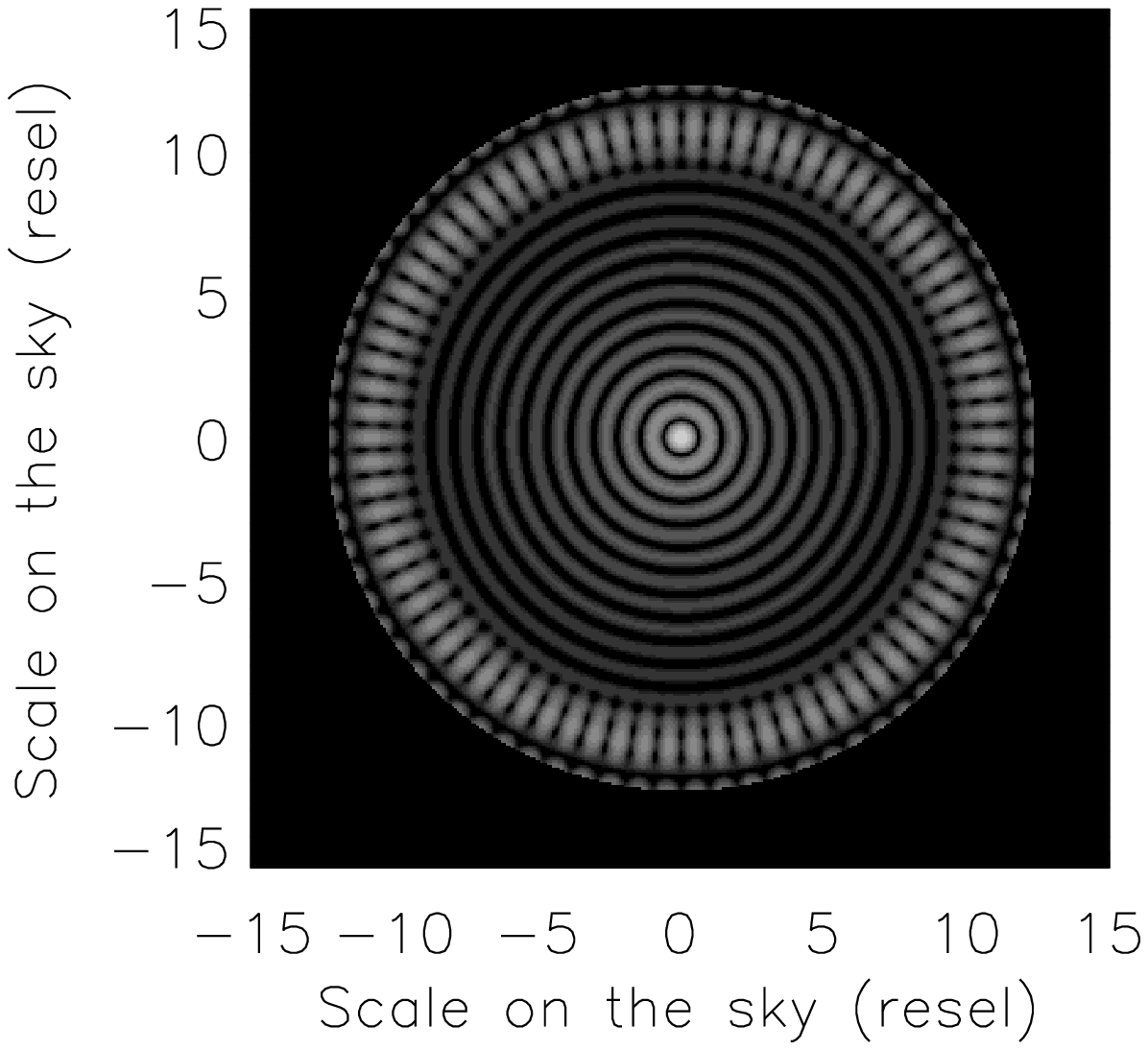} &
\includegraphics[width=39mm]{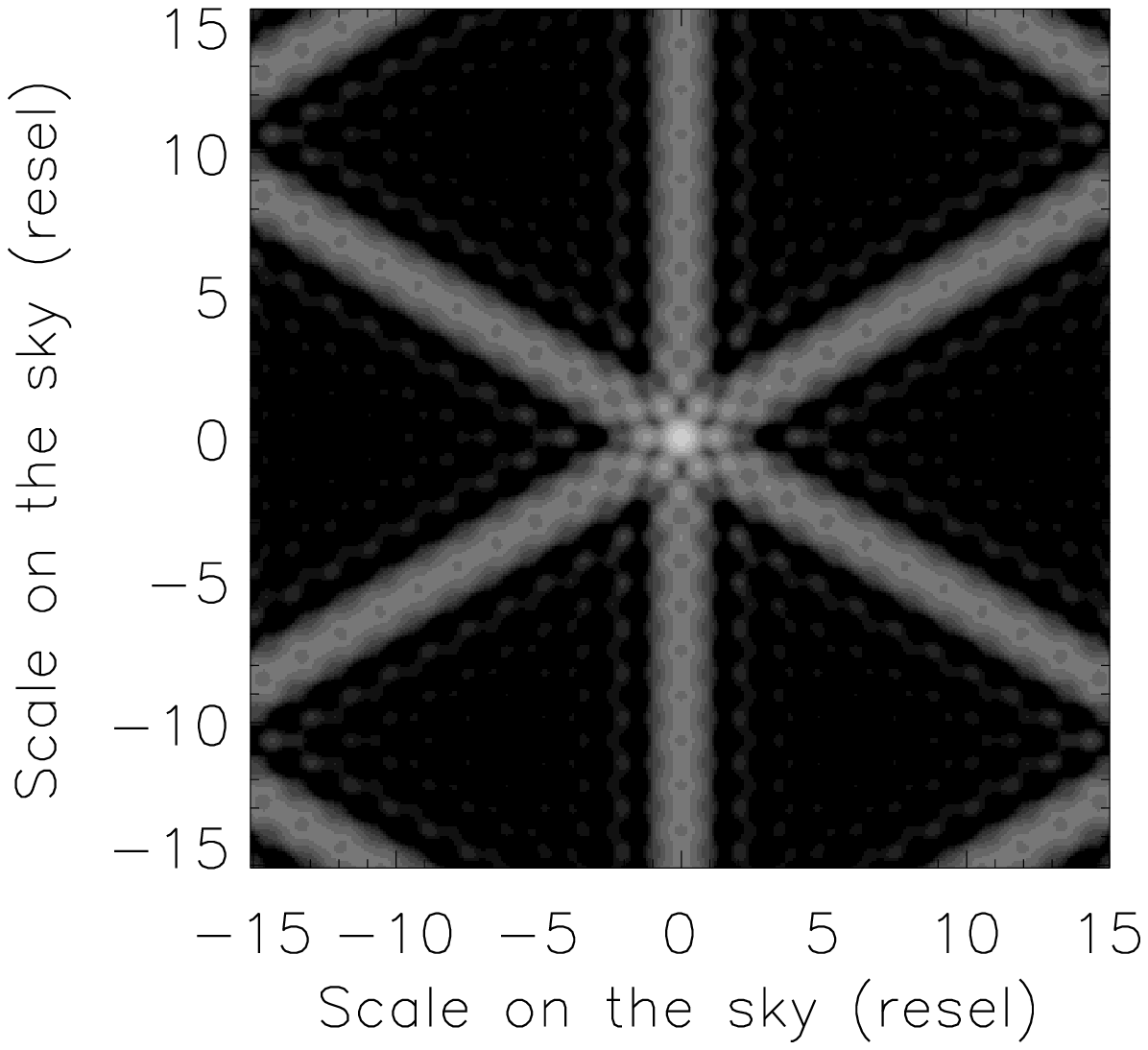} \\

\includegraphics[width=39mm]{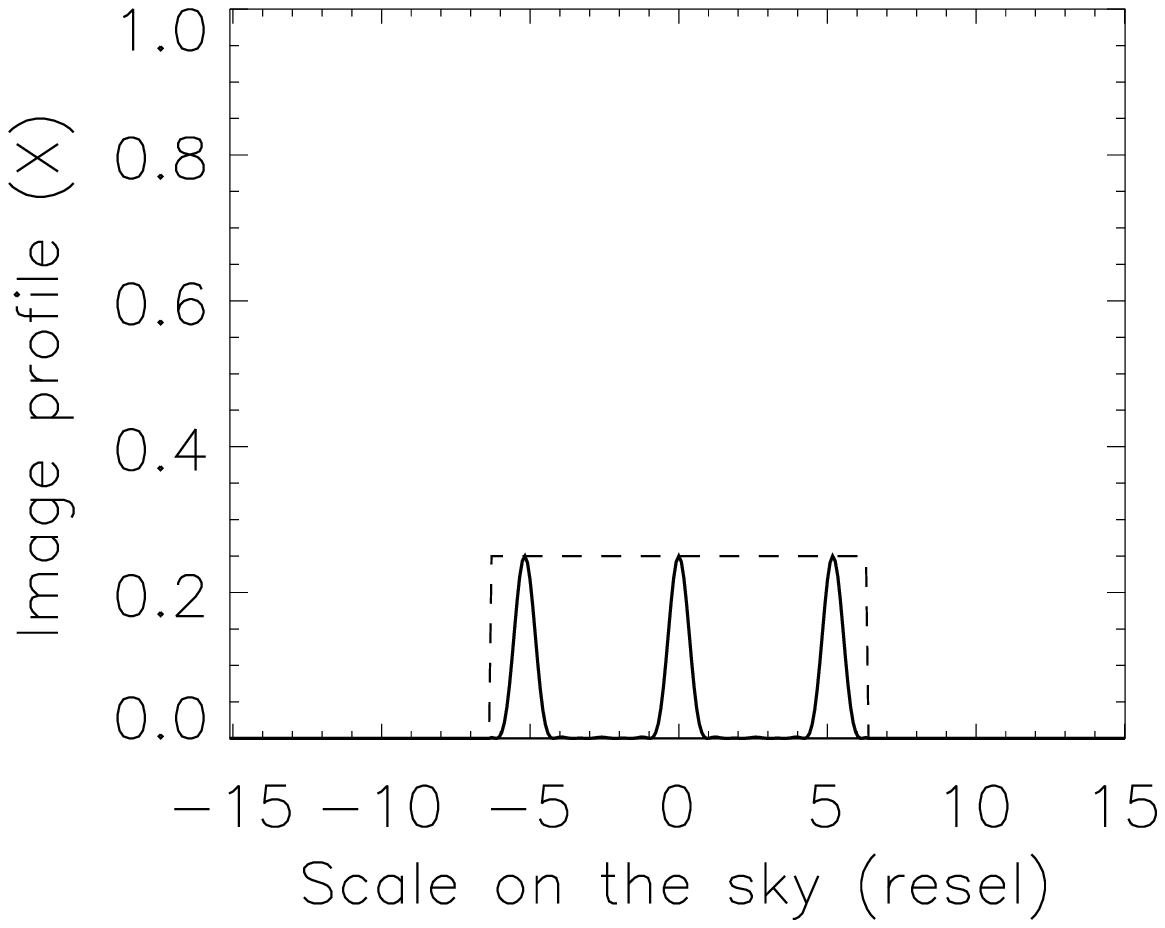} &
\includegraphics[width=39mm]{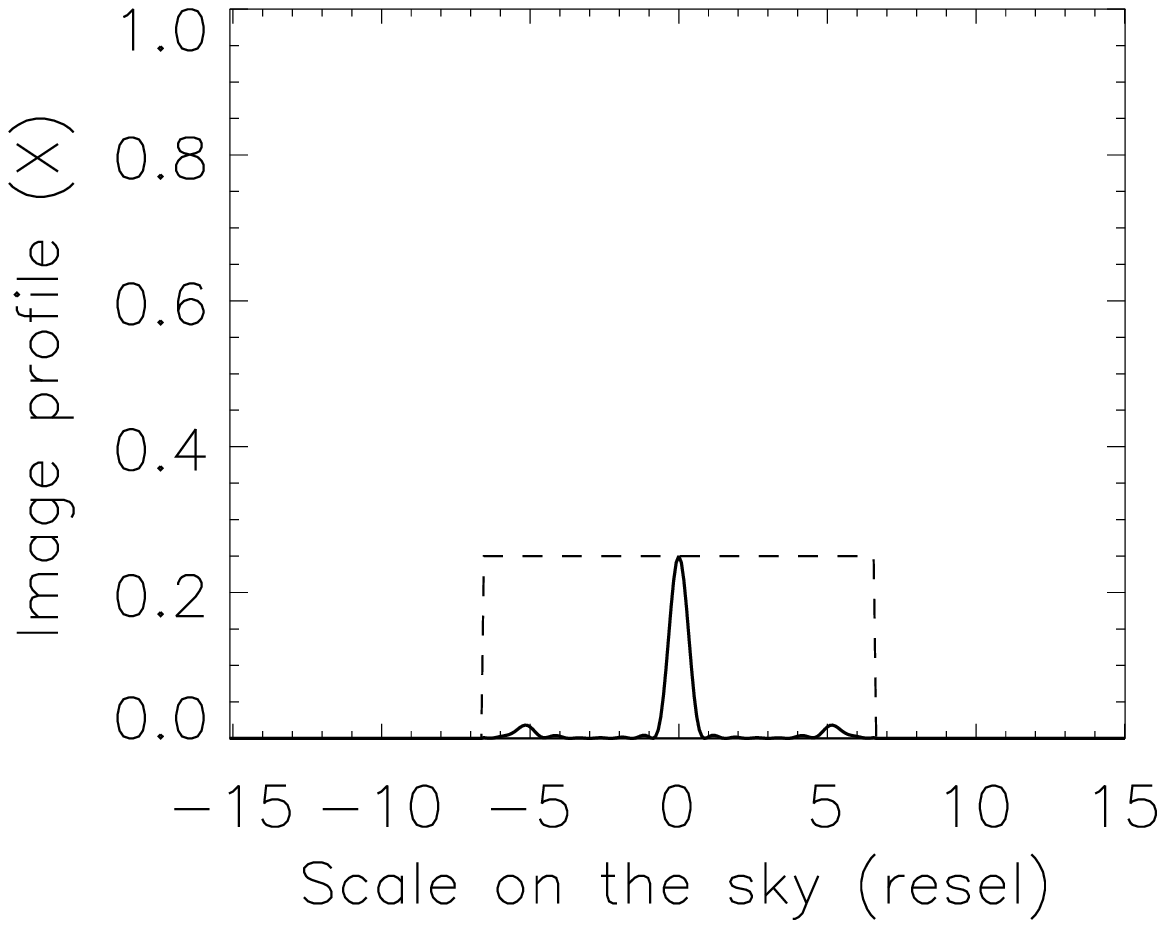} &
\includegraphics[width=39mm]{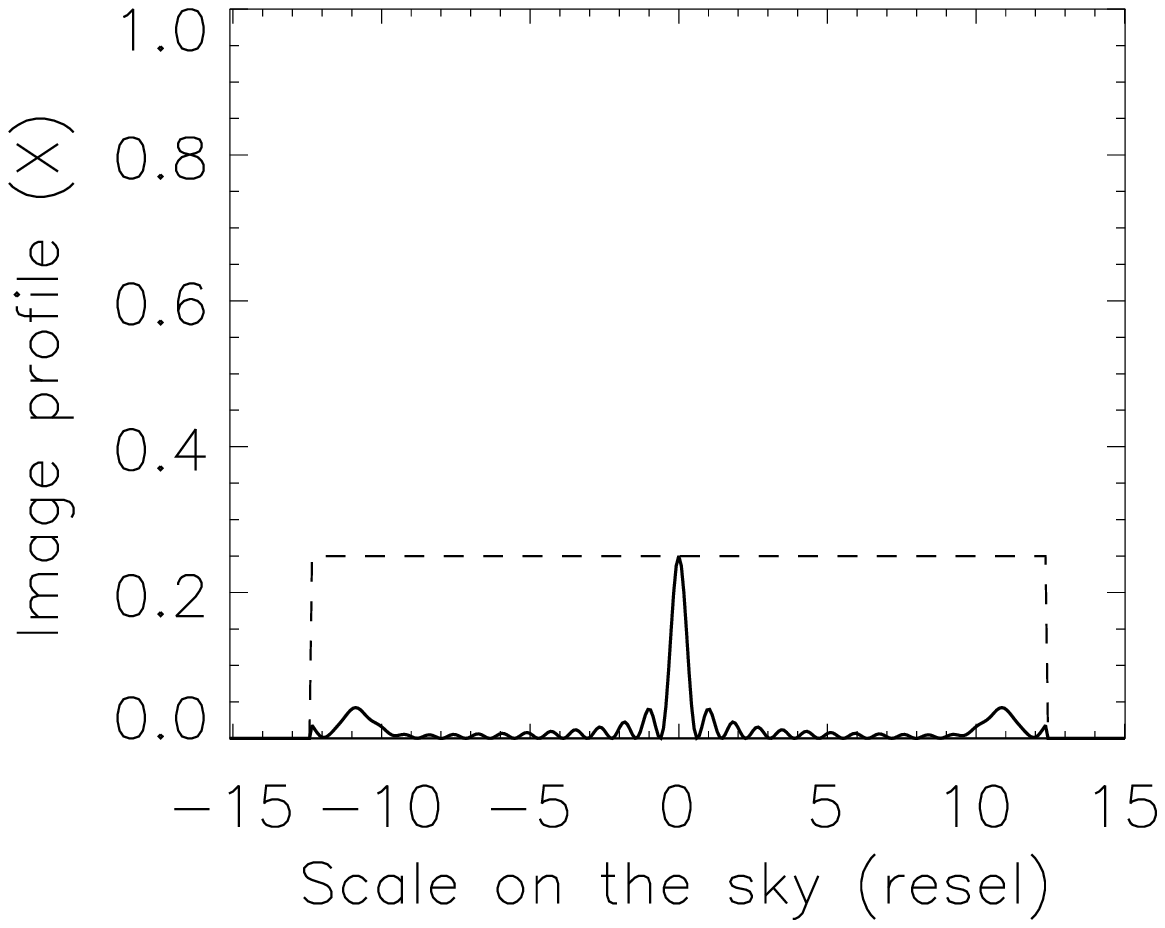} &
\includegraphics[width=39mm]{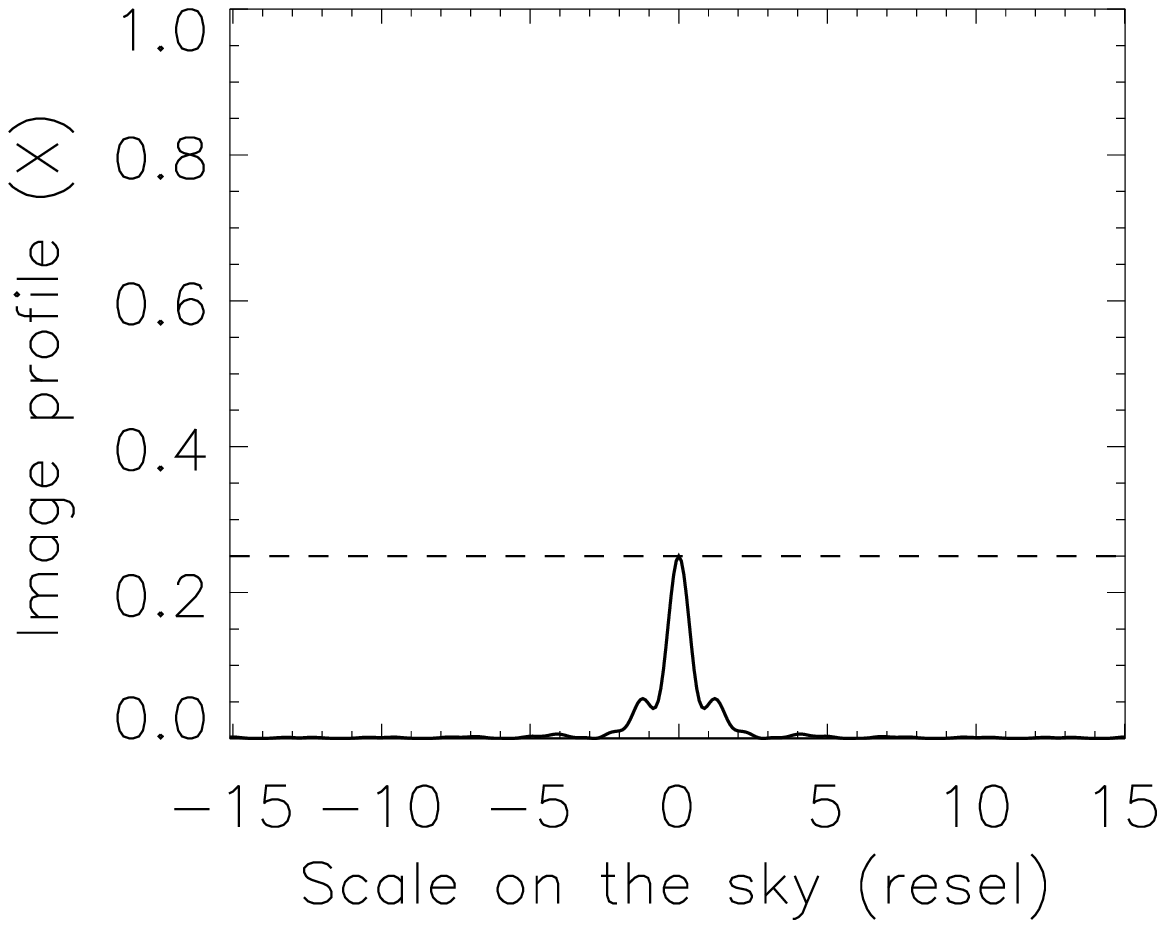} \\

\end{tabular}
\end{center}\caption{Imaging properties of 4 array configurations of 40 telescopes. The apertures with a diameter of $10m$ are laying out on an entrance pupil with an external diameter of $1 km$.
From top to bottom: Densified output pupil, (u,v) plan coverage,
image (logarithmic scale) and profile of the PSF in DP mode, image
and profile of the PSF in IRAN mode. The dashed lines
represent the profile of the diffraction envelope, its edge
corresponds to the CLean Field extent. The intensities in the
images are normalized to 1 for the central resel of each densified
image in DP mode. Due to a lower densification factor in the IRAN
mode ($\gamma_{max}/2$), the intensity of the central peak is only
$1/4$.} \label{psf_array40}
\end{figure*}

\subsubsection{Presentation of the simulations}

We consider four typical array configurations, made of 40
telescopes, 10m in diameter and distributed over a maximum baseline
of 1km with $\lambda$=0.6$\mu$m. The distribution of the
pupils is taken from ELSA \citep{Quirrenbach 2004}, OVLA
\citep{Labeyrie 1986}, KEOPS \citep{Vakili 2004b} and CARLINA
\citep{Labeyrie 2003}. Figure \ref{psf_array40} and Table~1 give the
characteristics of the PSFs and compare the cases of pupil
densification (DP mode) and image densification (IRAN mode
\citep{Vakili 2004a}).

OVLA and ELSA have an almost uniform coverage of the (u,v) plane and
a large clean field (10 and 18 resels respectively). OVLA has
diffraction rings and ELSA has diffraction spikes inside the clean
field, so that only $12\%$ of the energy is contained inside the
central peak. The maximum halo level corresponding to the
diffraction structures reaches $16\%$ (resp. $22\%$) of the
amplitude of the central peak for OVLA (resp. ELSA).

KEOPS and CARLINA have a uniform coverage of the input pupil, so that the coverage of the output pupil is maximized. The densified pupil filling rate reaches $75\%$ (resp. $69\%$) for KEOPS (resp. CARLINA). The minimal distance between the telescopes is also optimized, so that the clean field is reduced to
about $5$ resels. The advantage is an improvement of the image quality, so that the encircled energy reaches $71\%$ (resp. $65\%$) for KEOPS (resp. CARLINA), whereas the maximum halo level remains below $3\%$.

\begin{table*}
\begin{center}
\begin{tabular}{|ll|rrrr|}
\hline
    &       &   CARLINA-37 &   KEOPS-40   &   OVLA-39    &   ELSA-39    \\
\hline
Entrance pupil filling rate &   $\tau_i$    &   3.6e-3   &   3.9e-3   &   3.8e-3   &   3.8e-3   \\
Densified pupil filling rate    &   $\tau_o$    &   0.69   &   0.75   &   0.22   &   0.07   \\
Maximum densification level &   $\gamma_{max}$  &   15.8  &   15.8  &   8.1   &   4.4   \\
Clean field [mas]   &   CLF &   0.78  &   0.78  &   1.54  &   2.79 \\
Clean field [resel]     &   CLF &   5.18   &   5.18   &   10.18  &   18.45  \\
Direct imaging field [resel] &   DIF     &   5.53   &   5.53   &   11.62  &   23.82  \\
\hline
FWHM of the central peak [resel]    &   $FWHM$  &   0.73    &   0.70    &   0.55    &   0.89    \\
Encircled energy of the central peak    &   $E_0/E_{tot}$   &   0.71    &   0.65    &   0.12    &   0.12    \\
Maximum halo level in the CLF   &   $I_1/I_0$   &   0.03    &   0.02    &   0.16    &   0.22    \\
\hline
\end{tabular}
\caption{Imaging parameters of 4 array configurations of 40
telescopes. The aperture diameters are equal to $10m$ and the
maximum baseline is $1 km$, so that the $resel$ is $0.12~mas$ and
the Coupled Field $CF$ is $82~resels$.}
\end{center}
\label{tab:array_vs_ntel_gam_fov_39}
\end{table*}


\begin{figure*}
\begin{center}
\begin{tabular}{cccccc}

CARLINA & KEOPS & OVLA & ELSA\\

\includegraphics[width=40mm]{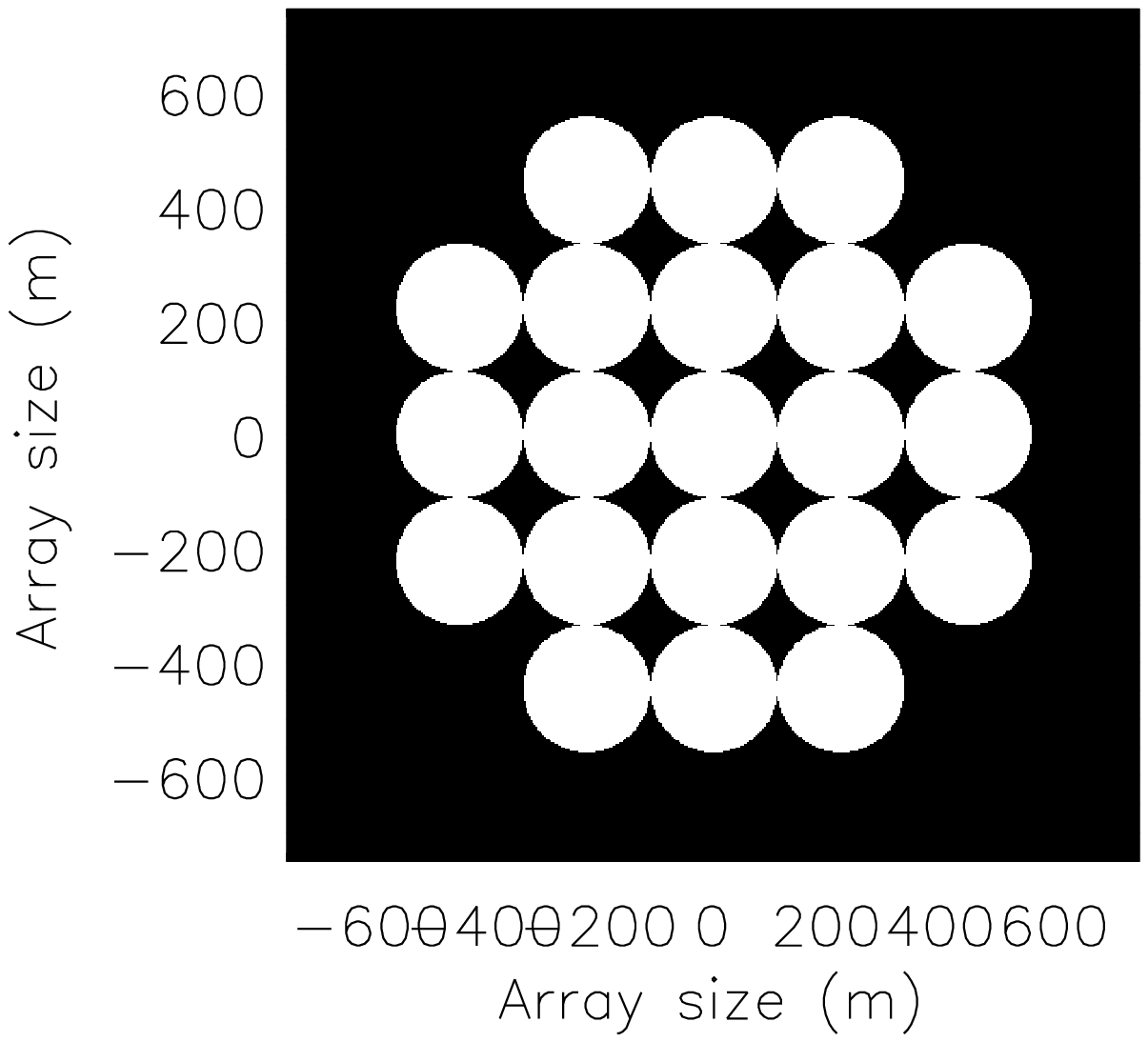} &
\includegraphics[width=40mm]{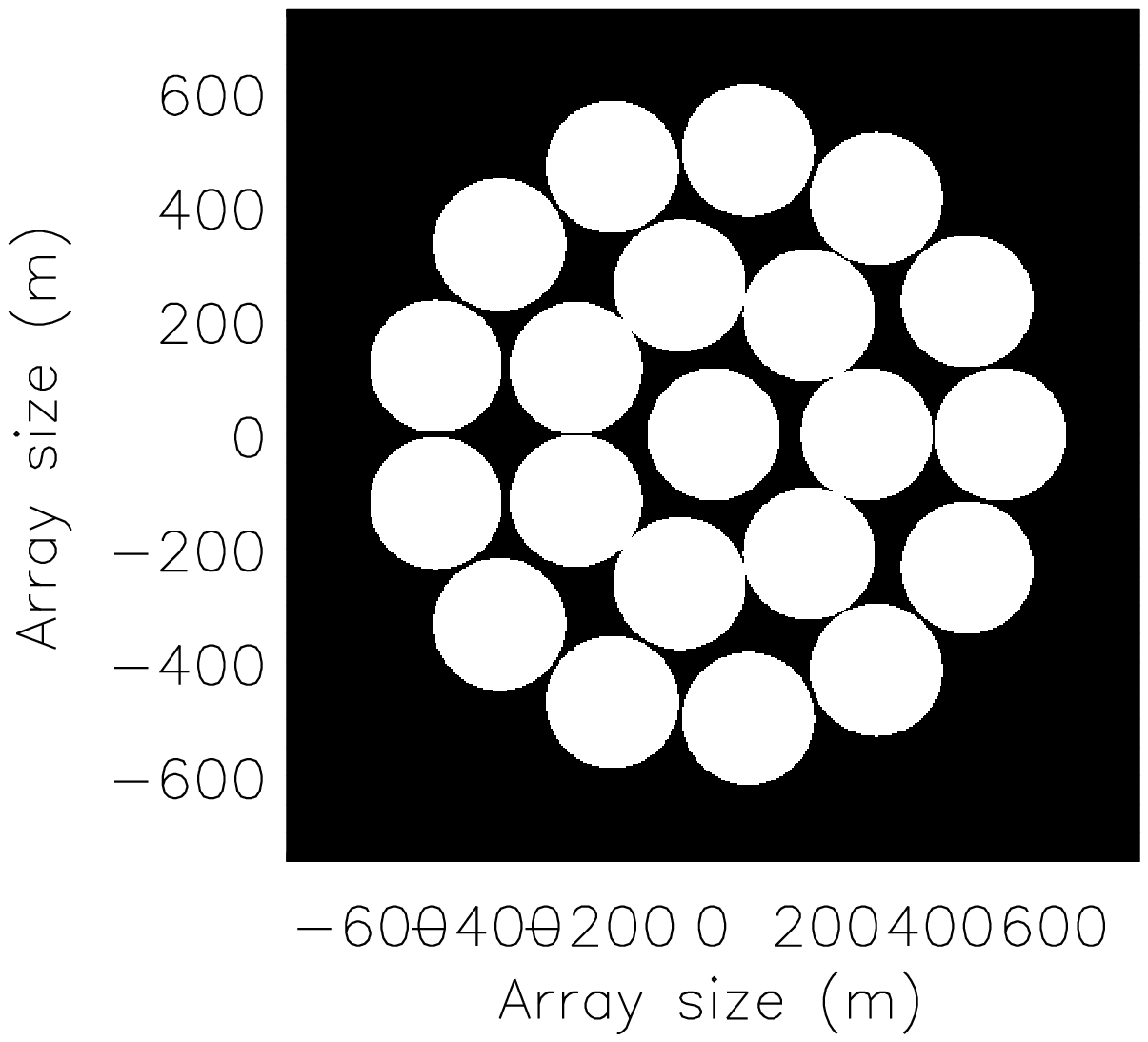} &
\includegraphics[width=40mm]{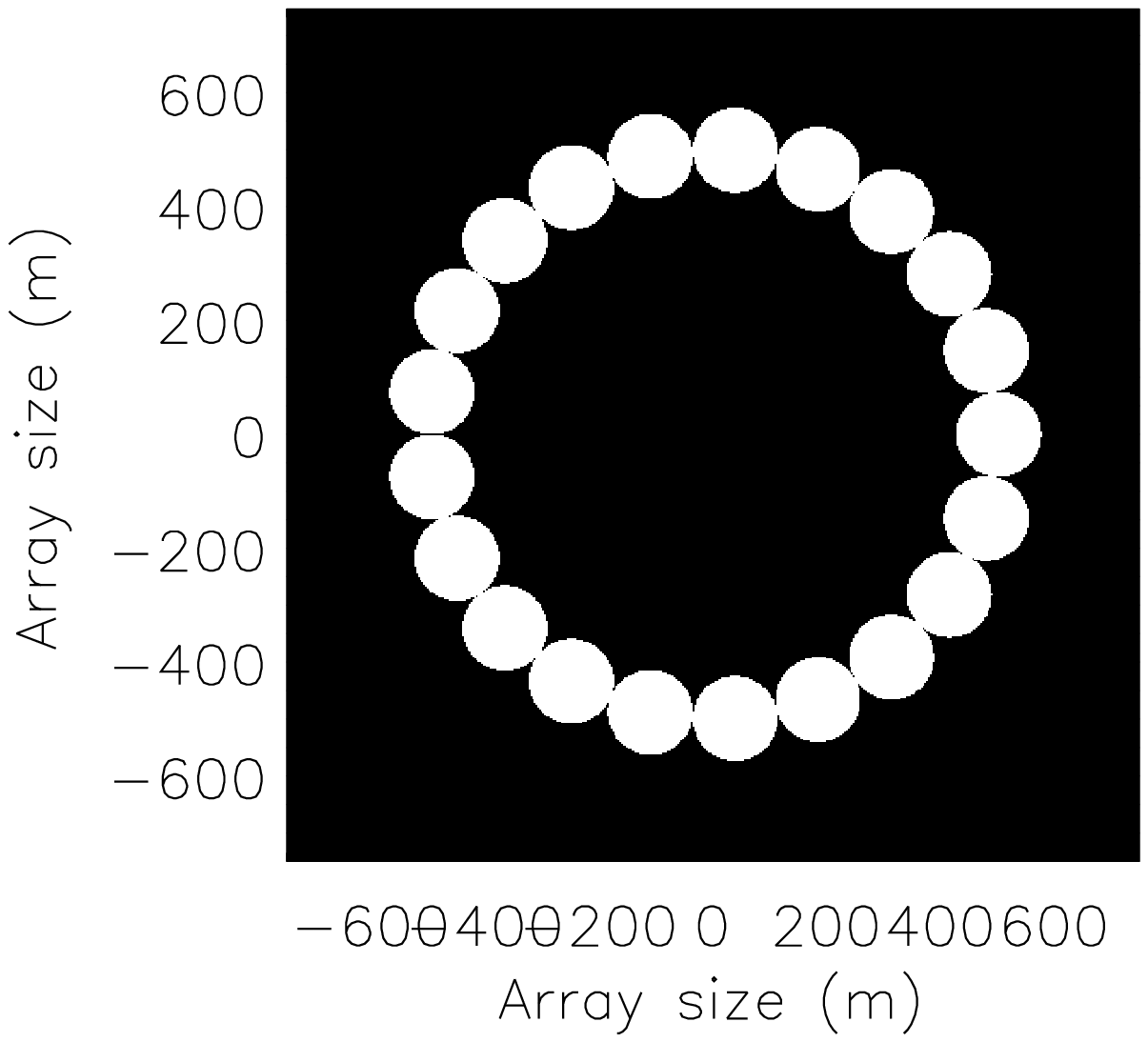} &
\includegraphics[width=40mm]{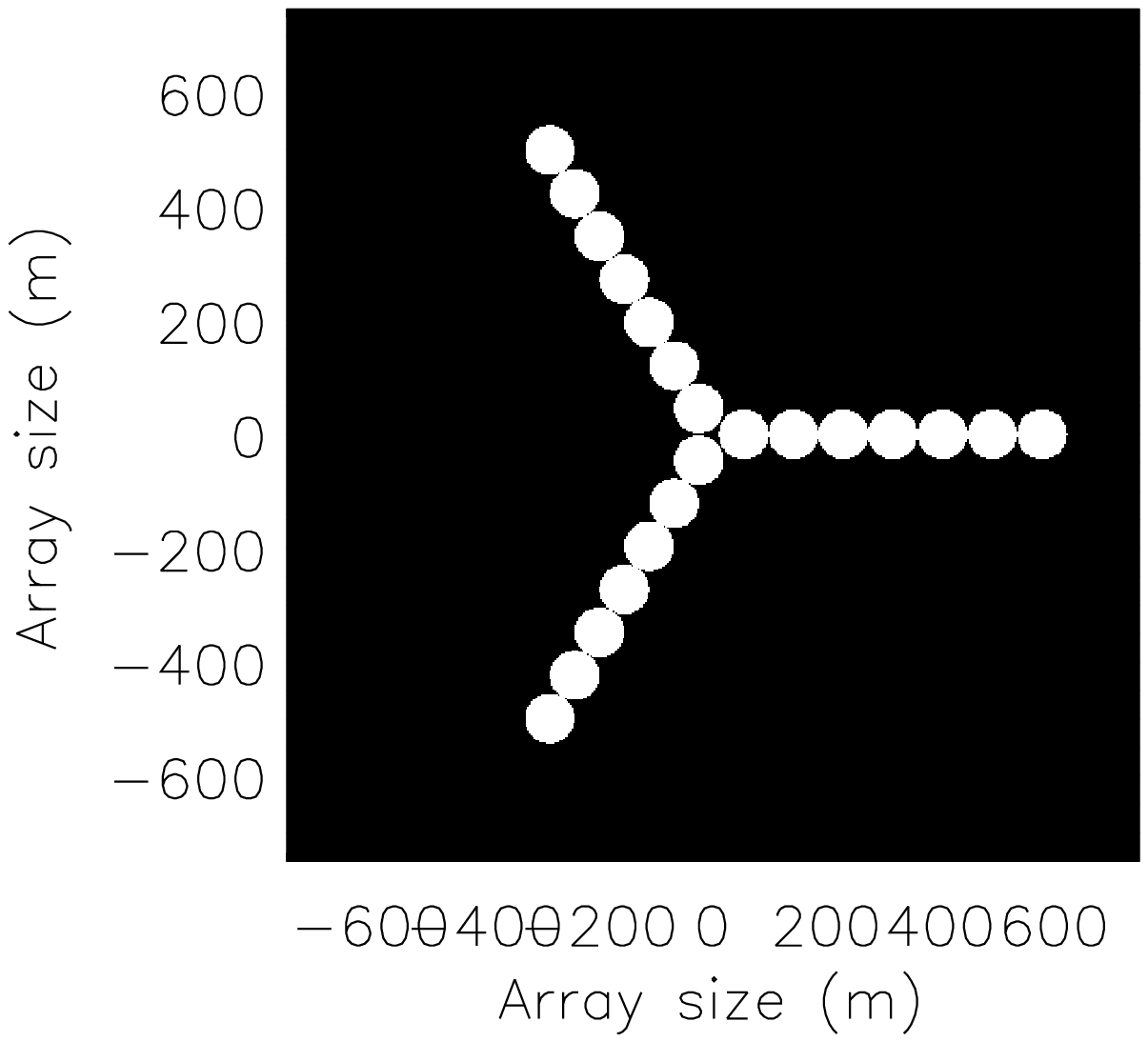} \\

\includegraphics[width=40mm]{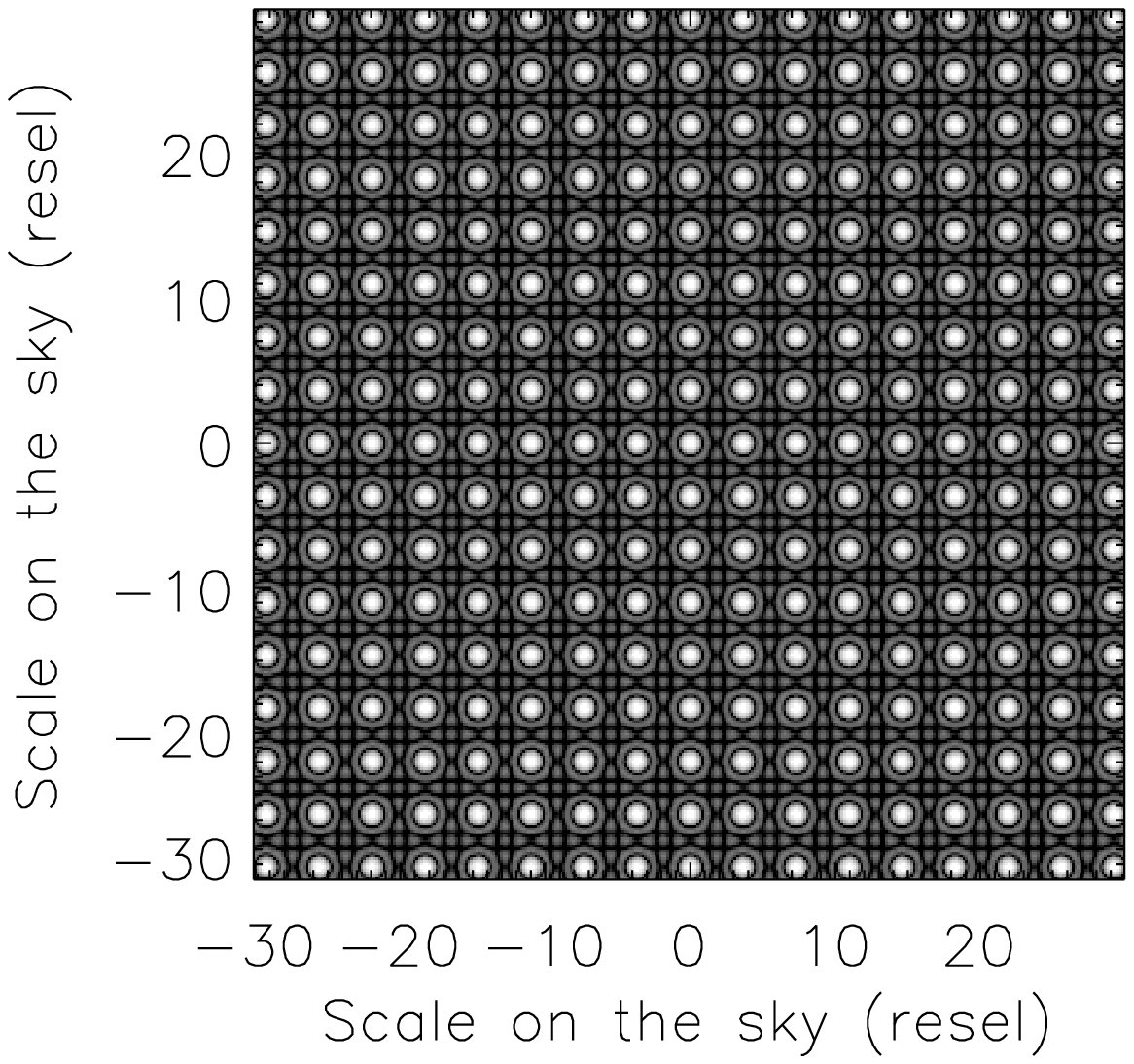} &
\includegraphics[width=40mm]{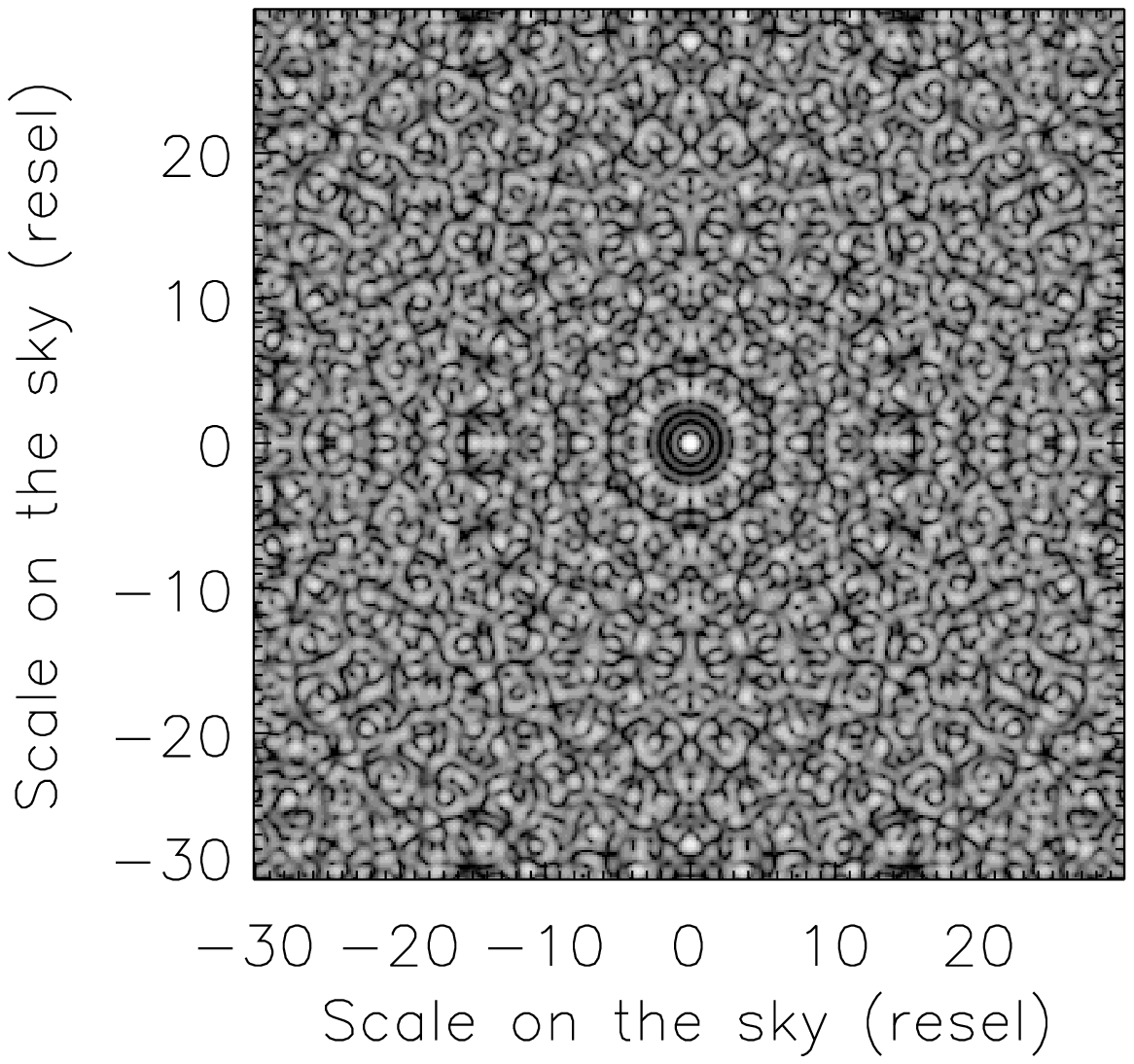} &
\includegraphics[width=40mm]{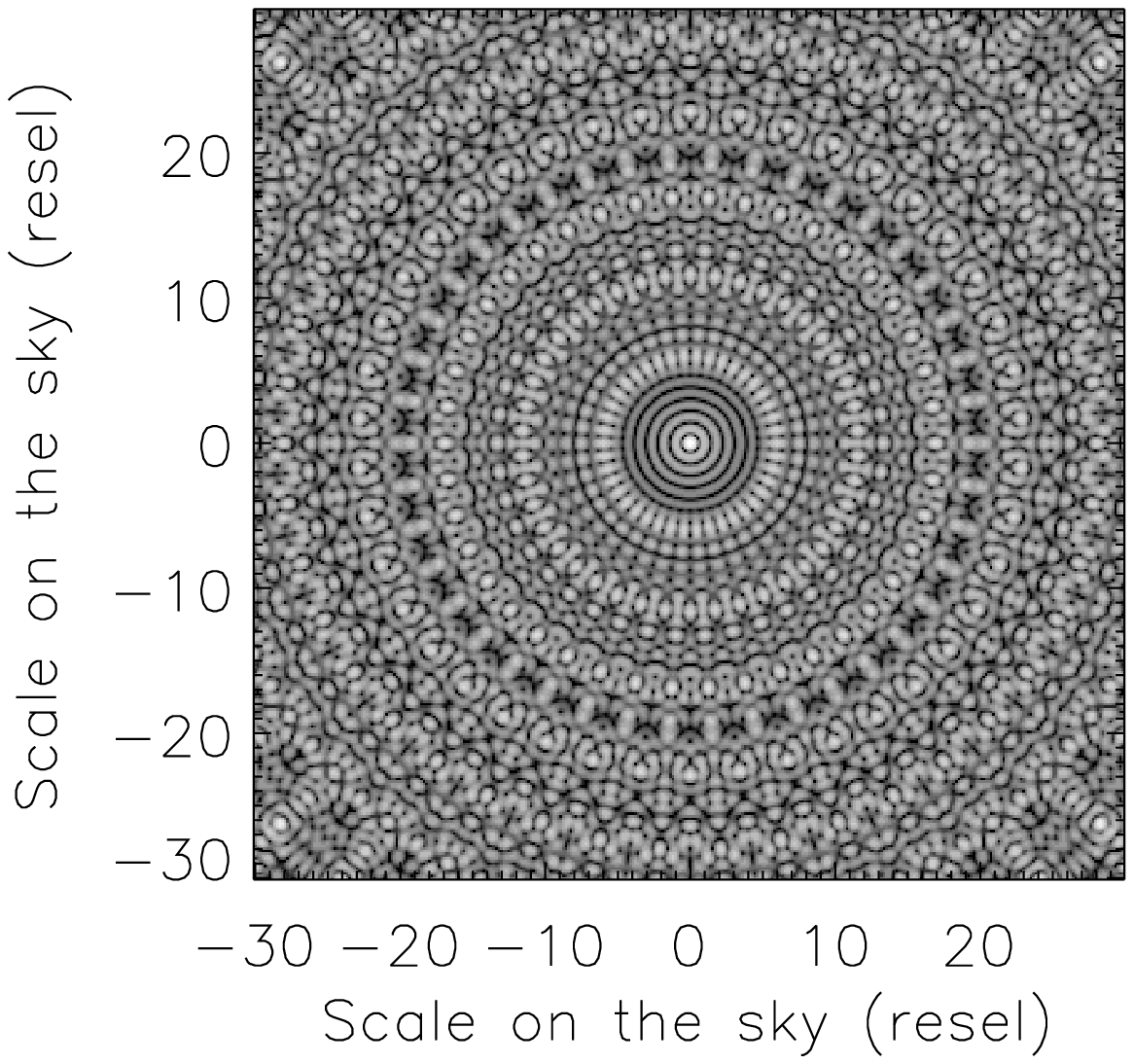} &
\includegraphics[width=40mm]{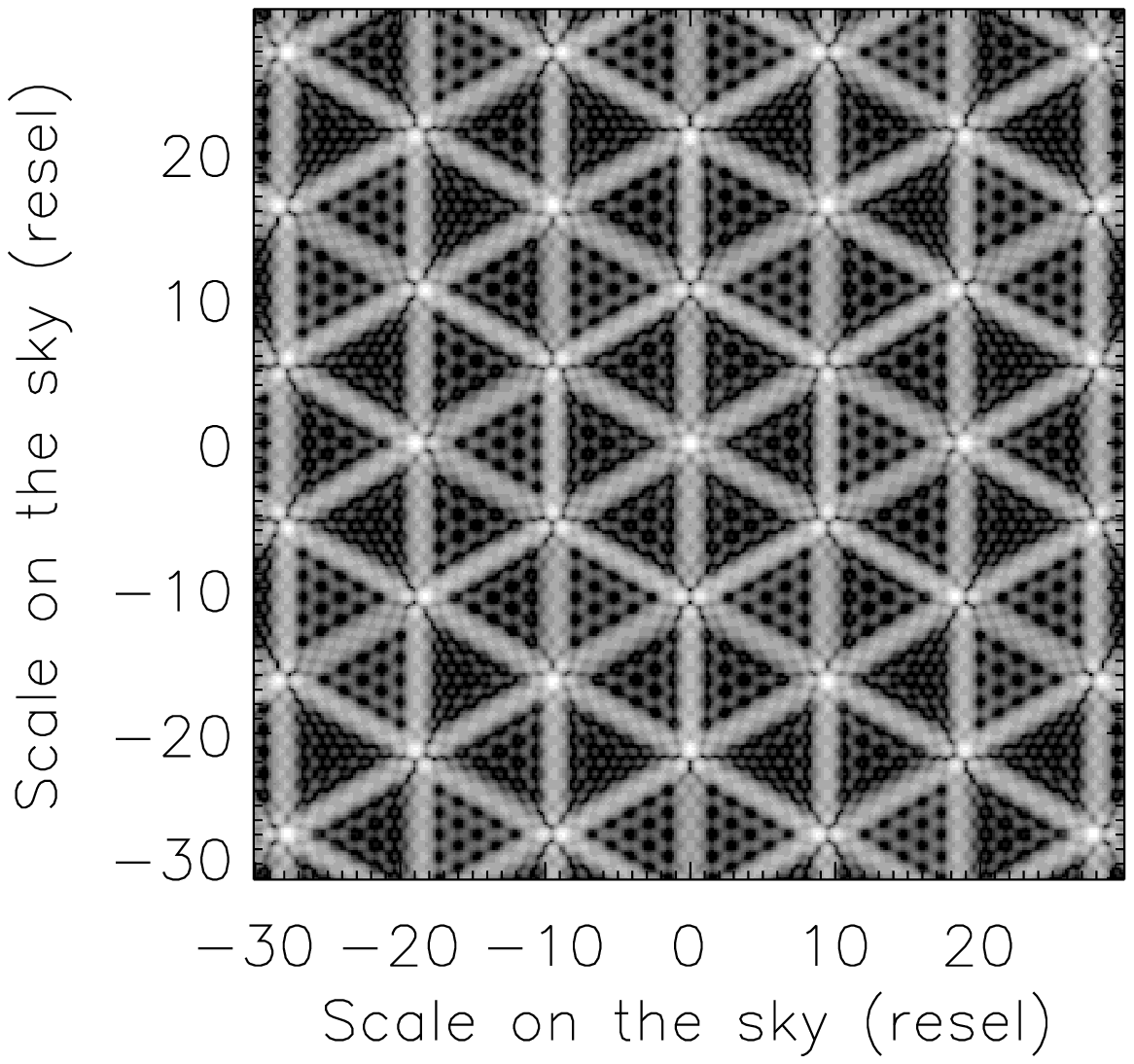} \\

\hline

\includegraphics[width=40mm]{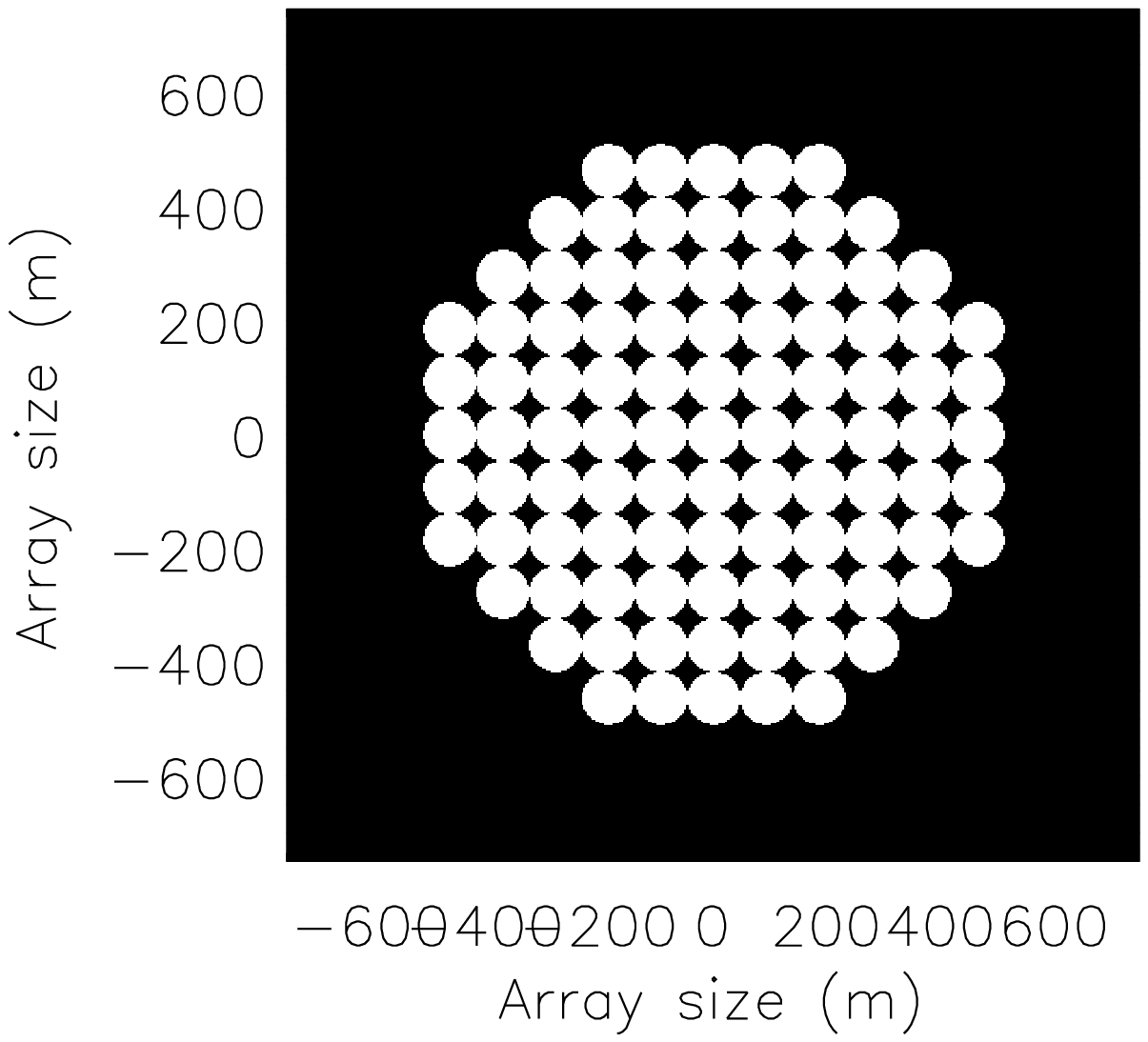} &
\includegraphics[width=40mm]{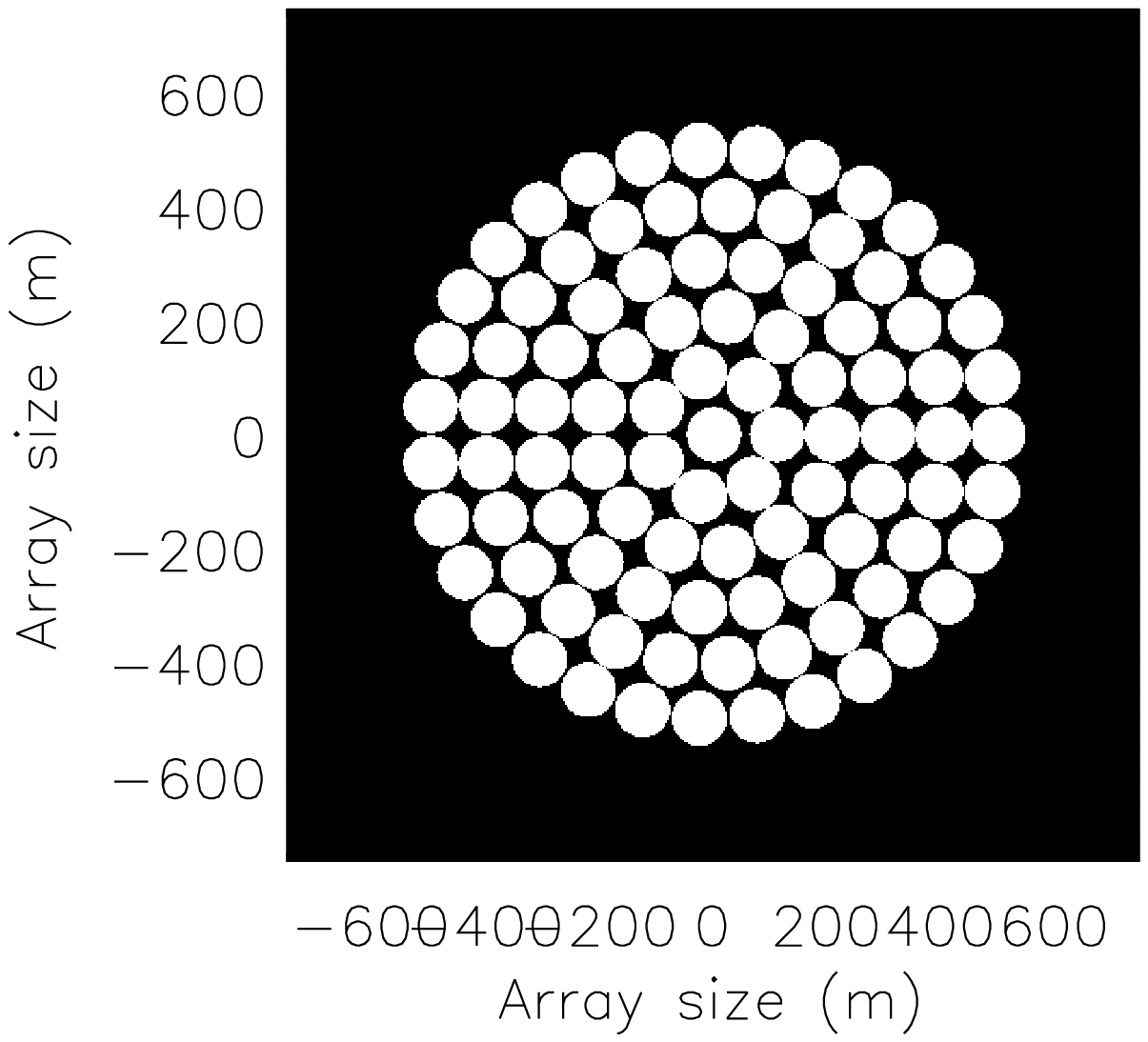} &
\includegraphics[width=40mm]{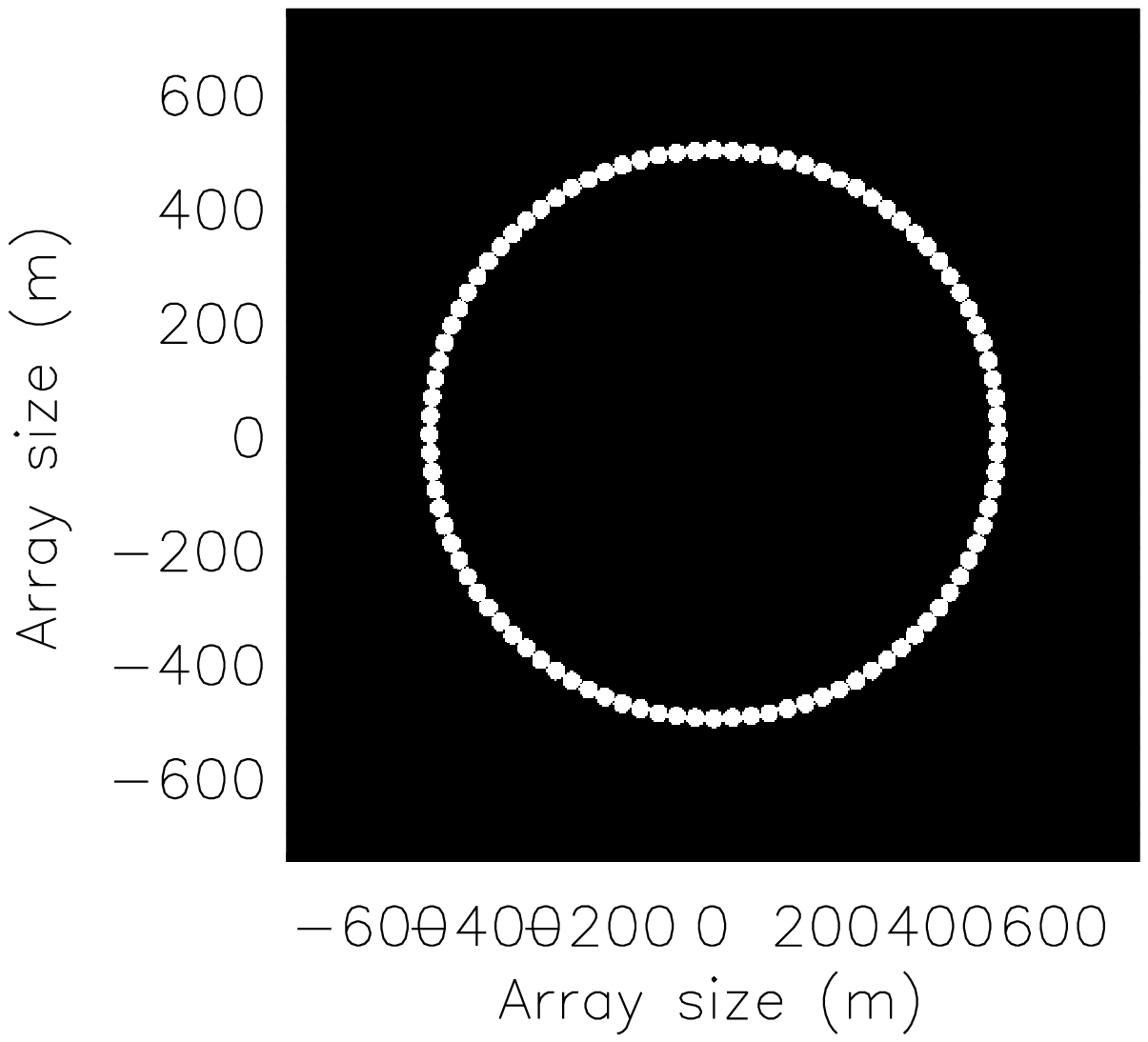} &
\includegraphics[width=40mm]{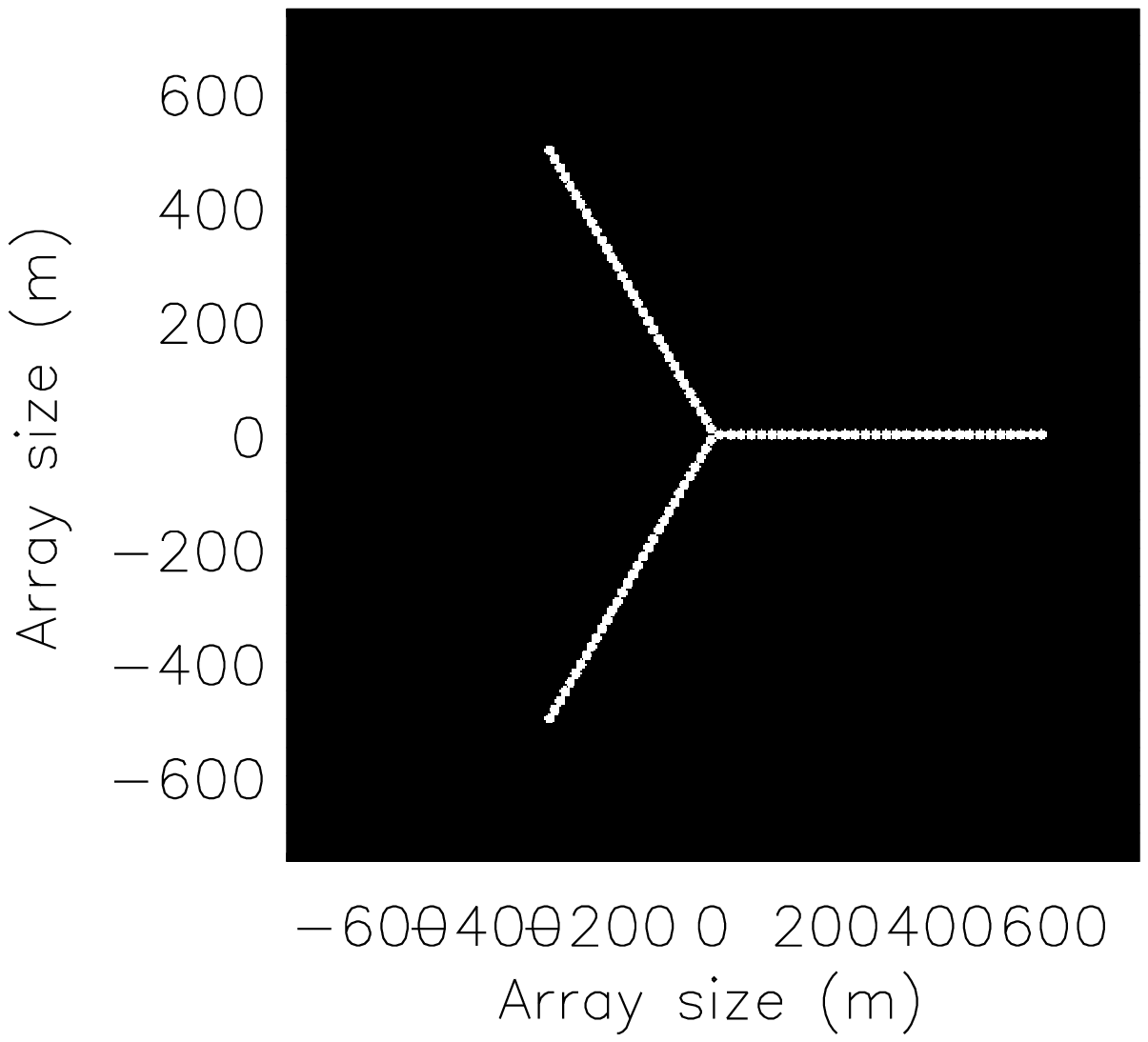} \\

\includegraphics[width=40mm]{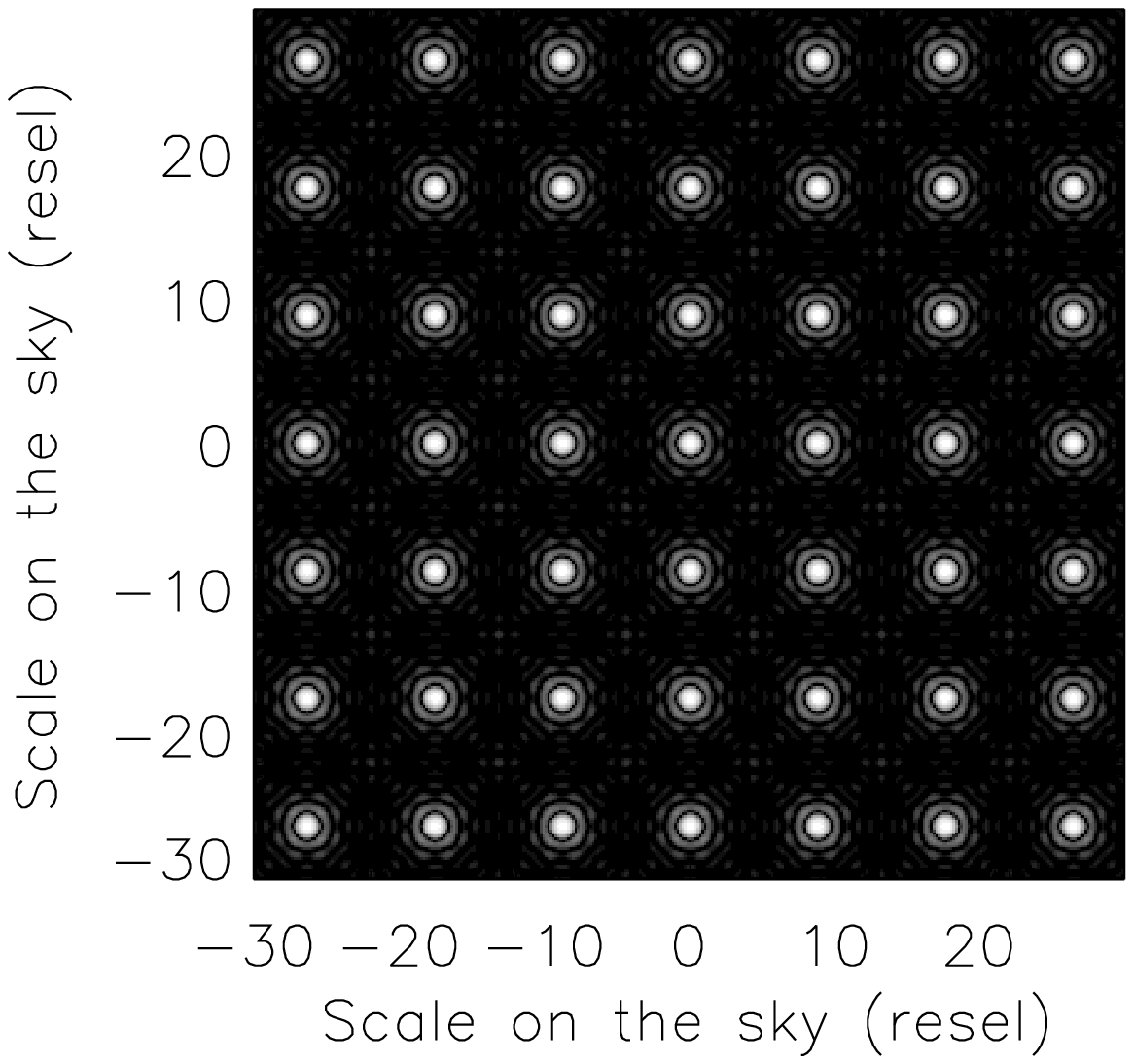} &
\includegraphics[width=40mm]{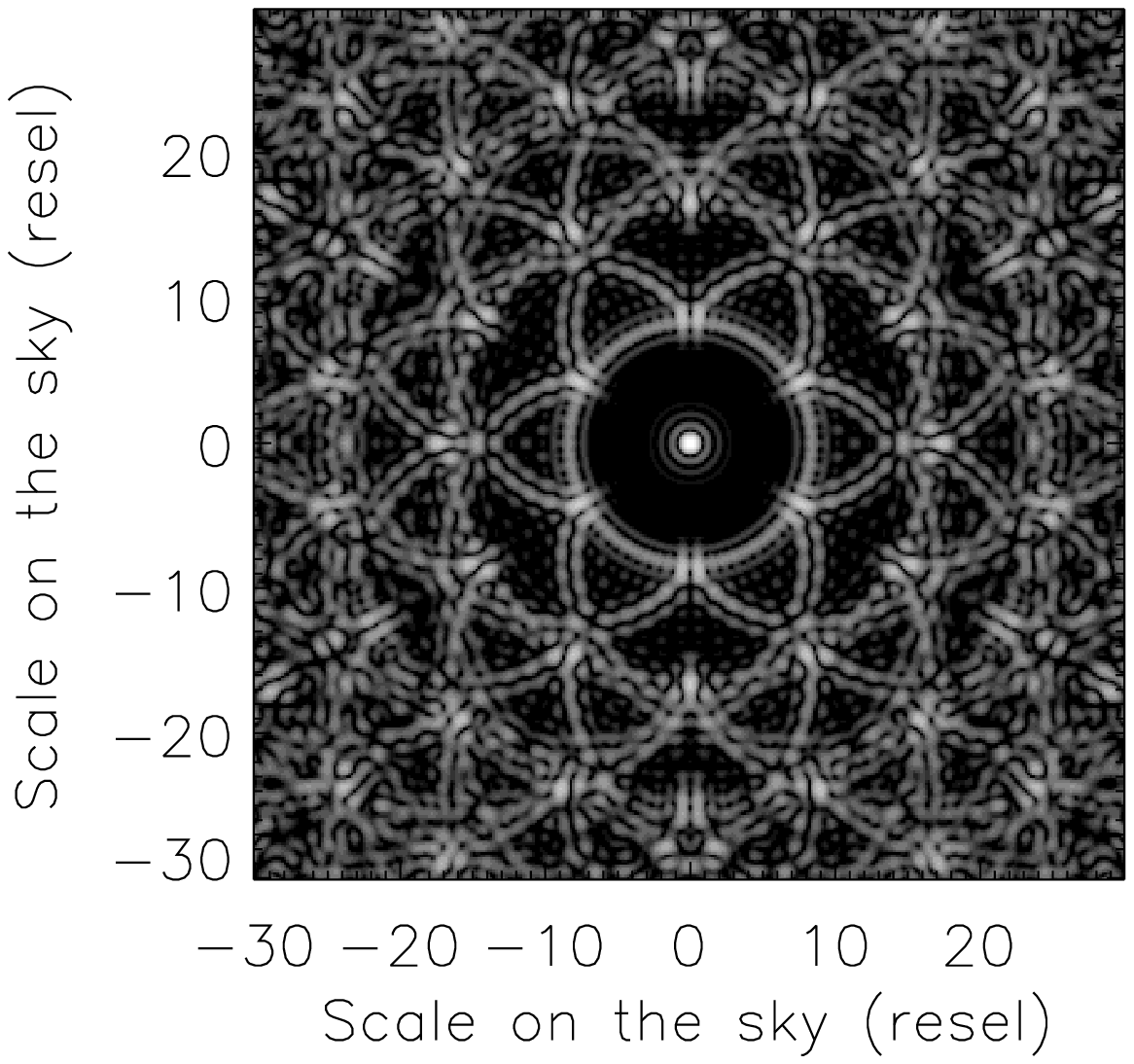} &
\includegraphics[width=40mm]{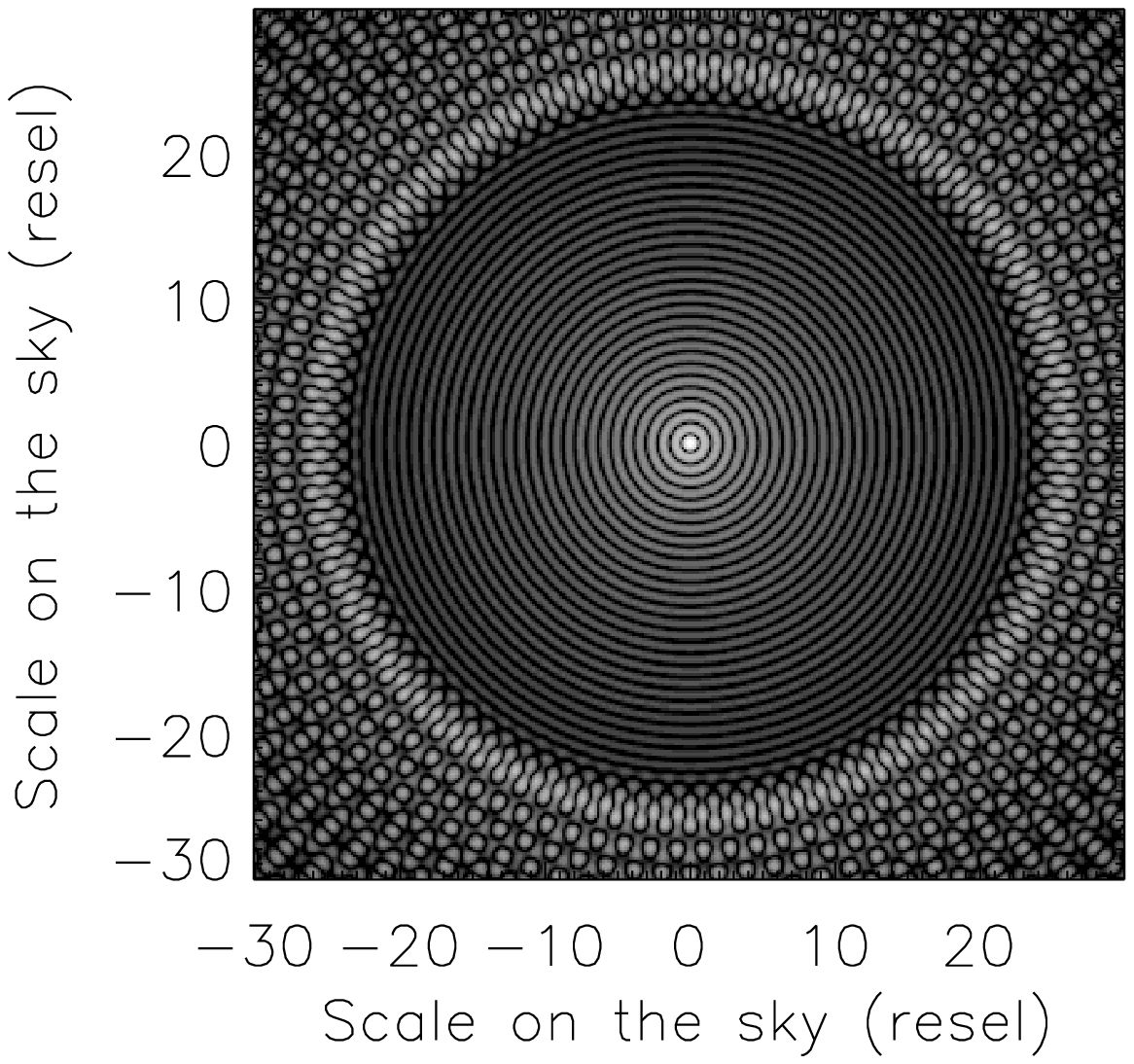} &
\includegraphics[width=40mm]{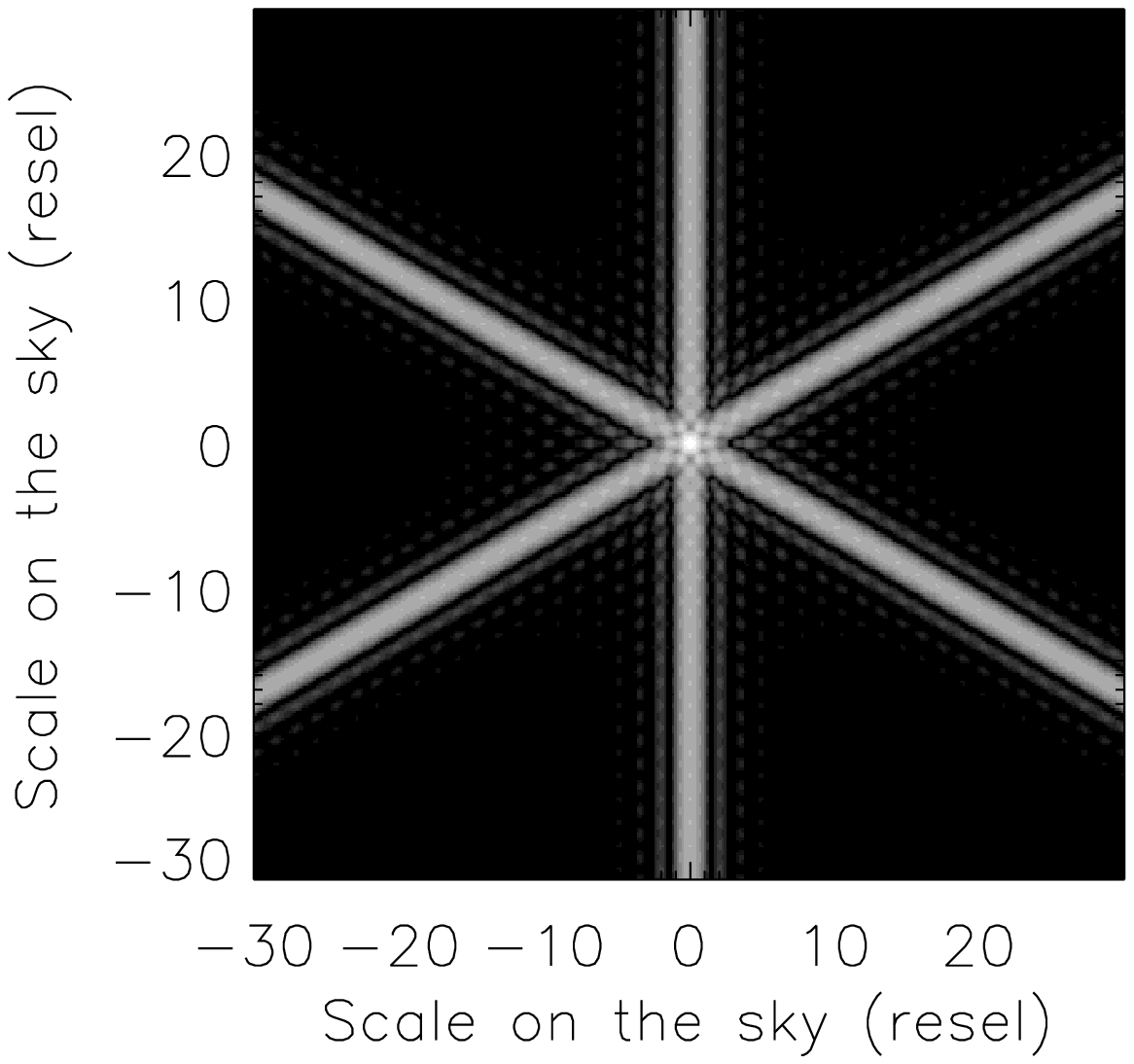} \\

\end{tabular}
\end{center}\caption{Imaging properties of 4 array configurations with $N_T$ telescopes.
Densified pupil and interference function (logarithmic scale) as
defined in Eq. \ref{equ:interf_ft} with an array of $N_T \simeq 20$
apertures (top) and $N_T \simeq 100$ apertures (bottom). The
apertures with a diameter of $10m$ are laying out on an entrance
pupil with an external diameter of $1 km$. For these images, the
intensity is normalized to the maximum on-axis intensity ($I_0=1$).
} \label{config_vs_ntel}
\end{figure*}


\subsubsection{Trade-off between halo level and field of view}

For a given resolution, it appears that, depending on the chosen
configuration, there is a trade-off between halo level and field of view.
OVLA is suitable to image large fields, since the
corresponding diffraction  envelope (dashed line) has a large full
width at half maximum. KEOPS and CARLINA are optimized for
high contrast imaging, thanks to a regular distribution of the
telescopes of the array.

The condition to reach a low halo level with a monolithic
telescope is to have an aperture without obstruction, or to use
apodization techniques \citep{Aime 2003a}. In these conditions, the
(u,v) plane coverage has a conic shape. In a similar way, the (u,v)
plane of an interferometer used for high contrast imaging must be
identical. It has been shown that the optimization of the filling of
the (u,v) plane is obtained by maximizing the integral of the
squared modulus of the Modulation Transfer Function \citep{Aime 2003b}. It consists in fact in maximizing the densified pupil
filling rate $\tau_o$ (Eq. \ref{tau}), by a regular distribution
of the sub-pupils.

\subsubsection{Trade-off between resolution and field of view}

For a given configuration, it appears also that there is a trade-off
between resolution and field. If one increases the global size of
the input pupil ($B$) with a constant number of apertures, the
resolution is improved whereas the clean field is decreased. A
compact array provides a large image with low resolution and a
diluted array provides a sharp image with high resolution. It is
interesting to benefit from a movable array with a fixed geometry
like KEOPS and with a sufficient number of telescopes. With a small
number of iterations, the telescopes could be moved keeping the same
geometry so as to finally adapt the CLean Field to the typical
dimension of the object.

\subsection{Impact of the number of apertures}\label{influ_aperture}
\begin{table*}
\begin{center}
\begin{tabular}{|ll|rrrrr|}
\hline
    &       &   KEOPS-8   &   KEOPS-21   &   KEOPS-40   &   KEOPS-65   &   KEOPS-96   \\
\hline
Entrance pupil filling rate &   $\tau_i$    &   0.78e-3   &   2.1e-3   &   3.9e-3   &   6.4e-3   &   9.4e-3   \\
Densified pupil filling rate    &   $\tau_o$    &   0.76   &   0.75   &   0.75   &   0.75   &   0.74   \\
Maximum densification level &   $\gamma_{max}$  &   44.5  &   23.3  &   15.8  &   12.0  &   9.6  \\
Clean field [mas]   &   $CLF$   &   0.28  &   0.53  &   0.78  &   1.03  &   1.28  \\
Clean field [resel]     &   $CLF$   &   1.84   &   3.52   &   5.18   &   6.83   &   8.49   \\
Direct imaging field [resel]   &   $DIF $  &   1.88   &   3.67   &   5.53   &   7.46   &   9.47   \\
\hline
Full-width half-maximum [resel] &   $FWHM$  &   0.56    &   0.66    &   0.70    &   0.72    &   0.74    \\
Encircled energy of the central peak    &   $E_0/E_{tot}$   &   0.66    &   0.67    &   0.69    &   0.72    &   0.73    \\
Maximum halo level (inside the CLF) &   $I_1/I_0$   &   0.03    &   0.02    &   0.02    &   0.02    &   0.02    \\
\hline \hline
    &       &   OVLA-9    &   OVLA-21    &   OVLA-39    &   OVLA-69    &   OVLA-96    \\
\hline
Entrance pupil filling rate &   $\tau_i$    &   0.88e-3   &   2.1e-3   &   3.8e-3   &   6.7e-3   &   9.4e-3   \\
Densified pupil filling rate    &   $\tau_o$    &   0.60   &   0.36   &   0.22   &   0.13   &   0.10   \\
Maximum densification level &   $\gamma_{max}$  &   34.8  &   14.9  &   8.1   &   4.6   &   3.3   \\
Clean field [mas]   &   CLF &   0.36  &   0.83  &   1.54  &   2.72 &   3.78 \\
Clean field [resel]     &   CLF &   2.36   &   5.48   &   10.18  &   18.0  &   25.1  \\
Direct imaging field [resel]   &   DIF     &   2.43   &   5.88   &   11.62  &   23.1  &   36.1  \\
\hline
Full-width half-maximum [resel] &   $FWHM$  &   0.53    &   0.55    &   0.55    &   0.55    &   0.55    \\
Encircled energy of the central peak    &   $E_0/E_{tot}$   &   0.42    &   0.20    &   0.11    &   0.07    &   0.05    \\
Maximum halo level (inside the CLF) &   $I_1/I_0$   &   0.13    &   0.16    &   0.16    &   0.16    &   0.16    \\

\hline
\end{tabular}
\caption{Imaging parameters of the configurations KEOPS (upper part)
and OVLA (lower part) as a function of the number of telescopes. The
aperture's diameter is $10m$ and the maximum baseline $1 km$, so
that the $resel$ is $0.12~mas$ and the Coupled Field $CF$ is
$82~resels$.}
\end{center}
\label{tab:array_vs_ntel_gam_fov_keops}
\end{table*}

\subsubsection{Presentation of the simulations}

We now consider the same four array configurations but with a
variable number of telescopes (up to 100) of diameter $10m$
distributed over a constant maximum baseline of $1km$. 

In order to
keep the geometry of each configuration, we use the following
principles for computing the different arrays:

\begin{itemize}
\item ELSA : at each step, we add one telescope on each arm of the Y, starting at 3 telescopes, then 6 and up to 99.
\item OVLA : at each step, 1 telescope is added on the ring with a diameter of 1 km, starting at 2 telescopes, then 3 and up to 100. The telescopes are regularly distributed in azimuth.
\item KEOPS : we start with a telescope at the centre and one
concentric ring of 7 telescopes. Then we add successively concentric
rings made of 13, 19, 25 and 31 telescopes. The array is
successively composed of 8 (1 ring), 21 (2 rings), 40 (3 rings), 65
(4 rings) and 96 (5 rings) telescopes. The diameter of the
largest ring always equals 1 km.
\item CARLINA : at each step, the array is build with $n^2$ telescopes
regularly distributed over a square grid, with $n$ from 2 to 11. The
telescopes outside the circle of diameter $B$ are removed. The array
is composed successively of 4, 9, 12, 21, 32, 37, 52, 69, 80 and 97
telescopes.
\end{itemize}

The intensity is normalized by the collecting surface area of each array so that the sensitivity considered here is independent of the number of telescopes.
\\

\subsubsection{PSF quality parameters}

\begin{figure}
\begin{center}
\begin{tabular}{cccccc}
\includegraphics[width=55mm]{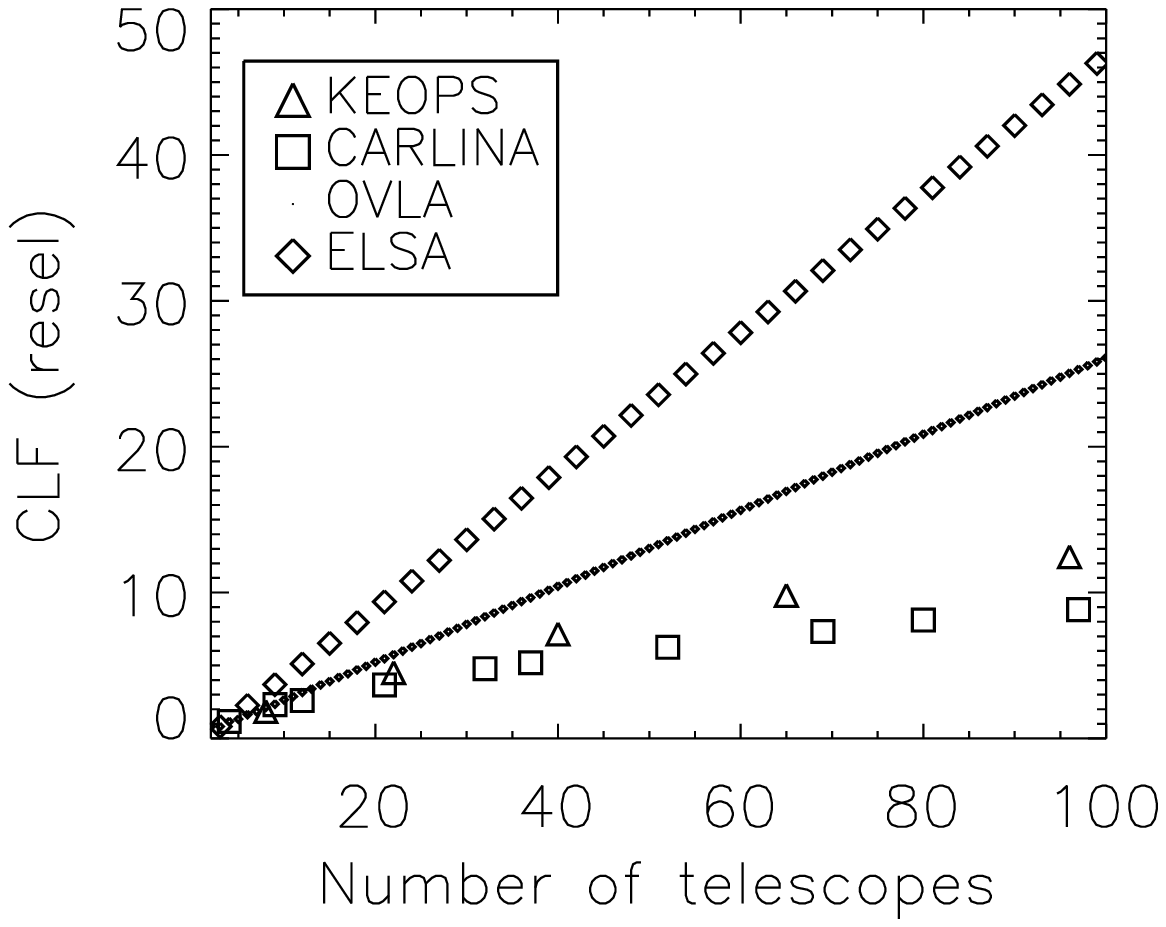} \\
\includegraphics[width=55mm]{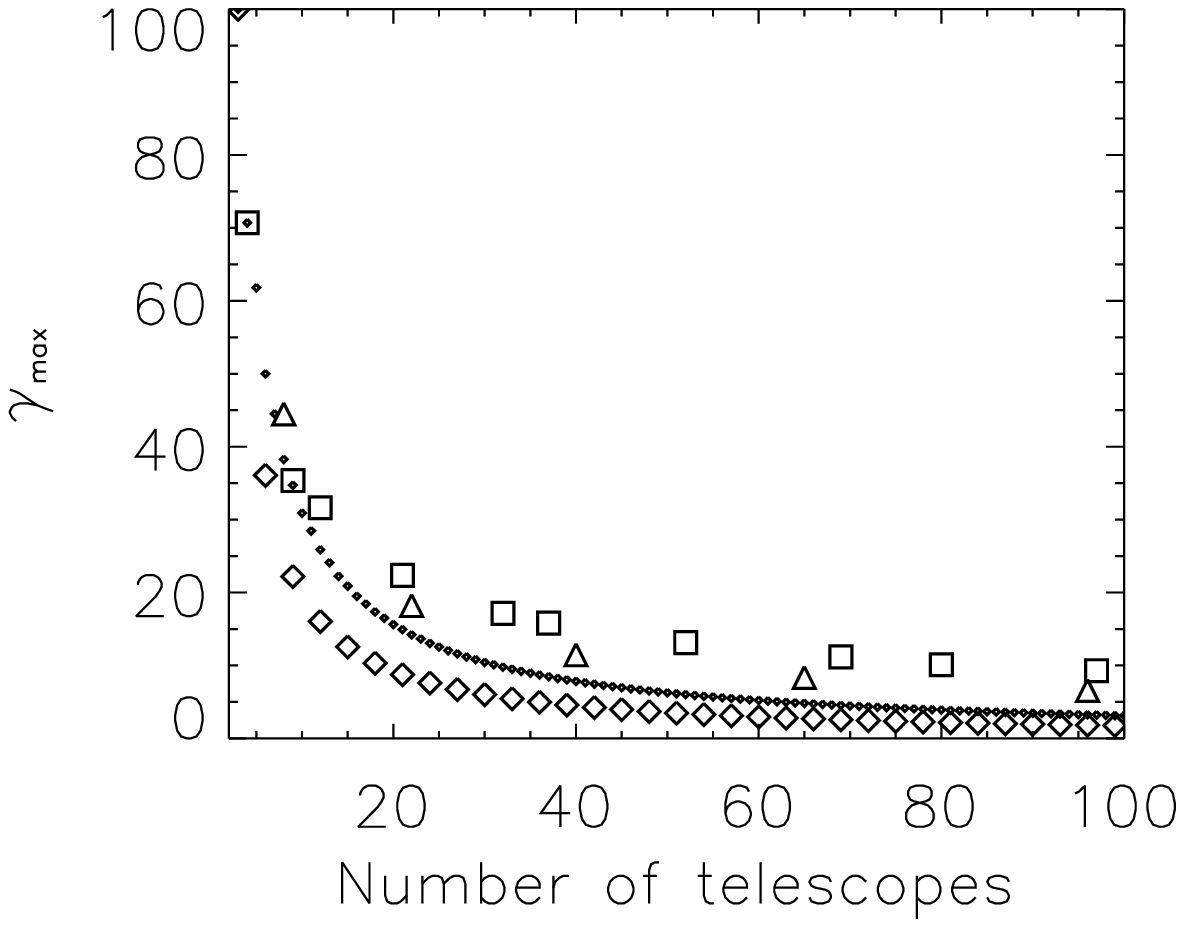} \\
\end{tabular}
\end{center}\caption{Clean Field (top) and densification level (bottom) of arrays as a function of the number of telescopes.}
 \label{fig:array_vs_ntel_gam_fov}
\end{figure}

\begin{figure}
\begin{center}
\begin{tabular}{cccccc}
\includegraphics[width=55mm]{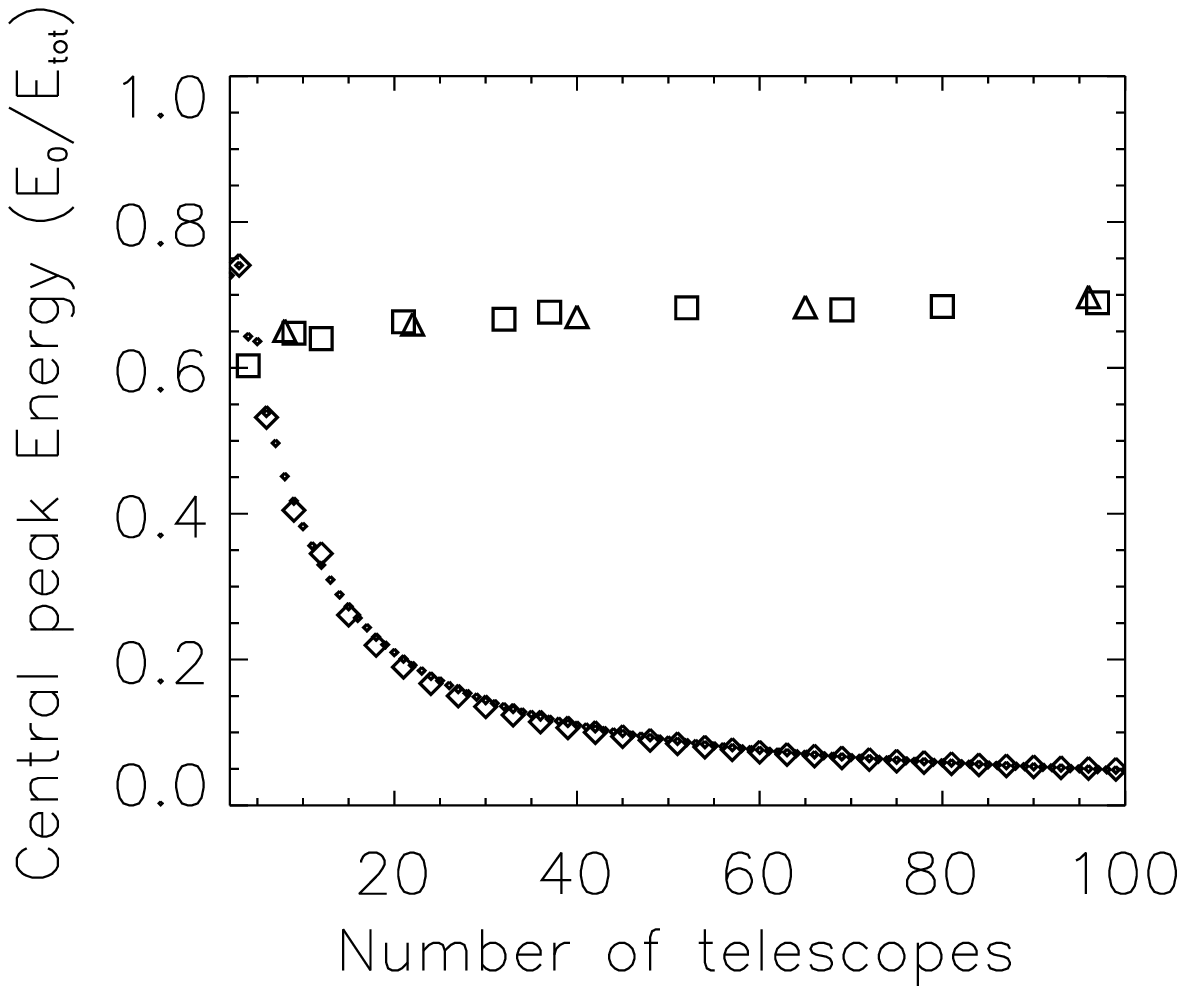} \\
\includegraphics[width=55mm]{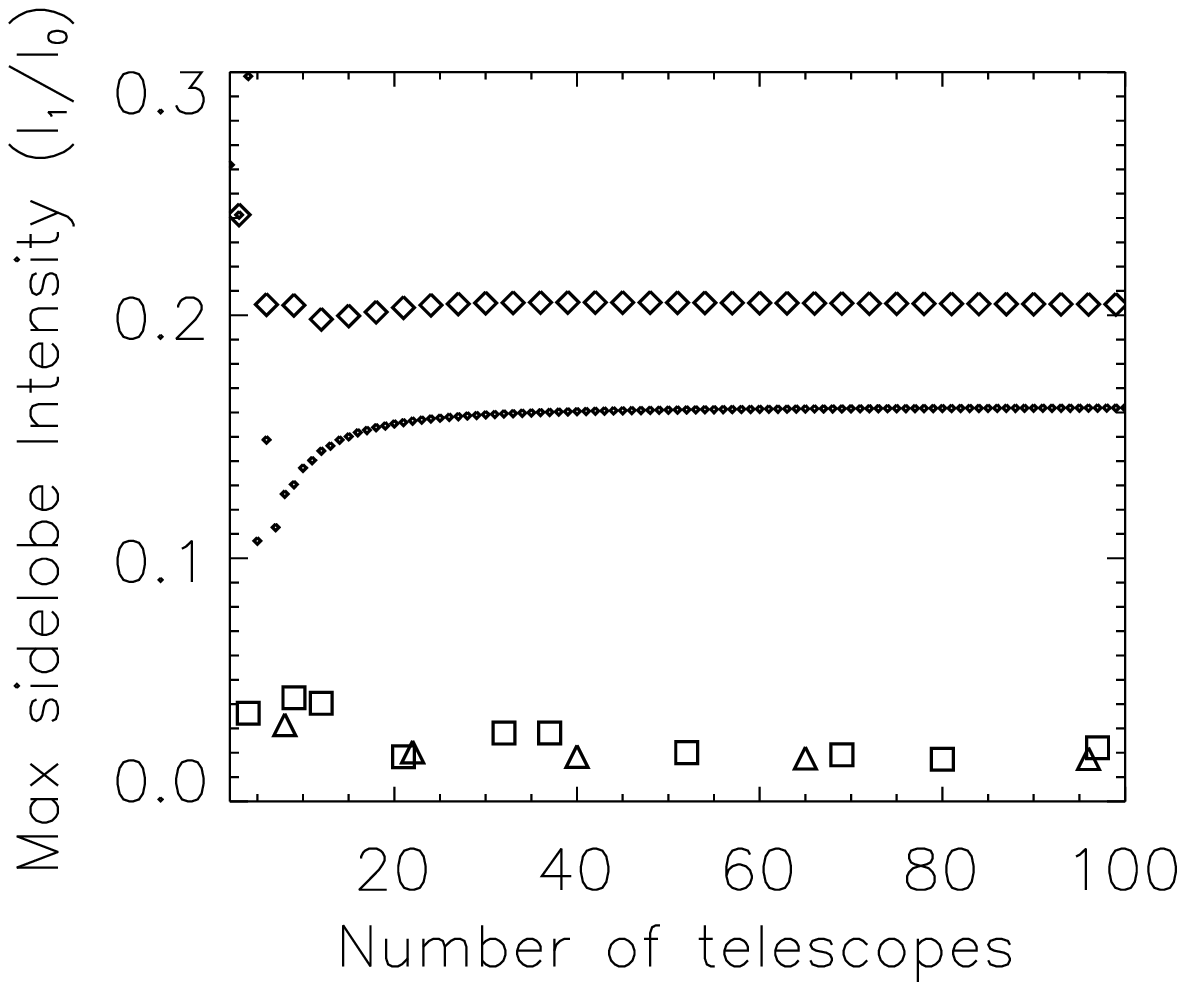} \\
\includegraphics[width=55mm]{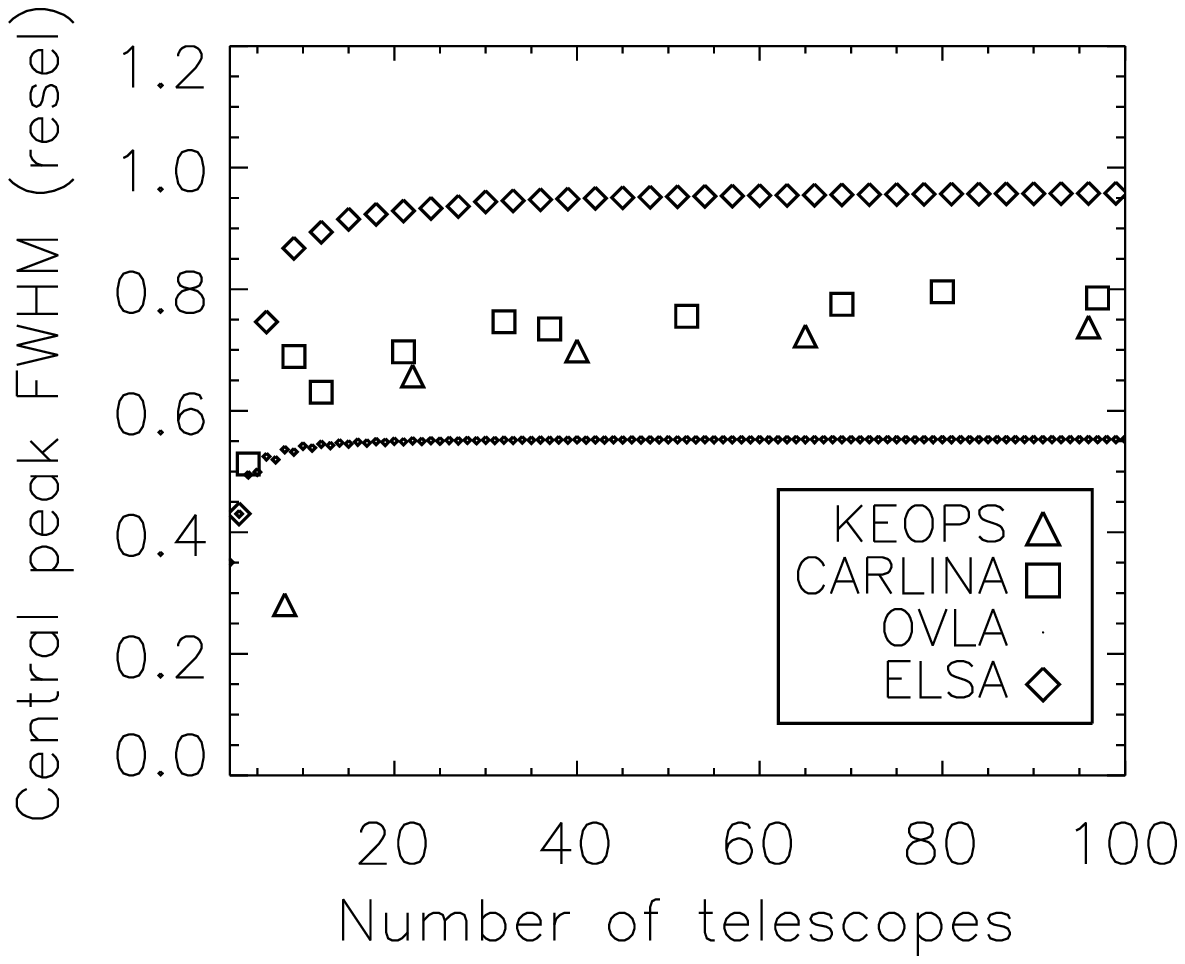} \\
\end{tabular}
\end{center}\caption{PSF parameters in DP mode as a function of the number of telescopes.
From top to bottom: Encircled energy, the maximum halo level, the FWHM of the
central peak} \label{fig:array_vs_ntel_fep}
\end{figure}

\begin{figure}
\begin{center}
\begin{tabular}{cccccc}
\includegraphics[width=55mm]{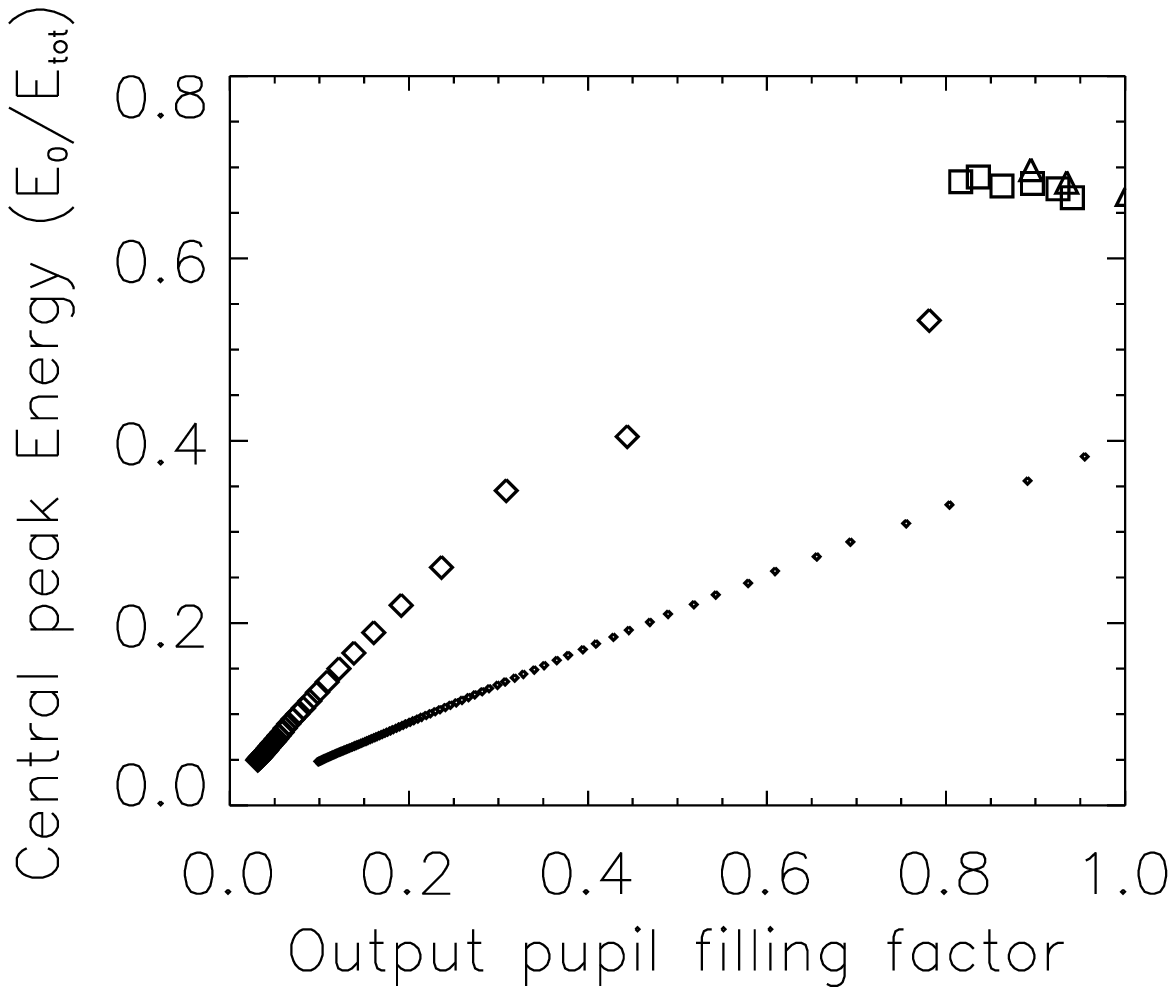} \\
\includegraphics[width=55mm]{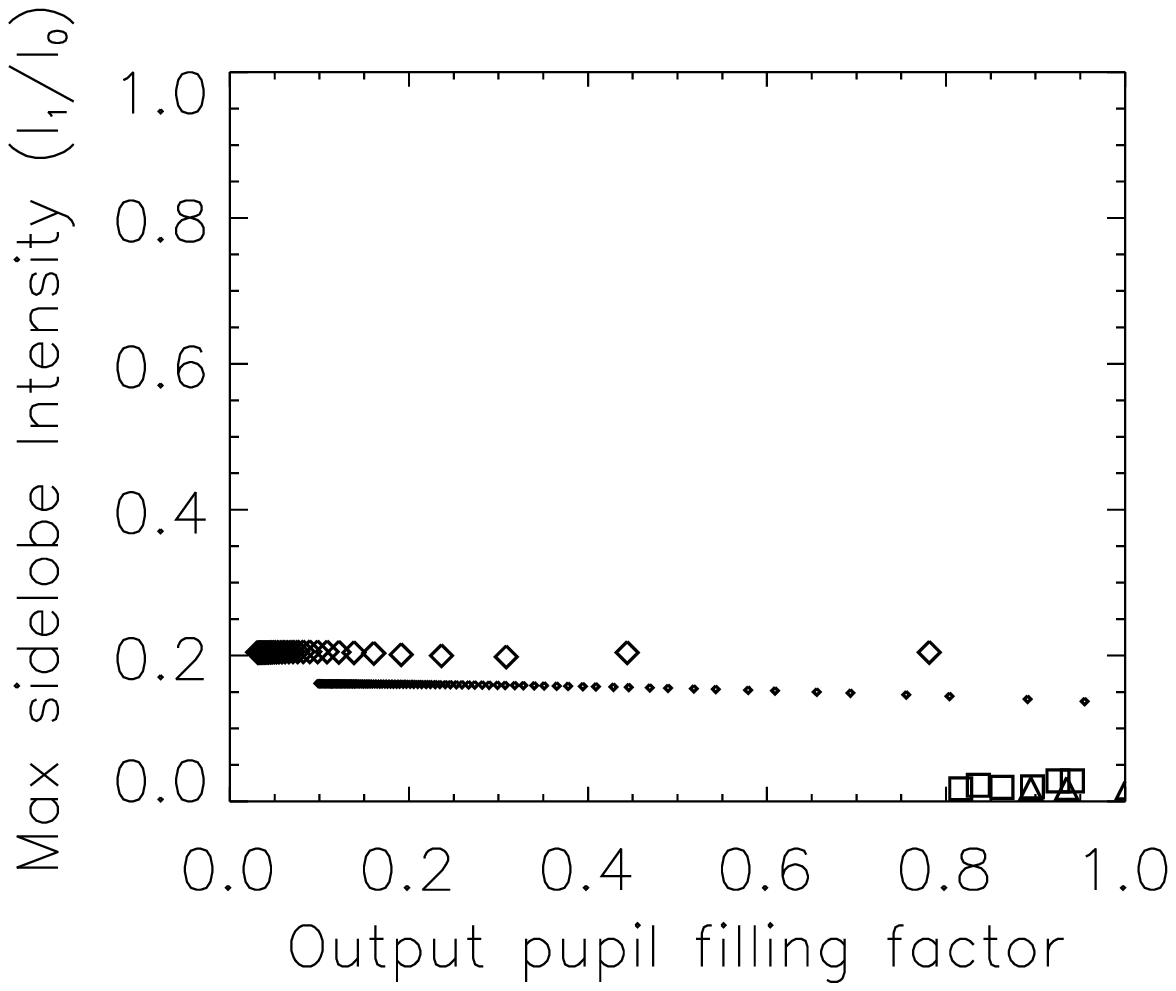} \\
\end{tabular}
\end{center}\caption{Encircled energy (top) and the maximum halo level (bottom) of the densified PSF in DP mode as a function of $\tau_o$. $B$ is constant and $N_T$ increases as on Fig.~\ref{fig:array_vs_ntel_gam_fov} or
Fig.~\ref{fig:array_vs_ntel_fep}. The symbols represent the
different configurations as on Fig.~\ref{fig:array_vs_ntel_gam_fov}
or Fig.~\ref{fig:array_vs_ntel_fep}.} \label{fig:array_vs_ro_fep}
\end{figure}

Figure~\ref{config_vs_ntel} shows the densified pupils and the interference function (as defined in Eq. \ref{equ:interf_ft}) of the four configurations with 20 and 100 telescopes.
Figure~\ref{fig:array_vs_ntel_gam_fov} and Table~2 give the evolution of the main parameters of the arrays as a function of the number of telescopes. Figure \ref{fig:array_vs_ntel_fep} gives the evolution of the densified PSF parameters as a function of the number of telescopes for each configuration.
Figure \ref{fig:array_vs_ro_fep} shows the correlation between the PSF
parameters and the densified pupil filling rate.
\\

Whatever the number of telescopes, CARLINA and KEOPS benefit from a
quasi-complete densified pupil filling rate, so that their imaging properties (inside the clean field) are very close to a monolithic telescope. The encircled energy in the central peak contains about $70\%$ of the luminous energy and the contribution of the halo remains below $3\%$ inside the clean field.
CARLINA and KEOPS are a priori equivalent in term of image quality, regarding to the halo level and the encircled energy.

For OVLA and ELSA, the output pupil shape is similar whatever the
number of telescopes. When the number of telescopes increases in
these arrays, the densification level remains low due to the
shortest baselines. The encircled energy in the central peak falls
from $40\%$ to only $5\%$, regarding the configurations from 10 to
100 telescopes. The halo level is not negligeable for ELSA
(20$\%$) and OVLA (15$\%$).

The sharpness of the image is characterized by the FWHM of the
central peak. The narrowest peak is provided by OVLA, thanks to the
huge central obstruction of the input pupil. ELSA exhibits strong
diffraction spikes but the PSF looks very sharp in the other
position angles. For KEOPS and CARLINA, the FWHM of the central peak
increases slowly with the number of telescopes.

\section{Biases induced on the PSF}\label{psf biases}

In the previous section, we have studied various
configurations of future arrays. This study has allowed us to
characterize them with quantitative parameters. It appears clearly
that, whatever the configuration, the imaging properties
will be degraded by difficulties in restoring the photometric
parameters in the field. Three main effects are identified:
\begin{itemize}
\item the bias of the interference function,
\item the space aliasing effect,
\item the bias of the diffraction envelope.
\end{itemize}

The two first depend on the array configuration, whereas the last bias is only related to the recombination mode.

\subsection{Bias of the interference function}

The photometric parameters of a source are biased by the halo of the interference function (as defined in Eq. \ref{equ:interf_ft}), which induces a contrast loss in the image.
The quality of the interference function of the array is simply related to the actual shape of the entrance pupil.

If the input sub-pupils are distributed regularly, the
densified pupil is almost complete, and the halo inside the clean
field reproduces the diffraction pattern of a large monolithic
telescope covering all the sub-pupils. If the densified pupil shows
gaps, additional diffraction figures are added to first ones.

Thus, the halo is minimized by maximizing the densified pupil
filling rate $\tau_o$, with a regular pattern of the sub-apertures
in the entrance pupil.


\begin{figure*}
\begin{center}
\begin{tabular}{cccccc}

CARLINA & KEOPS & OVLA & ELSA\\

\includegraphics[width=40mm]{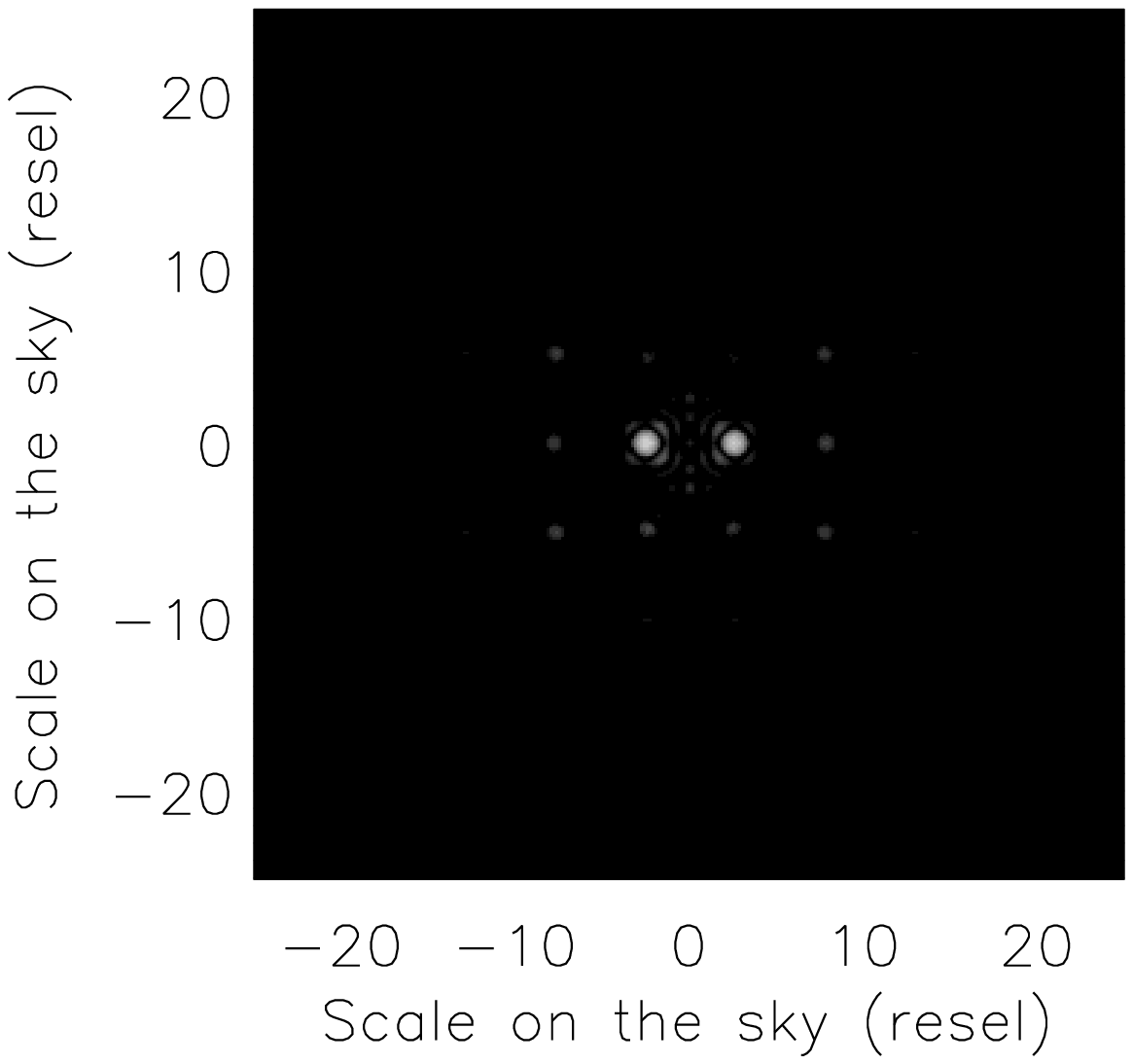} &
\includegraphics[width=40mm]{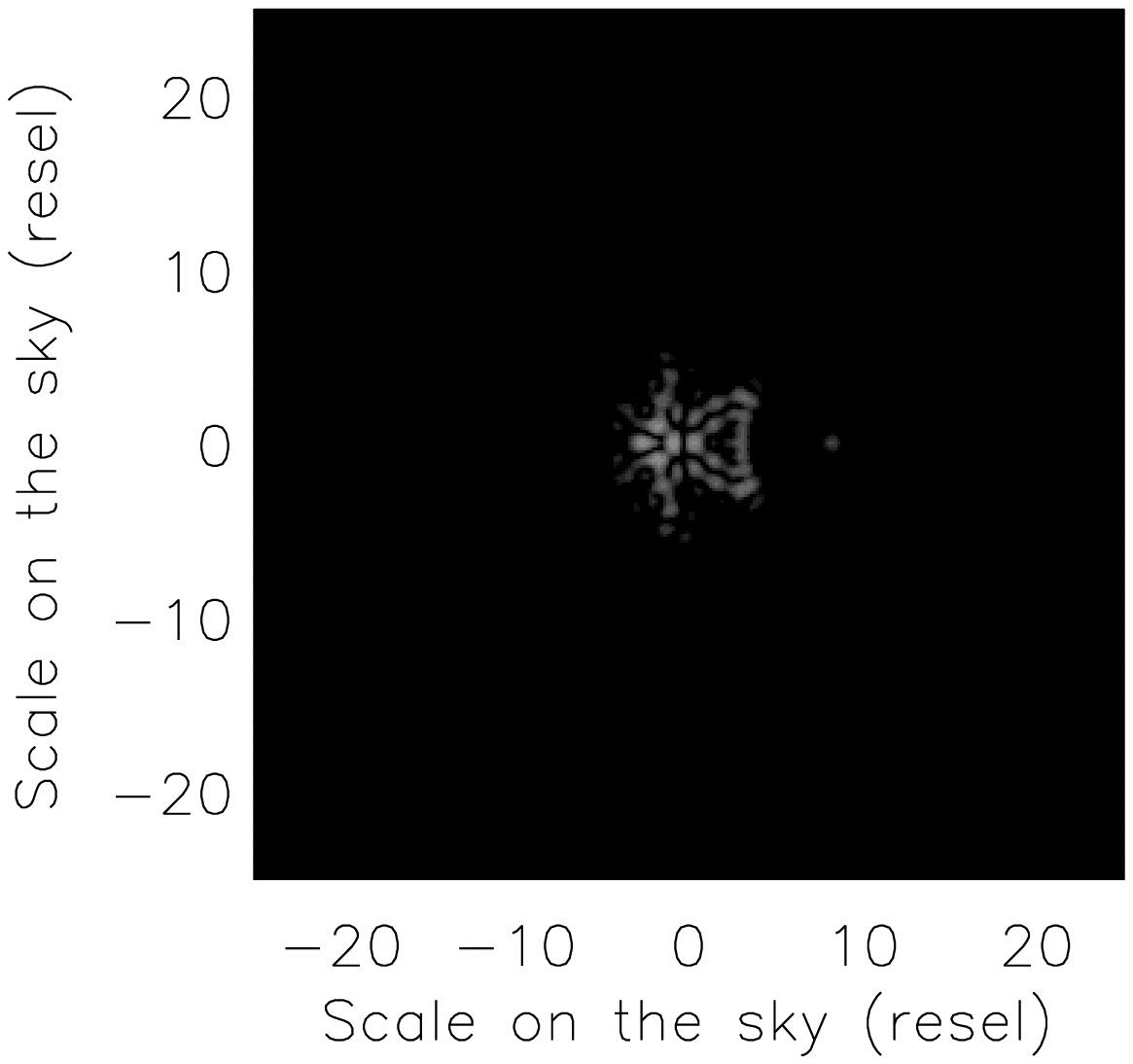} &
\includegraphics[width=40mm]{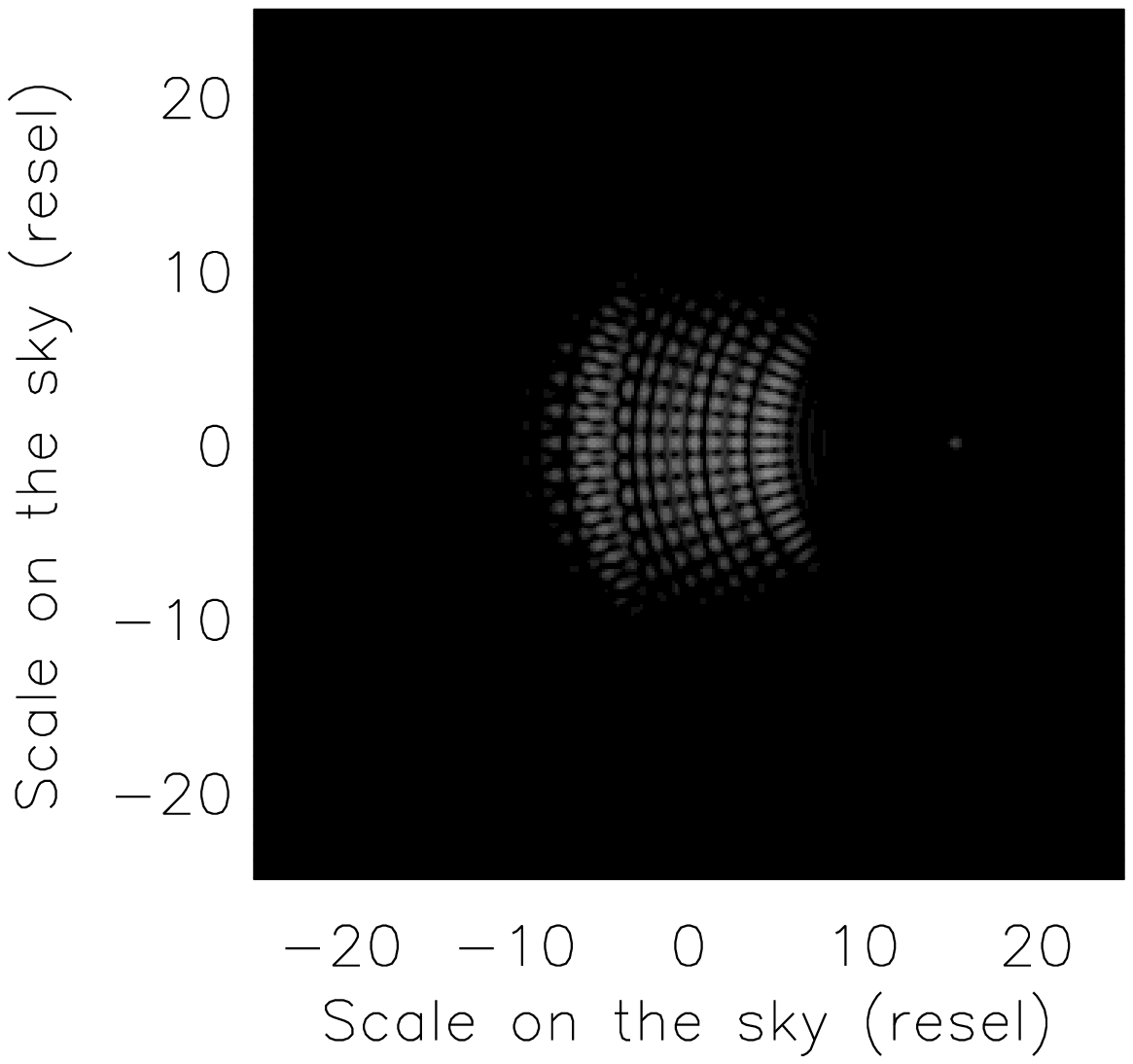} &
\includegraphics[width=40mm]{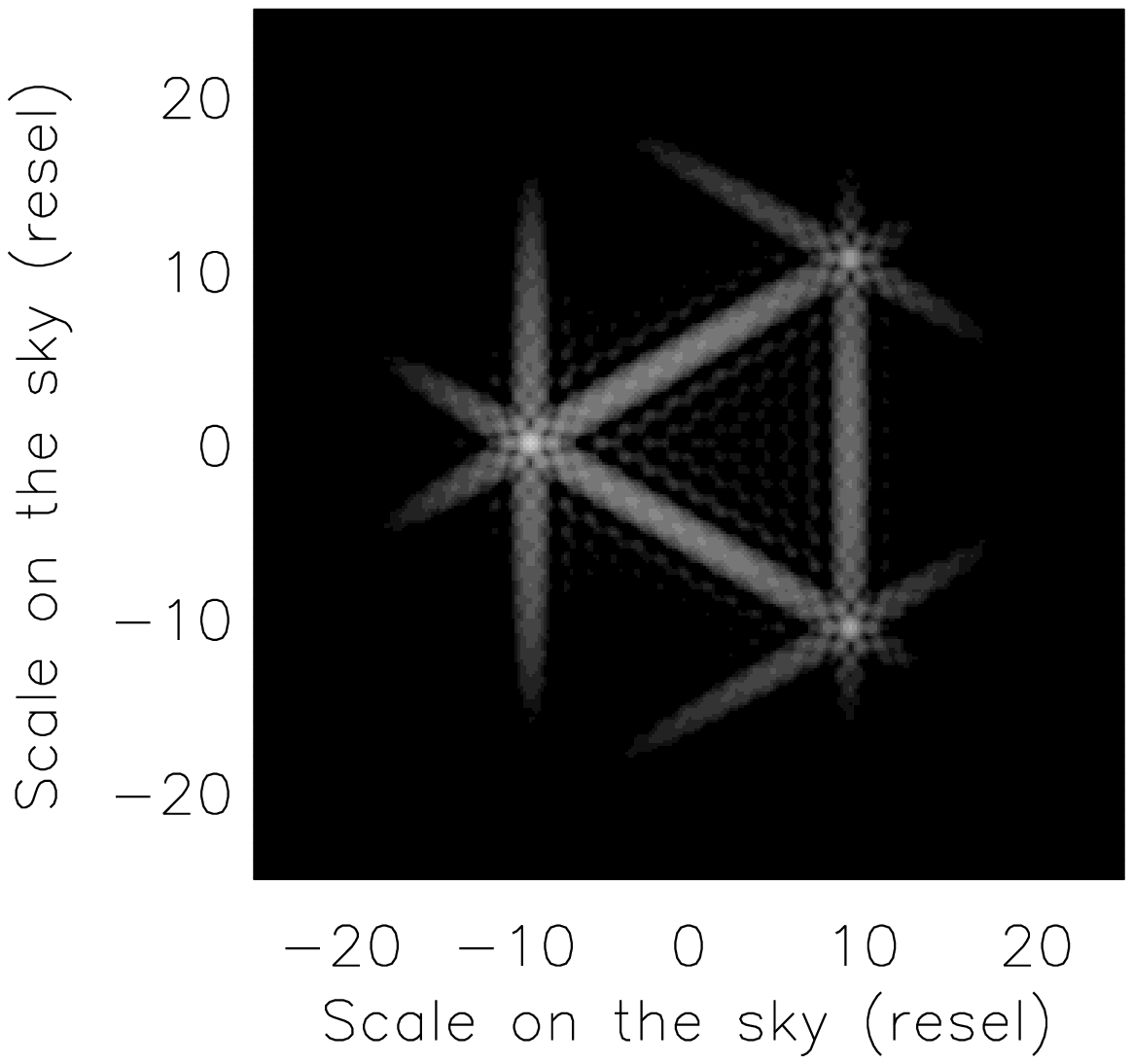} \\

\includegraphics[width=40mm]{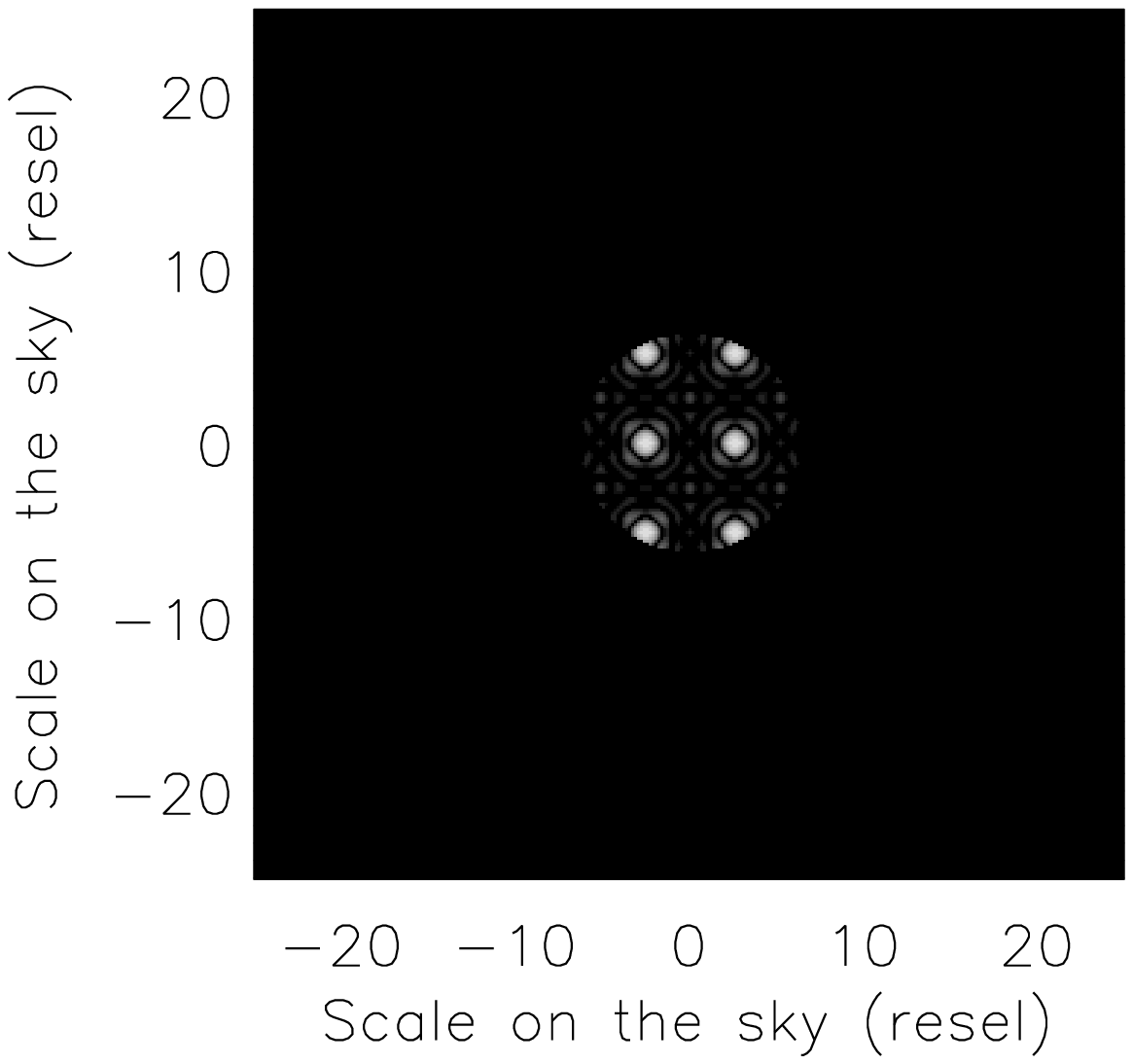} &
\includegraphics[width=40mm]{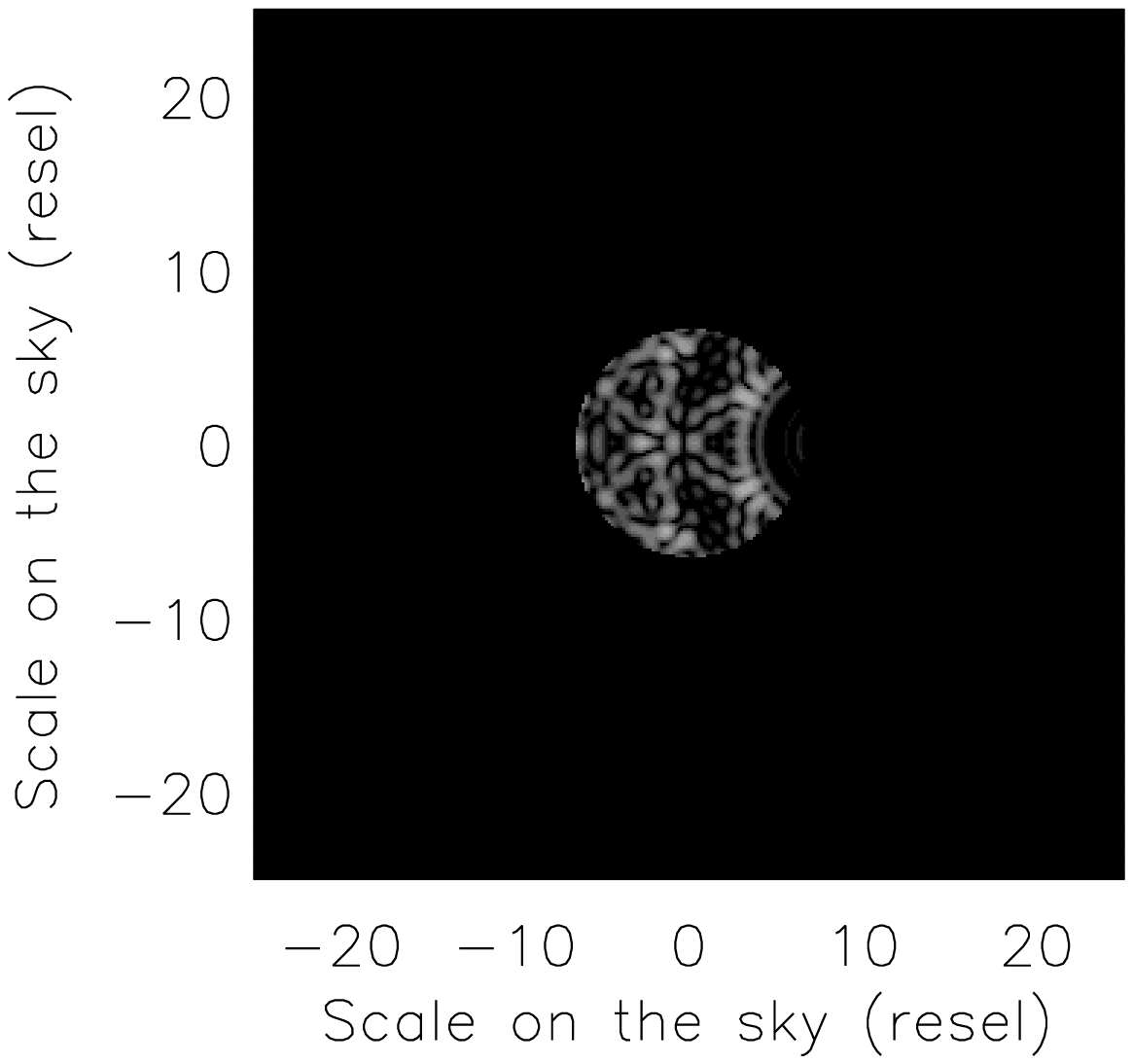} &
\includegraphics[width=40mm]{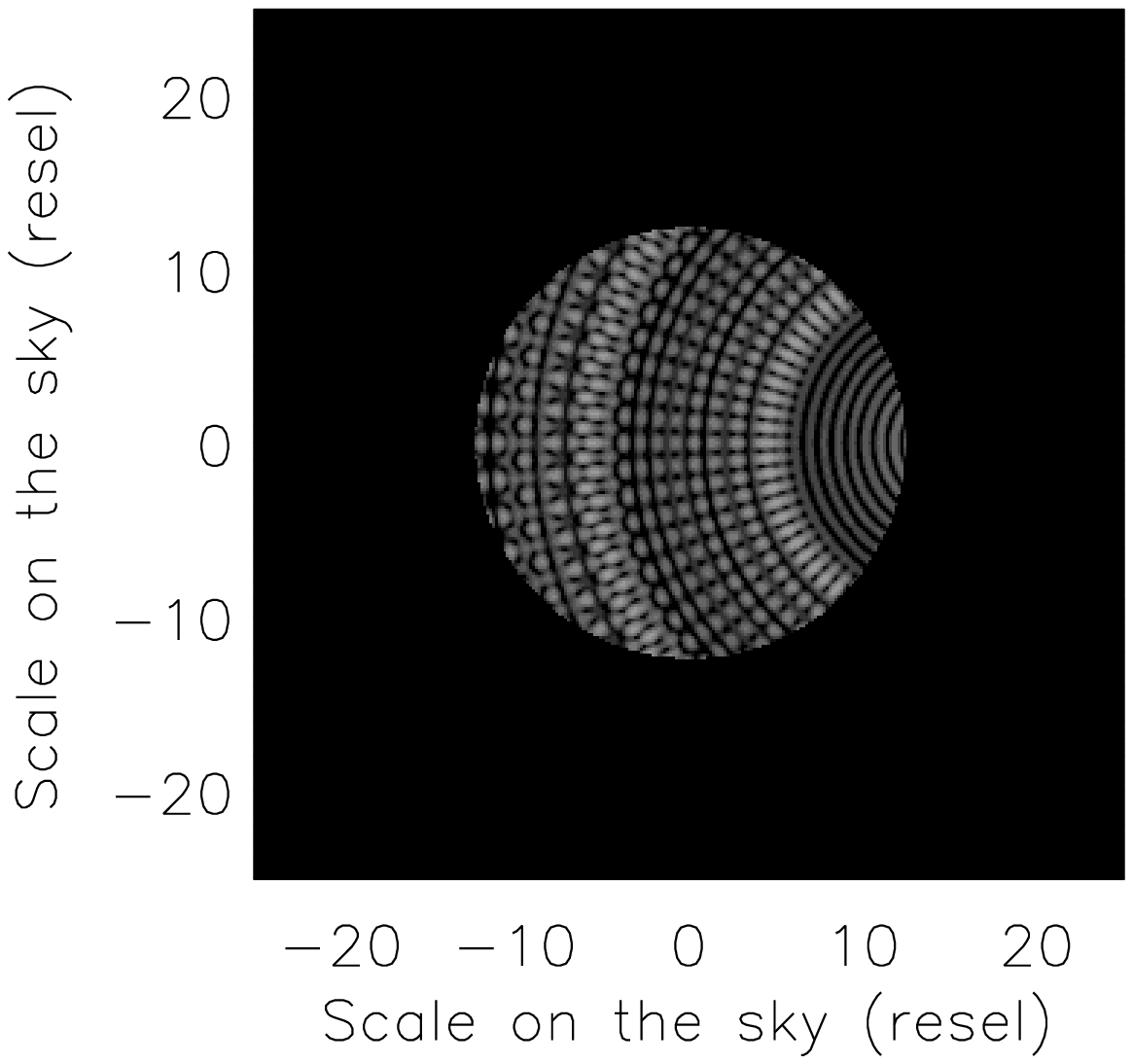} &
\includegraphics[width=40mm]{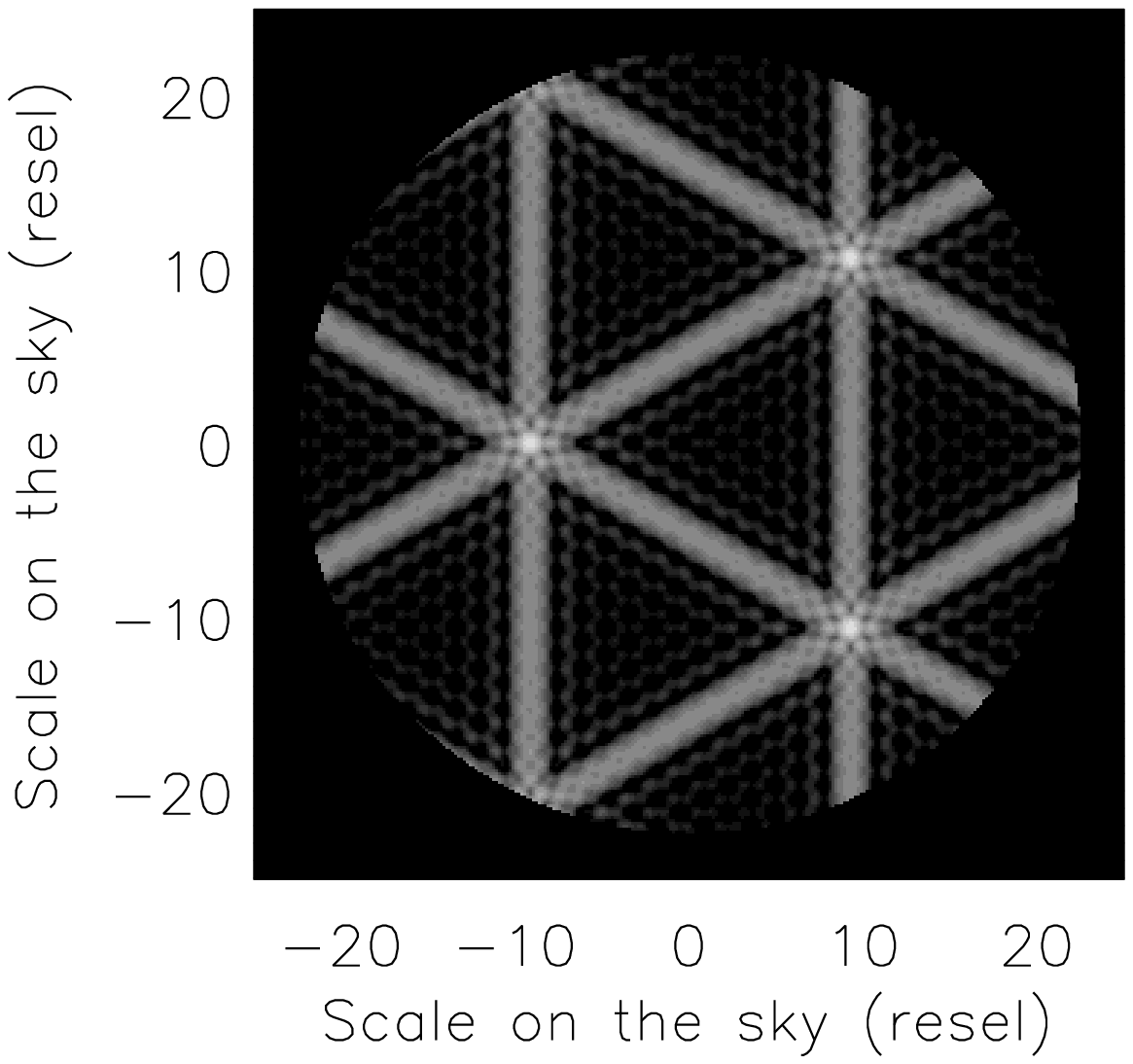} \\

\end{tabular}
\end{center}\caption{
Image (logarithmic scale) of an off-axis star simulated for 4 array configurations with 40 telescopes in DP mode (up) and in IRAN mode (down).
The off-axis position equals to 1.5 times the CLean Field, so that the star is outside the CLean field and inside the Coupled Field.
Due to the space aliasing effect, one or several ghost stars appear if the array configuration is redundant (CARLINA and ELSA), contrary to the non redundant cases (KEOPS and OVLA).
}\label{fig:offaxis}
\end{figure*}


\begin{figure*}
\begin{center}
\begin{tabular}{cccccc}

DP mode ($\gamma=\gamma_{max}$) & DP mode ($\gamma=\gamma_{max/2}$) & IRAN mode ($\gamma=\gamma_{max/2}$) & \\

\includegraphics[width=47mm]{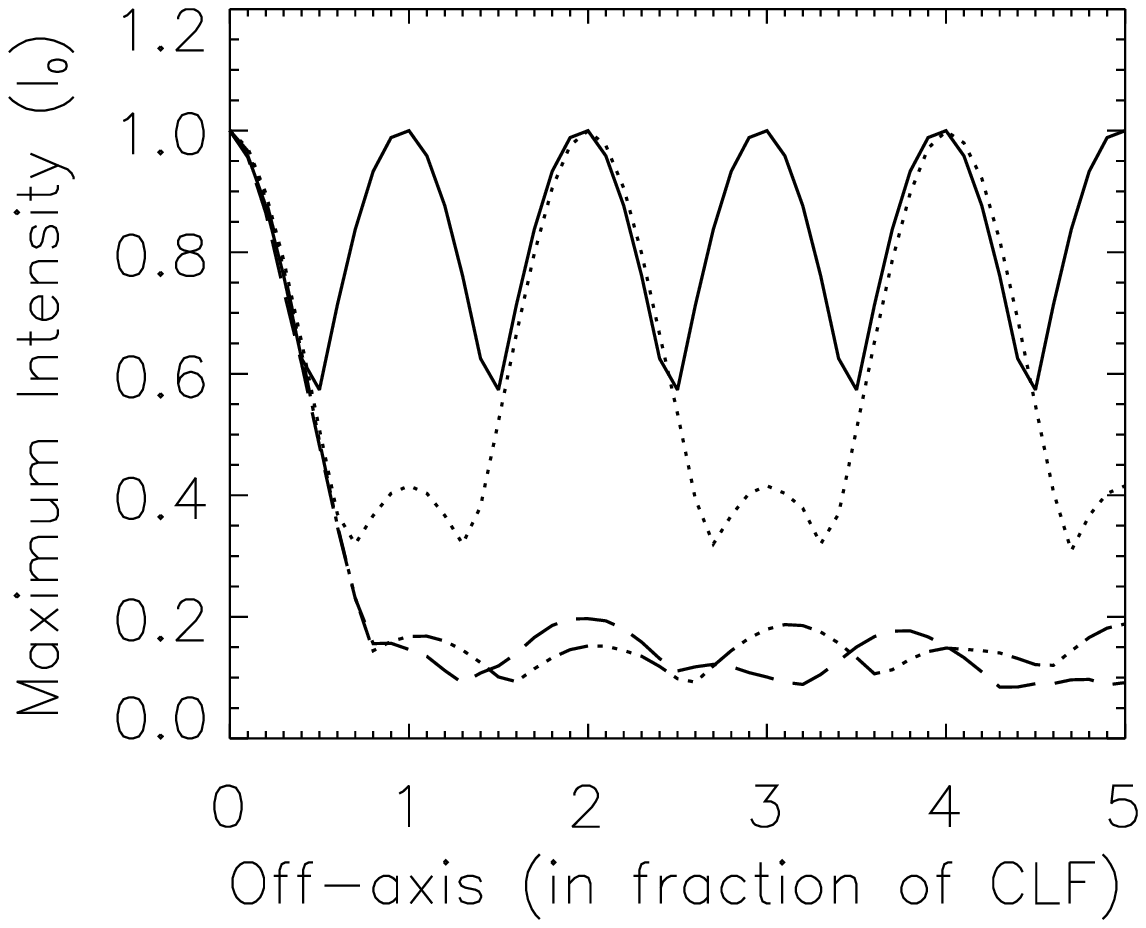} &
\includegraphics[width=47mm]{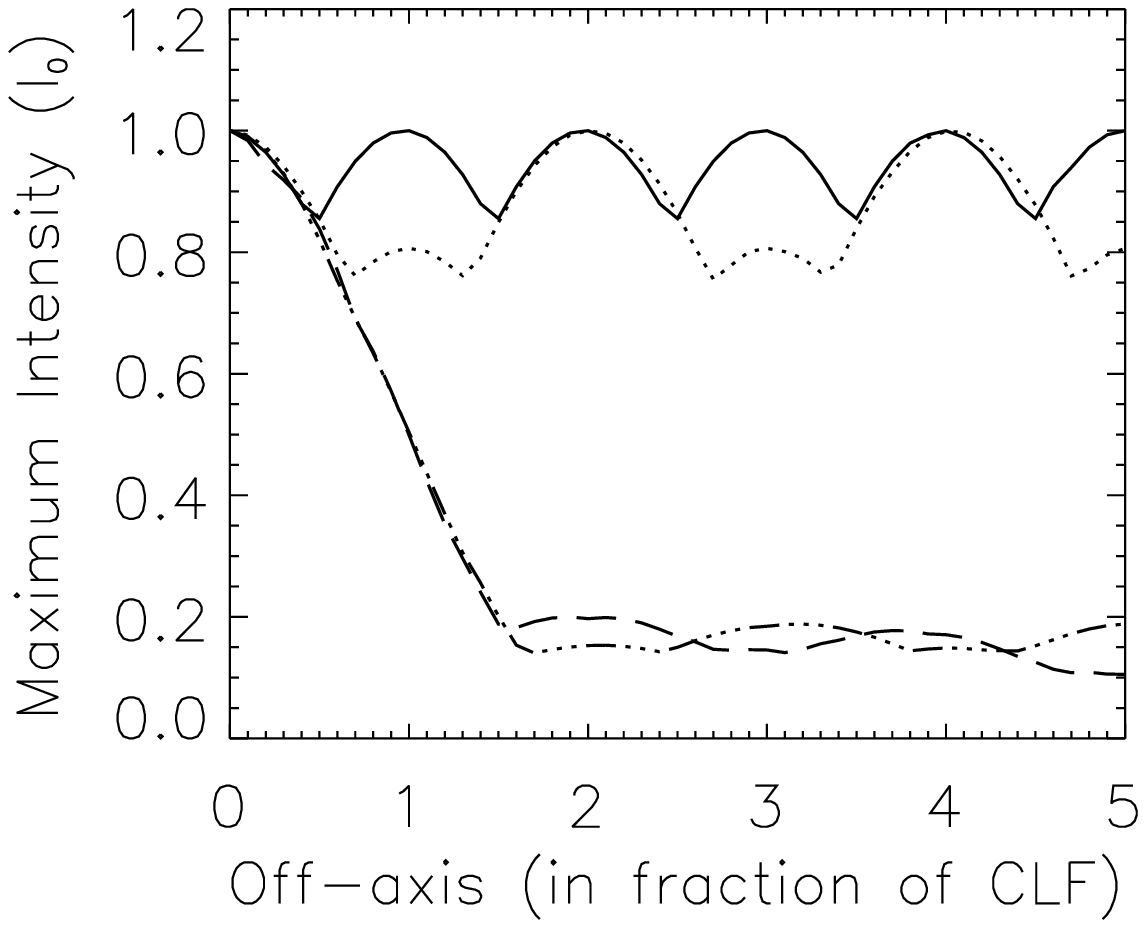} &
\includegraphics[width=47mm]{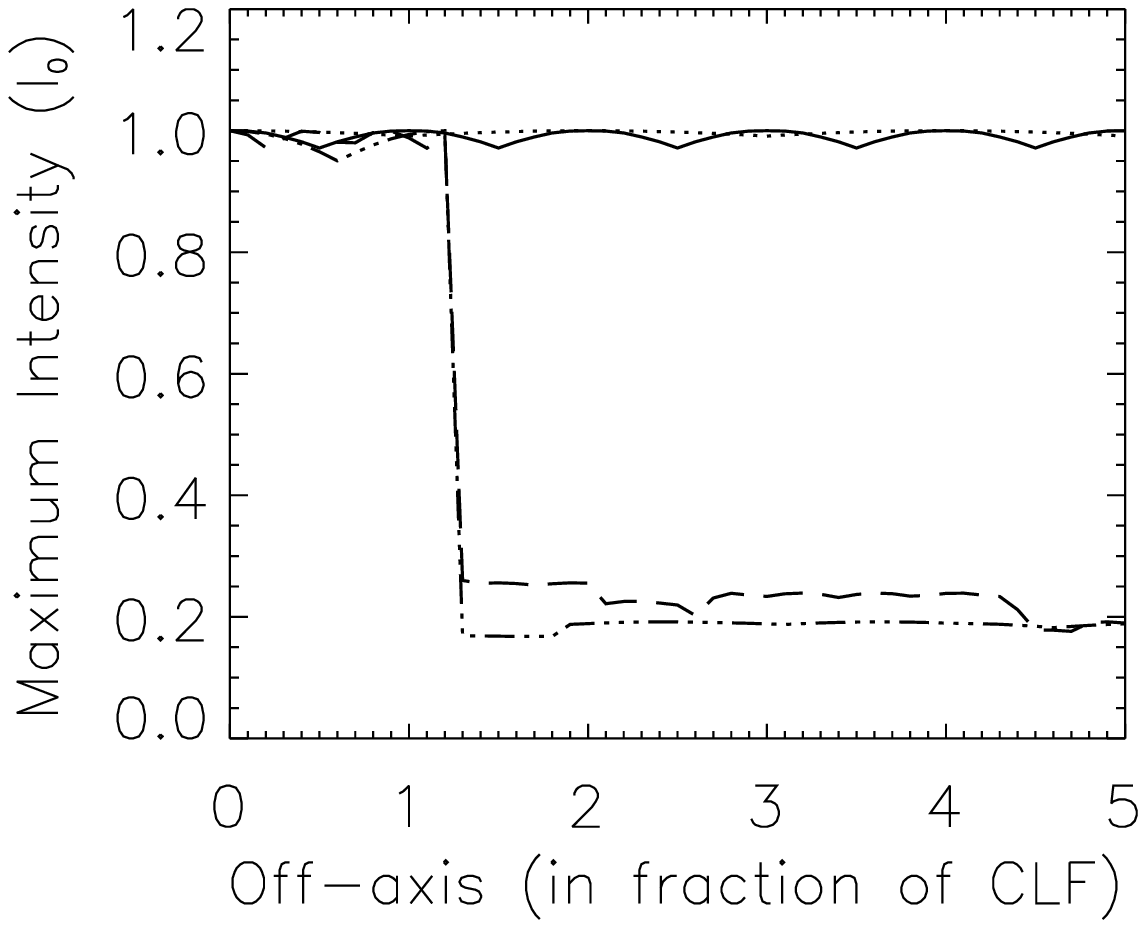} &
\includegraphics[width=20mm]{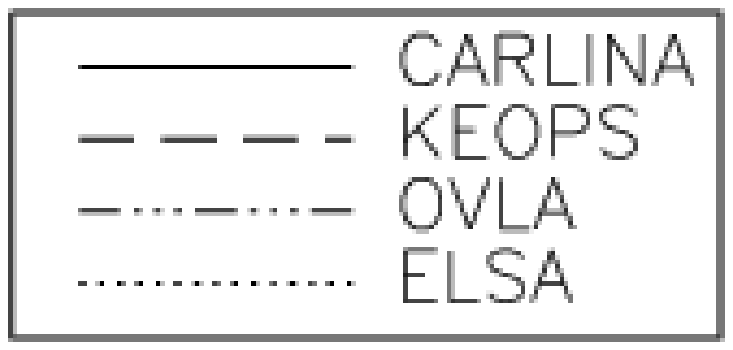} \\

\end{tabular}
\end{center}\caption{
Evolution of the maximum intensity in the CLean Field as a function of the radial position of an off-axis star. The intensity has been normalized to 1 when the star is on-axis, so that the intensity in DP$\gamma_{max}$ mode (left) is in practice 4 times higher than DP$\gamma_{max/2}$ mode (middle) and IRAN mode (right).
If the off-axis star equals to CLF/2, the intensity reach respectively $0.50$, $0.85$ and $1$ for the 3 modes (left to right), due to the diffraction enveloppe in DP mode (no effect in IRAN mode).
If the star is outside the CLF, a redundant array (CARLINA and ELSA) induces replication of the main lobe in the CLF. A non redundant array (KEOPS and OVLA) only induces a diffused halo lower than $0.25$ in the CLF.
} \label{fig:offaxis_var}
\end{figure*}

\subsection{Space aliasing effect}

The space aliasing effect \citep{Aime 2008} appears in the direct image when the science object is surrounded by sources outside the clean field but inside
the coupled field, or when the science object diameter is larger
than the clean field. Besides, all the sources in the CF
will contribute to the central image but only the sources in the CLF
will form a correct central peak plus a halo of sidelobes, whereas
the sources in the CF and outside the CLF will just form a halo of
sidelobes. These side-lobes induce photometric perturbations locally
distributed in the image.
\\

Redundant arrays are more affected by the space aliasing effect than non redundant arrays. Indeed, a redundant array provides an interference function with high level parasit peaks, whereas a non redundant one has a smooth interference function . In the image formation, these differences in the interference functions will introduce important space aliasing effects as shown on Fig. \ref{fig:offaxis} and \ref{fig:offaxis_var}.

In the redundant case, a source in the
Coupled Field but outside the Clean Field provides ghost images inside the Clean Field.
In the non redundant case, the same star will only induce a diffused halo reducing the contrast in the Clean Field.
\\

\subsection{Bias of the diffraction envelope}

Due to the diffraction envelope contribution, the quality of the
photometry restitution decreases from the axis to the edge of the
clean field (Fig. \ref{fig:offaxis_var}).
A partial densification restitutes a more homogeneous
photometry in the clean field, but decreases the sensitivity gain.
This bias does not exist in IRAN mode, where the envelope is flat.

\subsection{Discussion}

The main effects on the PSF  can be theoretically corrected
by image restoration or deconvolution.
Deconvolution techniques are required for complex objects.
In the case of a densified image, the problem is that the convolution
relationship is lost. The image and the PSF are in fact partially
truncated, which is a problem for the classical methods of
deconvolution. To overcome this problem, a hybrid method, based on
likelihood maximization, reconstructing simultaneously the object
and the PSF has been proposed \citep{Aristidi 2006}.

\begin{figure}
\begin{center}
\begin{tabular}{ccc}
\includegraphics[width=85mm]{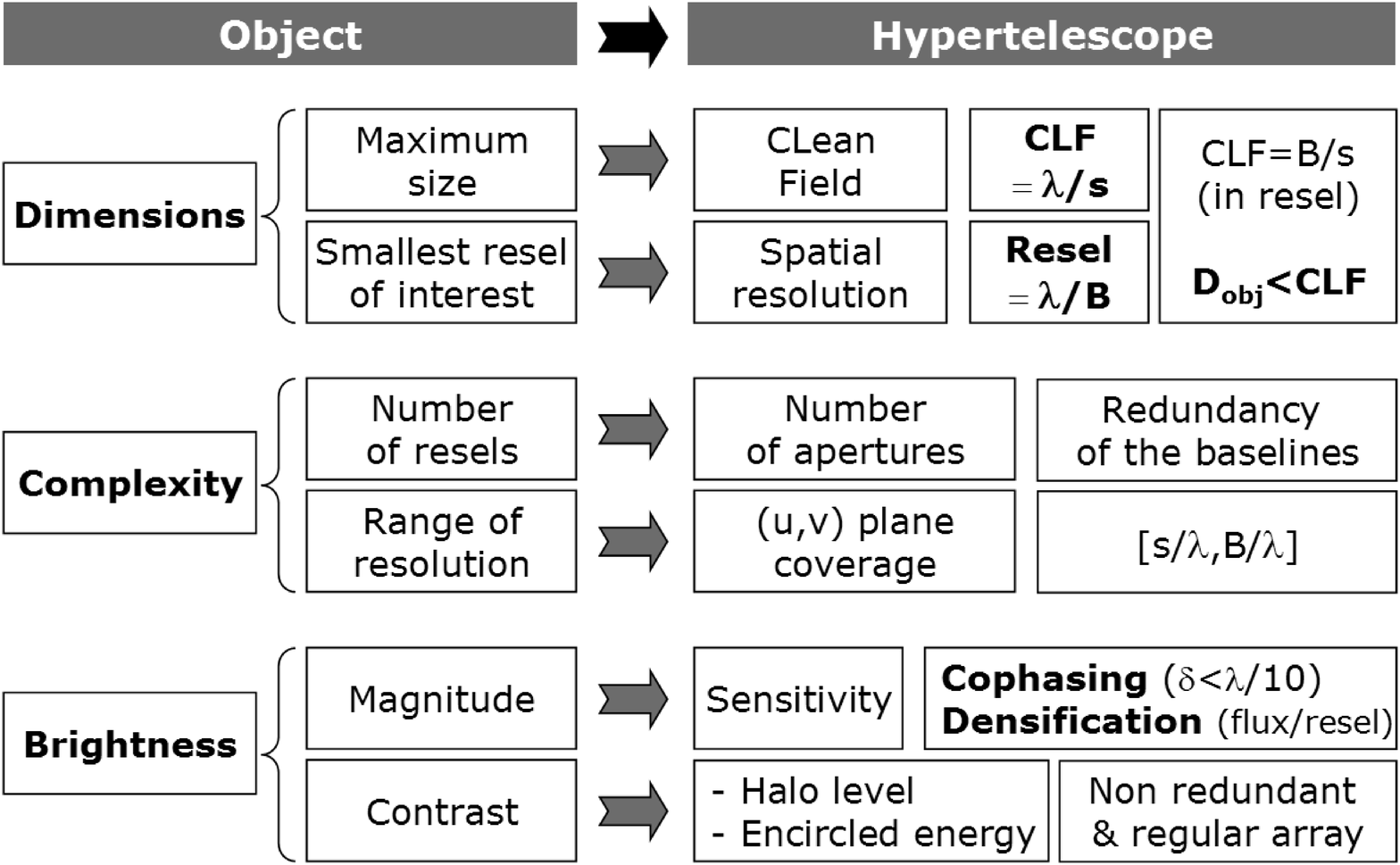}
\end{tabular}
\end{center}\caption{Instrumental parameters of a hypertelescope vs
astrophysical parameters of the science object.}
\label{fig:ht_vs_object}
\end{figure}

Figure \ref{fig:ht_vs_object} gives the links between the science
object and the hypertelescope's required characteristics. An
astrophysical object is characterized by its dimensions, its
complexity and its brightness. These characteristics are
linked to the main parameters of a hypertelescope: field,
resolution, pupil pattern.

The main dimensions of the object are the external diameter and the
smallest resel of interest. The maximum size of the object
should not exceed the diameter of the clean field, which leads to
the value of the minimum baseline of the array ($CLF=\lambda/s$).
The smallest resel corresponds to the required resolving power,
which imposes the largest baseline B of the array
($resel=\lambda/B$). The clean field can also be expressed in number
of resels ($CLF=B/s$).

The complexity of the object determines the required number of
resels in the image, the number of telescopes and the array
geometry. The other aspect to be considered is the resolution range
of the object on the interval $[\lambda/B,\lambda/s]$.

The limiting magnitude of an array is directly related to the
performances of the cophasing device, allowing long exposures.
The image quality (highest encircled energy in the central
peak and lowest halo level in the CLean Field) is directly related
to the densified pupil filling rate.

\section[]{Conclusion}

Simulations have shown that the choice of the array configuration
(array pattern and number of apertures) is a trade-off between the
resolution, the halo level and the useful field. The spatial
resolution is given by the largest baseline ($resel=\lambda/B$). The
clean field is a function of the smallest baseline
($CLF=\lambda/s$). The halo level and the encircled energy in the central peak are a function of the densified pupil filling rate $\tau_o$.
The sine qua none condition to image a complex
source without space aliasing effect is that the object diameter should
not exceed the clean field width ($\theta_{obj}<\lambda/s$).

Concerning the beam combiner, it has been shown that the maximum densification
is optimal in term of sensitivity, by equalizing the direct imaging
field (DIF) with the clean field.

Simulations have shown that the configurations KEOPS and
CARLINA are equivalent as regards the image characteristics.
However, KEOPS is less sensitive to the space aliasing effect, contrary to
CARLINA where ghost stars are introduced by the pollution of the
surrounding sources. Thus, a non redundant array is required for
Direct Imaging to minimize the space aliasing effect.

Finally, the best configuration seems to be the one proposed by
KEOPS, which has a regular and non redundant layout of the
telescopes.
This configuration provides the best quality of the interference function, in
comparison with the interference function of the equivalent giant
telescope. Indeed, it provides the lowest halo level (inside the
Clean Field), at the limit of the diffraction of such kind of array.
Moreover, minimizing the halo level improves the signal to noise
ratio, which should simplify the deconvolution process.

Thus, a KEOPS configuration seems to be suited for
high-contrast imaging of compact sources. An
OVLA configuration can be used for wide field imaging, providing a
larger clean field and the best resolving power.

This paper was mainly focused on future large arrays with a
large number of sub-apertures. However, the direct imaging technique
already has a great interest for current operating interferometers,
using an efficient cophasing system. Indeed, the densification may
provide the ultimate sensitivity.

It will be also interesting in the future to compare the
performances of direct imaging and aperture synthesis. The
introduction of the fundamental and instrumental noises is also
mandatory for a correct evaluation of the scientific performances.
Finally this work will be
developed in two main directions: the study of the
instrumental performances of direct imaging arrays when coupled
with integral field spectrometers or coronagraphs, and the
comparison of the practical imaging performances of various array
configurations when deconvolution is applied.

\section*{Acknowledgments}
The authors wish to thank the referee, Chris Haniff, for
important suggestions and clarifications.

\label{lastpage}

\end{document}